\documentclass[12pt]{article}

\pdfoutput=1
\usepackage{ifpdf}
\usepackage[a4paper,top=2.8cm,width=17cm, bottom=3cm, left=2cm]{geometry}
\usepackage{graphicx}
\usepackage{microtype}
\usepackage[utf8x]{inputenc}
\usepackage[T1]{fontenc}
\usepackage{ae}
\usepackage{aecompl}
\usepackage{amsmath}
\usepackage{amssymb}
\usepackage{mathdots}
\usepackage{setspace}
\usepackage{cite}
\def\bfseries{\fontseries \bfdefault \selectfont \boldmath}

\usepackage[english]{babel}
\usepackage{datetime}
\shortdate 
\usepackage{bm}
\usepackage{subfigure}

\usepackage[normalem]{ulem} 
\usepackage{soul} 
\usepackage[titletoc,title]{appendix}
\usepackage{authblk}
\usepackage{xcolor,soul}
\usepackage[colorlinks=true,linktocpage=true,linkcolor=blue,citecolor=blue]{hyperref}

\usepackage{titlesec}

\titleformat*{\section}{\large\bfseries}
\titleformat*{\subsection}{\bfseries}
\titleformat*{\subsubsection}{\bfseries}

\newcommand{\bes}{\begin{eqnarray*}}
\newcommand{\ees}{\end{eqnarray*}}

\newcommand{\bel}[1]{\begin{eqnarray}\label{#1}}
\newcommand{\be}{\begin{eqnarray}}
\newcommand{\ee}{\end{eqnarray}}

\newcommand{\rf}[1]{Eq.~(\ref{#1})}

\newcommand{\rfs}[1]{Section~\ref{#1}}
\newcommand{\rfss}[1]{Subsection~\ref{#1}}
\newcommand{\rff}[1]{Fig.~\ref{#1}}
\newcommand{\rfc}[1]{Ref.~\cite{#1}}
\newcommand{\rfApp}[1]{Appendix~\ref{#1}}

\newcommand{\f}[2]{\frac{#1}{#2}}

\newcommand{\rmi}{\mathrm{i}}

\newcommand{\sym}{${\mathcal N}=4$}
\newcommand{\symm}{${\mathcal N}=4$ SYM}

\begin{document}

\title{{\bf The large proper-time expansion of Yang-Mills plasma as a resurgent transseries}}

\renewcommand\Authfont{\scshape\small}

\author[1,2]{In\^{e}s Aniceto\thanks{I.Aniceto@soton.ac.uk}}

\author[3]{Jakub Jankowski\thanks{Jakub.Jankowski@fuw.edu.pl}}

\author[4]{Ben Meiring\thanks{Ben.Meiring@physics.ox.ac.uk}}

\author[5,6]{Micha\l{} Spali\'nski\thanks{M.Spalinski@uwb.edu.pl}}

\affil[1]{Institute of Physics, Jagiellonian University, ul. \L{}ojasiewicza 11, 30-348  Krak\'{o}w, Poland}
\affil[2]{Mathematical Sciences, University of Southampton, Highfield, Southampton SO17 1BJ, UK}
\affil[3]{Faculty of Physics, University of Warsaw, ul. Pasteura 5, 02-093 Warsaw, Poland}
\affil[4]{Rudolf Peierls Centre for Theoretical Physics, University of Oxford,
Parks Rd, Oxford OX1 3PJ, UK}
\affil[5]{Physics Department, University of Bia\l{}ystok, PL-15-245 Bia\l{}ystok, Poland}
\affil[6]{National Center for Nuclear Research, PL-00-681 Warsaw, Poland}
\vspace{15pt}

\date{}

\maketitle
\thispagestyle{empty}

\begin{abstract}

We show that the late-time expansion of the energy density of $\mathcal{N}=4$
supersymmetric Yang-Mills plasma at infinite coupling undergoing Bjorken flow
takes the form of a multi-parameter transseries. Using the AdS/CFT
correspondence we find a gravity solution which supplements the well known large
proper-time expansion by exponentially-suppressed sectors corresponding to
quasinormal modes of the AdS black-brane. The full solution also requires the
presence of further sectors which have a natural interpretation as  couplings
between  these modes. The exponentially-suppressed sectors  represent
nonhydrodynamic contributions to the energy density of the plasma. We use
resurgence techniques on the resulting transseries to show that all the
information encoded in the nonhydrodynamic sectors can be recovered from the
original hydrodynamic gradient expansion.

\end{abstract}

\vspace{5pc}
\noindent{\it Keywords}: quark-gluon plasma, relativistic viscous hydrodynamics, holography, resurgence.


\newpage

\tableofcontents

\section{Introduction}

The discovery of quark-gluon plasma (QGP) at RHIC and the ongoing studies of
its properties both there and at the LHC have lead to a burst of activity
related to the theoretical description of this new state of matter. Tiny drops
of QGP created in these heavy-ion collision experiments appear initially in
highly non-equilibrium states. Nevertheless, after a short time this system
reaches a state amenable to an effective description formulated in the language
of hydrodynamics, which continues up until the medium hadronizes as the local
effective temperature drops below the confinement scale. Recent activity aimed
at understanding the equilibration of QGP has lead to significant progress
concerning the foundational aspects of relativistic hydrodynamics (for reviews
see \textit{e.g.} Refs.~\cite{Florkowski:2017olj,Romatschke:2017ejr}), such as
the regime of applicability of
hydrodynamics~\cite{Chesler:2009cy,Heller:2011ju,Jankowski:2014lna,Spalinski:2016fnj,Romatschke:2016hle},
the role and meaning of the hydrodynamic gradient expansion~\cite{Heller:2013fn}
and attractor behaviour far from
equilibrium~\cite{Heller:2015dha,Romatschke:2017vte,Spalinski:2017mel}. The
difficulties of treating real-time evolution far from equilibrium in QCD have
provided strong motivation to look for other systems where such an analysis can
be more tractable. Since many of the fundamental questions concern relativistic
hydrodynamics itself rather than its specific application to QCD, the study of
related model systems has proved both fruitful and practical.

An important framework where many of these ideas were developed is \sym\
supersymmetric Yang-Mills theory (SYM), where the AdS/CFT correspondence
provides an effective method of carrying out ab-initio calculations of highly
non-equilibrium phenomena in a strongly-coupled quantum system by relating
non-trivial observables at infinite coupling to solvable problems in classical
Einstein gravity~\cite{Maldacena:1997re,CasalderreySolana:2011us}. While
supersymmetric Yang-Mills theory is very different from QCD, these differences
are less pronounced at finite temperature and some of the results obtained using
AdS/CFT appear to give a qualitatively useful picture even when interpreted in
the context of QGP. Most importantly, since this approach has allowed for
reliable calculations of equilibration, it has provided a priceless theoretical
laboratory where the emergence of a hydrodynamic behaviour could be explored.

A kinematic situation of phenomenological interest is that of boost-invariant
longitudinal expansion -- Bjorken flow, which mimics the behaviour of matter
during the early stages of an ultra-relativistic heavy ion
collision~\cite{Bjorken:1982qr}. Here, under the assumption of conformality,
symmetry constraints impose that the expectation value of the $3+1$ dimensional
energy-momentum tensor can be expressed in terms of the energy density
$\mathcal{E}$ as a function of a single parameter, the proper time $\tau$
elapsed since the collision event.  This leads to an amazingly simple model, yet
one which preserves much of the essential complexity of the original problem,
making it possible to build a detailed picture of the transition from a highly
non-equilibrium initial state to hydrodynamics.

A key approximation method which makes analytic calculations possible on the
gravity side of the duality is the large proper-time
expansion~\cite{Janik:2005zt}. The bulk geometry obtained in this way implies
that the energy density of \symm\ {\em at late times} takes the form
\bel{gradex}
\mathcal{E}\left(u\right) \equiv \Phi_{\boldsymbol{0}}\left(u\right)=u^{-2}\sum_{k=0}^{+\infty}\varepsilon_{k}^{(\boldsymbol{0})}\,u^{-k}\, ,
\ee
where $u=\tau^{2/3}$, with $\tau$ the proper time. Following
\rfc{Heller:2013fn}, here and in the following we have suppressed a dimensionful
parameter, effectively choosing to measure the proper-time in units set by the
scale of the energy density. This expansion is directly related to the
hydrodynamic gradient expansion of Bjorken flow, which is the expansion in
powers of $\tau\mathcal{E}^{1/4}$.  In fact, given the coefficients of the late
proper-time expansion one can calculate those of the gradient expansion and
vice-versa -- in other words, these two expansions contain the same information.
With this understanding we will refer to \rf{gradex} as the hydrodynamic
expansion.

The dimensionless expansion coefficients $\varepsilon_{k}^{(\boldsymbol{0})}$
are calculable from the gravity solution. The leading 240 coefficients were
computed in Ref.~\cite{Heller:2013fn}, and further 140 were obtained in
Ref.~\cite{Casalderrey-Solana:2017zyh}.  The series appearing in \rf{gradex} is
asymptotic,  and it is natural to ask how to best interpret it.

Already in Ref.~\cite{Heller:2013fn} it was observed that the analytic
continuation of the Borel transform of this series has branch-point
singularities at locations  corresponding to frequencies of the least-damped
nonhydrodynamic quasinormal modes of the AdS black-brane. This connection was
subsequently studied at the level of hydrodynamics, where series expansions such
as \rf{gradex} can be generated as asymptotic solutions to MIS-type differential
equations such as those introduced in Refs.~\cite{Baier:2007ix,Heller:2014wfa}.
These studies~\cite{Heller:2015dha,Aniceto:2015mto} lead to the realisation that
the proper setting for the hydrodynamic gradient expansion is in fact a
transseries~\cite{Edgar:2008tr,Aniceto:2018bis}, which systematically
incorporates the contributions from nonhydrodynamic modes. These contributions
take the form of terms which are exponentially suppressed at large times, but
become significant at earlier times, where they encode the initial state
data.\footnote{It is worth noting that different asymptotic analysis of   QNM
frequencies of black holes can be found in the context of stability of black
holes and their infinitely damped QNMs, see \textit{e.g.}
\cite{Motl:2003cd,Andersson:2003fh,Natario:2004jd}.}

In this paper we present the first calculations which directly reveal the
transseries structure at the level of the microscopic theory. This is done by
finding a generalisation of the late proper-time expansion guided by the form of
asymptotic solutions of the hydrodynamic model studied in \rfc{Aniceto:2015mto}.
However, in contrast to this hydrodynamic model which features just two
conjugate nonhydrodynamic modes, in the case of \symm\ there is an infinite
number of nonhydrodynamic modes, naturally ordered by the rate of exponential
suppression following from the complex values of the black-brane  quasinormal
mode (QNM) spectrum. Each such mode introduces a separate sector -- a separate
asymptotic series weighted by an independent exponential involving the QNM
frequencies. The bulk solution will be described in detail in
\rfs{sec:bulksolution}. Here we will present the ensuing form of the energy
density of \symm.

In order to describe this structure explicitly let us introduce some notation
which we will employ throughout this paper. With each pair of nonhydrodynamic
QNMs (characterised by a pair of complex frequencies $A_k, \overline{A_k}$ with
$k>0$) we associate a pair of unit vectors
$\boldsymbol{e}_{k},\,\overline{\boldsymbol{e}}_{k}$, whose components vanish
apart from a one in slots $2k-1, 2k$ respectively. These unit vectors form a
basis of an infinite-dimensional semi-lattice space $\mathbb{N}_0^{\infty}$. Each QNM
is labelled by a  vector $\boldsymbol{n}$ which is equal to   one of these basis
vectors and introduces an asymptotic expansion of the form,
\bel{phin}
\Phi_{\boldsymbol{n}}\left(u\right) = u^{-\beta_{\boldsymbol{n}}}\sum_{m=0}^{+\infty} \varepsilon_{m}^{(\boldsymbol{n})}\,u^{-m}\,,
\ee
with some {\em characteristic exponent} $\beta_{\boldsymbol{n}}$, and expansion coefficients $\varepsilon_{m}^{(\boldsymbol{n})}$, with the understanding that the leading coefficient  $\varepsilon_{0}^{(\boldsymbol{n})}\neq 0$.
One can view this series as a hydrodynamic ``dressing'' of the individual QNMs.

The contribution of each pair of QNMs comes with a pair of complex normalisation constants, the {\em transseries parameters}, which will be denoted by $\sigma_k, \sigma_{\bar{k}}$. These numbers can be interpreted as integration constants -- they contain information about the initial state.
We find that the energy density at large proper time $\tau$ (or, equivalently, for large $u$) has an expansion as a transseries of the form:
\bel{edens}
\mathcal{E}\left(u,\boldsymbol{\sigma}\right)=\sum_{\boldsymbol{n}\in\mathbb{N}_{0}^{\infty}}\boldsymbol{\sigma}^{\boldsymbol{n}}\,\mathrm{e}^{-\boldsymbol{n}\cdot\boldsymbol{A}\,u}\,\Phi_{\boldsymbol{n}}\left(u\right)~.
\label{eq:trans-en-dens}
\ee
where
\be
\boldsymbol{\sigma}^{\boldsymbol{n}} \equiv \sigma_{1}^{n_{1}}\,\sigma_{\overline{1}}^{n_{\overline{1}}}\,\sigma_{2}^{n_{2}}\,\sigma_{\overline{2}}^{n_{\overline{2}}}\,\cdots
\label{eq:sigmas-def}
\ee
The {\em exponential weights} appearing in \rf{edens} are expressed in terms of the vector of QNM frequencies\footnote{The $A_k,\overline{A_k}$ and the QNM frequencies are simply related by a proportionality constant (see \rfs{sec:res-fund-sectors}).}
\begin{equation}
\boldsymbol{A}=\left(A_{1},\overline{A_1},A_{2},\overline{A_2},\cdots\right)~,
\label{eq:vector-QNM-freq}
\end{equation}
and vectors
\bel{ndef}
\boldsymbol{n}=\left(n_{1},n_{\overline{1}},n_{2},n_{\overline{2}},\cdots\right)
\ee
with non-negative integer entries. Note that the vector of QNM frequencies in Eq.~\eqref{eq:vector-QNM-freq} determines
the form of Eq. \eqref{eq:sigmas-def}, which encodes the initial data of the boundary theory through the values of the
transseries parameters.

The collections of contributions to \rf{edens} with $\boldsymbol{n}$ equal to one of the basis vectors
$\boldsymbol{e}_{k},\,\overline{\boldsymbol{e}}_{k}$ will be referred to as {\em fundamental sectors} -- each such
sector corresponds to a specific QNM. Note however that the sum appearing in \rf{edens} also includes terms with vectors
$\boldsymbol{n}$ corresponding to linear combinations of the basis vectors with non-negative coefficients. These
contributions will be referred to as {\em mixed sectors}. They can be thought of as a reflection of QNM coupling. We
will be referring to all of these as nonhydrodynamic sectors. The contribution to \rf{edens} corresponding to
$\boldsymbol{n}=\boldsymbol{0}$  (with $\beta_{\boldsymbol{0}}=2$) is the original hydrodynamic expansion seen in
\rf{gradex}.

This type of (resurgent) transseries structure appears often in studies of asymptotic  expansions, and the expansion coefficients appearing in different sectors are known to be related by intricate consistency conditions, the so-called \textit{large-order relations}, thoroughly described by Écalle's theory of resurgence \cite{Ecalle:8185} (see also \textit{e.g.} the review \cite{Aniceto:2018bis} and references therein).  These remarkable relations in principle allow us to extract the full content of the nonhydrodynamic sectors from the original hydrodynamic series. In practice, one can use them as a consistency check of the nonhydrodynamic sectors directly determined from the bulk solution. Such relations provide conclusive evidence that the transseries solution, supplemented by the relevant initial conditions, captured the full non-perturbative behaviour of the fluid. These large-order relations have successfully been used to predict novel non-perturbative physics \cite{Garoufalidis:2010ya,Aniceto:2011nu,Schiappa:2013opa,Aniceto:2015rua,Dorigoni:2015dha,Arutyunov:2016etw,Couso-Santamaria:2016vcc} as well as in consistency checks of perturbative and non-perturbative sectors appearing in transseries solutions \cite{Bender:1990pd,Marino:2007te,Basar:2013eka,Couso-Santamaria:2014iia,Basar:2015ava,Aniceto:2015mto,Aniceto:2017me,Codesido:2016dld,Demulder:2016mja,Gukov:2016njj,Dorigoni:2017smz,Codesido:2017jwp}.

Our findings are 
reminiscent of what is known about such series in the context of coupling-constant perturbative expansions in quantum mechanics, where these non-perturbative, exponentially suppressed sectors correspond to specific instanton solutions and are usually referred to as instanton sectors.
It is well known that one needs more than just the instanton sectors as non-perturbative corrections to perturbative calculations of energy levels in quantum-mechanical models -- one also needs multi-instanton sectors, which correspond to nonlinear effects arising from the quantisation condition (see \textit{e.g.} Refs. \cite{Balian:1978pv,ZinnJustin:1980uk} and Refs. within \cite{Aniceto:2018bis}). In the case at hand, the analogue of multi-instanton sectors are the mixed nonhydrodynamic sectors described above. Their content corresponds to going beyond the linearised approximation which gives rise to the QNM spectrum.
From the point of view of resurgence, these additional sectors are unavoidable and are constrained by the aforementioned large-order relations. Verifying that they are satisfied gives us a very high level of confidence in the consistency of the transseries ansatz in the present context.

In \rfs{sec:bulksolution} we construct a transseries solution for the gravitational dual to the Bjorken expansion of $\mathcal{N}=4$ SYM plasma by supplementing the large proper-time expansion with exponentially suppressed contributions, and in \rfss{subSec:enden} we extract the coefficients of the dual gauge theory for several transseries sectors of Eq. \eqref{eq:trans-en-dens} to high orders. The asymptotic expansions of these sectors  show similarities with those arising in the context of the hydrodynamic model of Ref.~\cite{Heller:2014wfa}, whose asymptotic behaviour was investigated in Ref.~\cite{Aniceto:2015mto}. In particular, the large-order behaviour of the coefficients in these series involve  various factorial and exponential growths with complex parameters (related to the exponential weights and characteristic exponents), which make it impossible to apply standard tools used in many recent studies of asymptotic series. The study of the large-order behaviour of these coefficients can be found in \rfs{sec:transeries}, where we introduce a novel method for reliably extracting resurgent information from an asymptotic series whose leading large-order relations depend on multiple complex factorial growths. The numerical checks showing the necessity of the full transseries solution  Eq.~\eqref{eq:trans-en-dens} to describe the energy density of the $\mathcal{N}=4$ SYM plasma can be then found in Secs. \ref{sec:res-fund-sectors} (for evidence of inclusion of fundamental sectors) and \ref{sec:res-mixed-sector} (for the inclusion of mixed sectors).

This work offers a full non-perturbative picture of the late time expansion of the $\mathcal{N}=4$ SYM plasma's energy density,  paving the way to use these new techniques in more general situations. In \rfs{sec:outlook} we summarise our main results and discuss future applications of our work.

\section{The bulk gravity solution and energy density}
\label{sec:bulksolution}


The late proper-time expansion for Bjorken flow in $\mathcal{N}=4$ SYM was studied for the first time analytically in \cite{Janik:2005zt} (see also Refs.~\cite{Janik:2006ft,Heller:2007qt,Booth:2009ct}) and calculated numerically to high orders in Refs.~\cite{Heller:2013fn,Casalderrey-Solana:2017zyh}. We adopt the Eddington-Finkelstein (EF) coordinate system which implements the symmetries of Bjorken flow with the following ansatz~\cite{Heller:2008mb,Kinoshita:2008dq,Chesler:2009cy,Chesler:2013lia}
\begin{equation} \label{eq:CYmetric}
ds^2 = - H(r,\tau)d\tau^2 + 2 dr d \tau + S(r,\tau)^2 \left( e^{-2 B(r,\tau)} d y^2 + e^{B(r,\tau)} d x_{\perp}^2 \right).
\end{equation}
Here $r$ is the holographic radial co-ordinate with $0\le r \le \infty$ and $(\tau, y, x_1, x_2)$ reduce to proper time, rapidity and transverse spatial dimensions on the $r \rightarrow \infty$ conformal boundary.

Einstein's equations can be written in the compact form \cite{Chesler:2013lia},\footnote{Einstein's equations with cosmological constant $\Lambda = -6$ are given by
$R_{\mu\nu} + 4 g_{\mu\nu} = 0$.}
\begin{align}
 S'' & = -\frac{1}{2} S \left(B'\right)^2 \, , \label{eq:CY1}\\
 S \dot{S}' & = 2 S^2- 2\dot{S} S' \, , \label{eq:CY2} \\
 S \dot{B}' & = -\frac{3}{2} \left( \dot{S} B'+\dot{B} S'\right) \, , \label{eq:CY3} \\
 H'' & = -3 \dot{B} B'-4 + 12 \frac{\dot{S} S'}{S^2} \, , \label{eq:CY4} \\
 \ddot{S} & = \frac{1}{2}\left(\dot{S} H'-\dot{B}^2 S\right) \, , \label{eq:CY5}
\end{align}
where $f' = \partial_r f$ and $\dot{f} = \left(\partial_{\tau} + \frac{1}{2} H(r,\tau) \partial_{r} \right) f$. Through the re-definitions
\begin{align}
H(r,\tau) & = r^2 h(r,\tau) \label{eq:metric-capH} \\
B(r,\tau) & = \frac{1}{3}\left( \log{\left(\frac{r^2}{(r \tau+1)^2}\right)} + d(r,\tau) -\frac{1}{2} b(r,\tau) \right) \\
S(r,\tau) & = r^{2/3} (1+r \tau)^{1/3} \exp{\left(\frac{1}{3} d(r,\tau) \right)}
\end{align}
we can express Eq. (\ref{eq:CYmetric}) in a form convenient
for the large-$\tau$ expansion
\cite{Kinoshita:2008dq}
\begin{equation} \label{eq:convenientMetric}
ds^2 = - r^2 h d\tau^2 + 2d\tau dr + (r\tau+1)^2 e^b dy^2 + r^2 e^{-\frac{1}{2} b + d} dx_{\perp}^2 ~.
\end{equation}
For Eq. (\ref{eq:convenientMetric}) to reduce to a flat, boost-invariant solution at the boundary we enforce the conditions
\begin{equation}
\lim_{r\rightarrow\infty} h(r,\tau)  = 1 ~, \hspace{22pt}
\lim_{r\rightarrow\infty} b(r,\tau)  = 0 ~, \hspace{22pt}
\lim_{r\rightarrow\infty} d(r,\tau)  = 0 ~.
\end{equation}
For ideal Bjorken flow, the local effective temperature of the plasma at the $4$-dimensional Minkowski space boundary should behave as $T \sim \tau^{-\frac{1}{3}}$ at late times.
We therefore fix $s = \frac{1}{r} \tau^{-\frac{1}{3}}$
so that in the late time limit the naive location of the horizon in the $s$
co-ordinate will remain finite. Further, we note that gradient corrections
should go like $\frac{1}{\tau T} \sim \tau^{-\frac{2}{3}}$, so we fix $u =
\tau^{\frac{2}{3}}$ and expand the metric functions in inverse powers of $u$.
Note also that $0 < s <1$, $u > 0$, and that late times correspond to large $u$.

Having in mind the hydrodynamic expansion as well as presence
of the transient modes
we consider a transseries ansatz for the metric functions of the form
\begin{align}
h(r,\tau) & = \sum_{\boldsymbol{n}\in\mathbb{N}_{0}^{\infty}}\,\Omega_{\boldsymbol{n}}(u) \sum_{i=0}^\infty u^{-i} h_{i}^{(\boldsymbol{n})}(s)~,
\label{eq:Aexp} \\
b(r,\tau) & = \sum_{\boldsymbol{n}\in\mathbb{N}_{0}^{\infty}}\,\Omega_{\boldsymbol{n}}(u) \sum_{i=0}^\infty u^{-i} b_{i}^{(\boldsymbol{n})}(s)~,
\label{eq:bexp} \\
d(r,\tau) & = \sum_{\boldsymbol{n}\in\mathbb{N}_{0}^{\infty}}\,\Omega_{\boldsymbol{n}}(u) \sum_{i=0}^\infty u^{-i} d_{i}^{(\boldsymbol{n})}(s)~~.
\label{eq:dexp}
\end{align}
where
\begin{equation}
\Omega_{\boldsymbol{n}}(u) \equiv u^{- {\boldsymbol{n}} \cdot \boldsymbol{\alpha}} \,\mathrm{e}^{-\boldsymbol{n}\cdot\boldsymbol{A}\, u}~.
\end{equation}
Here (as already described in the introduction)
$\boldsymbol{n}=\left(n_{1},n_{\overline{1}},n_{2},n_{\overline{2}},\cdots\right)\in\mathbb{N}_{0}^{\infty}$ is an infinite dimensional vector of non-negative, integer
components which will be used to define the space of solutions. We will use the field equations to determine the quantities $\boldsymbol{A}=\left(A_{1},\overline{A_1},A_{2},\overline{A_2},\cdots\right)$ and  $\boldsymbol{\alpha}$ which correspond to particular nonhydrodynamic sectors.\footnote{Note that we will assume that each component of $\boldsymbol{A}$ is not a rational multiple of another component.} We will denote the hydrodynamic sector by
$\boldsymbol{0}$ and define special unit vectors  $\boldsymbol{e}_{k}$ which have all zero entries except the $2k-1$ entry which will be set equal to 1. Similarly, $\overline{\boldsymbol{e}}_{k}$ is defined to have only the $2k$ entry non-zero (and equal to 1). These are defined so that the scalar products with $\boldsymbol{A}$ select the corresponding elements, in the sense that  $\boldsymbol{e}_{k}\cdot\boldsymbol{A}=A_k$, and $\overline{\boldsymbol{e}}_{k}\cdot\boldsymbol{A}=\overline{A_k}$.

After substituting our ansatz into Eqs. (\ref{eq:CY1}) to (\ref{eq:CY5}), the linear independence of $\Omega_{\boldsymbol{n}}(u)$ will imply an infinite hierarchy of equations. These equations can again be expanded in inverse powers of $u$ to find linear ODEs for the functions $h_{i}^{(\boldsymbol{n})}$, $b_{i}^{(\boldsymbol{n})}$ and $d_{i}^{(\boldsymbol{n})}$.

We use residual gauge invariance to set the radial co-ordinate $r$ to keep the horizon fixed at $s = 1$ at every order. This will amount to choosing $h_{i}^{(\boldsymbol{n})}(s=1) = 0$ for all $i$. By imposing regularity in the bulk and flatness at the boundary for the metric functions we are able to solve for functions $f_{i}^{(\boldsymbol{n})}$ at each order.

\subsection{Hydrodynamic and nonhydrodynamic sectors}

The case of $\boldsymbol{n} = \boldsymbol{0}$ will be referred to as the
hydrodynamic sector. Any power of $\Omega_{\boldsymbol{n}\neq
\boldsymbol{0}}(u)$ that enters into the equations of motion  will remain
linearly independent from terms proportional to $\Omega_{\boldsymbol{0}}(u)$.
Therefore $h_{i}^{(\boldsymbol{0})}$, $b_{i}^{(\boldsymbol{0})}$ and
$d_{i}^{(\boldsymbol{0})}$ can be found independent of the solutions to any
other sector. The zero-th order solution (in our chosen gauge) that preserves
flatness and bulk regularity will be given by a boosted black-brane written as
\cite{Janik:2006gp},
\begin{equation}
d_{0}^{(\boldsymbol{0})}(s)  = 0~,\hspace{14pt}
h_{0}^{(\boldsymbol{0})}(s)  = 1-s^4 ~,\hspace{14pt}
b_{0}^{(\boldsymbol{0})}(s)  = 0 ~.
\end{equation}
The cases of $\boldsymbol{n} = \boldsymbol{e}_k$ and $\boldsymbol{n} = \overline{\boldsymbol{e}}_k$ will be called the fundamental nonhydrodynamic sectors.\footnote{In the discussion that follows one can exchange  $\boldsymbol{e}_k$ for $\overline{\boldsymbol{e}}_{k}$ where appropriate.} Equations linear in $\Omega_{\boldsymbol{e}_k}(u)$ can depend on the hydrodynamic sector and solutions within its own nonhydrodynamic sector. The zero-th order solution in each case is given by,
\begin{equation}
d_{0}^{(\boldsymbol{e}_k)}(s)  = 0~,\hspace{14pt}
h_{0}^{(\boldsymbol{e}_k)}(s)  = 0~,\hspace{14pt}
b_{0}^{(\boldsymbol{e}_k)}(s)  = Z_{\boldsymbol{e}_k}(s)~,
\end{equation}
where $Z_{\boldsymbol{e}_k}(s)$ satisfies
\begin{equation}
\left(s(1-s^4) \partial_{s}^{2}-(3+s^4 - 2\, \rmi \, \boldsymbol{e}_{k}\cdot \boldsymbol{\omega} \, s)\partial_{s} - 3\,  \rmi \, \boldsymbol{e}_{k}\cdot \boldsymbol{\omega} \right) Z_{\boldsymbol{e}_{k}}(s) = 0~, \label{qnmEqn}
\end{equation}
and where we have defined $\boldsymbol{\omega}=-\frac{2 \, \rmi }{3}\boldsymbol{A}$ to put Eq. (\ref{qnmEqn}) into the standard form of the Quasinormal Mode (QNM) equation given in infalling EF co-ordinates \cite{Janik:2015waa,Kovtun:2005ev}. Imposing flatness and regularity in
$Z_{\boldsymbol{e}_{k}}(s)$ turns Eq. (\ref{qnmEqn}) into an eigenvalue problem with an infinite number of solutions $\boldsymbol{\omega}$. We take the eigenvalues $\boldsymbol{\omega}$ with negative real part to correspond to the unit vector $\boldsymbol{e}_{k}$, and those with positive real part to correspond to $\overline{\boldsymbol{e}}_{k}$, with index $k$ ordering them naturally by the negative imaginary part of each eigenvalue. The infinite dimensional vectors $\boldsymbol{e}_{k}$ and $\overline{\boldsymbol{e}}_{k}$ can now be understood as spanning the space of QNM frequencies.

Each eigenfunction $Z_{\boldsymbol{e}_{k}}(s)$ is determined up to an arbitrary
integration constant associated with the choice of
$Z_{\boldsymbol{e}_{k}}(s=1)$. This overall normalisation we will later identify
as the transseries parameter. This freedom comes from the fact that (along with
imposing flatness at the boundary) we only require $Z_{\boldsymbol{e}_{k}}(s)$
to be regular in the bulk, allowing a family of solutions which satisfy this
condition. We can interpret each associated sector as an independent
nonhydrodynamic excitation. The vector $\boldsymbol{\alpha}$ is fixed 
through a similar eigenvalue
problem at order $i=1$. Our numerically computed values of $\boldsymbol{\alpha}$ and $\boldsymbol{A}$ satisfy $\boldsymbol{\alpha} = \frac{\boldsymbol{A}}{6}$ with high precision.\footnote{A heuristic argument for this value is suggested by Refs.~\cite{Heller:2013fn, Janik:2006gp}. At late times $\Omega_{\boldsymbol{e}_{k}}(u)$ is expected take the
form of a decaying mode with frequency $\omega_k$ in a slowly evolving plasma of
effective temperature $T \sim \mathcal{E}^{1/4}$ given by,
\begin{equation}
\Omega_{\boldsymbol{e}_{k}}(u) \sim \exp{\left(  \rmi \, \omega_k   \int d \tau
\, T \right)} \, = \exp{\left(  \rmi \, \frac{3}{2} \omega_k
\left(u - \frac{1}{6} \log{u} + \mathcal{O}(u^{-1}) \right) \right)} \, .
\end{equation}
Equating $A_{k} = \frac{3 \, \rmi}{2} \omega_k$ for every $k$, we
can identify $\boldsymbol{\alpha} = \frac{\boldsymbol{A}}{6} \, .$ }
The cases of $\boldsymbol{n} \neq \boldsymbol{e}_{k}$ and $\boldsymbol{n} \neq
\overline{\boldsymbol{e}}_{k}$ will be called the mixed sectors.
Once we have fixed $\boldsymbol{\omega}$ in \rf{qnmEqn}, if we replaced $\boldsymbol{e}_k$ by a general vector $\boldsymbol{n}$, then this equation would have no non-trivial solutions obeying the chosen boundary conditions for $\boldsymbol{n}\ne\boldsymbol{e}_k$.
All functions $h_{i}^{(\boldsymbol{n})}$,
$b_{i}^{(\boldsymbol{n})}$ and $d_{i}^{(\boldsymbol{n})}$ for
$\boldsymbol{n}\neq\boldsymbol{e}_{k},\,\overline{\boldsymbol{e}}_{k}$ and $i
\ge 0$ will be fully determined by solutions in the hydrodynamic and the
fundamental  nonhydrodynamic sectors, and so will contain no free integration
constants. In this way these sectors can be viewed as a cascade of interactions
of the fundamental nonhydrodynamic modes rather than independent solutions.

\subsection{Series solution and numerical implementation}

Subsequent equations of motion for $i \ge 1$ where $\boldsymbol{n} = \boldsymbol{0}$, $\boldsymbol{e}_k$ or $\overline{\boldsymbol{e}}_k$, (and for $i \ge 0$ where $\boldsymbol{n}\neq\boldsymbol{e}_k$ and $\boldsymbol{n}\neq\overline{\boldsymbol{e}}_k$) can be found by directly substituting Eqs. (\ref{eq:Aexp}) to (\ref{eq:dexp}) into Eqs. (\ref{eq:CY1}) to (\ref{eq:CY3}). They can be written respectively as
\begin{align}
\mathcal{L}_{\boldsymbol{n}}^{d} d_{i}^{(\boldsymbol{n})} & = j_{i}^{d,\boldsymbol{n}} \, , \label{dEOM} \\
\mathcal{L}^{h}_{\boldsymbol{n}} h_{i}^{(\boldsymbol{n})} & = j_{i}^{h,\boldsymbol{n}} \, , \label{AEOM} \\
\mathcal{L}^{b}_{\boldsymbol{n}} b_{i}^{(\boldsymbol{n})} & = j_{i}^{b,\boldsymbol{n}} \, , \label{bEOM}
\end{align}
where $j_{i}^{d,\boldsymbol{n}}$, $j_{i}^{h,\boldsymbol{n}}$ and $j_{i}^{b,\boldsymbol{n}}$ are source terms, which are sequentially given in terms of solutions at lower orders in $i$, and solutions at the same order which are determined when solving the equations in the order (\ref{dEOM}) to (\ref{bEOM}). The linear operators above have relatively simple forms,
\begin{align}
\mathcal{L}_{\boldsymbol{n}}^{d} & = \partial_{s}^{2} \, , \label{dlinearOperator} \\
\mathcal{L}_{\boldsymbol{n}}^{h} & = s \partial_{s}- 4 \, , \label{AlinearOperator} \\
\mathcal{L}_{\boldsymbol{n}}^{b} & = s(1-s^4) \partial_{s}^{2}-(3+s^4 - 2 i \, \boldsymbol{n}\cdot \boldsymbol{\omega} \, s)\partial_{s} - 3 i \, \boldsymbol{n}\cdot \boldsymbol{\omega} \, . \label{blinearOperator}
\end{align}
In our chosen gauge, Eq.~(\ref{eq:CY5}) can be written as a constraint at $s=1$,
\begin{equation} \label{eq:constr}
d_{i}^{(\boldsymbol{n})}(1) = J_{i}^{(\boldsymbol{n})} \, ,
\end{equation}
with $J_{i}^{(\boldsymbol{n})}$ a number determined by lower order solutions. Eq. (\ref{eq:CY4}) is redundant, being implied by the other four equations.


The implementation of the numerical methods used in this paper are a continuation of the techniques used in Ref. \cite{Heller:2013fn}. To integrate Equations (\ref{dEOM}) to (\ref{bEOM}) we use Chebyshev spectral methods \cite{Grandclement:2007sb} with $400$ grid points over the $s \in [0,1]$ radial co-ordinate with a minimum of $240$ digits of numerical precision.\footnote{In each case the computation could be done on a typical laptop to high orders within several hours.}

As mentioned before, in our gauge we have chosen the warp factor $h(r,t)$ to vanish at the naive location of the horizon ($s=1$) which results in the boundary condition that $h_{i}^{(\boldsymbol{n})}(s=1) = 0$. Imposing the constraint given by Eq. (\ref{eq:constr}), and standard flatness and regularity conditions (that all functions $d_{i}^{(\boldsymbol{n})}$, $h_{i}^{(\boldsymbol{n})}$ and $b_{i}^{(\boldsymbol{n})}$ vanish at $s=0$ and are regular at $s=1$) fully determines the system for $i\ge 1$ for the hydrodynamic sector, $i \ge 2$ for the fundamental nonhydrodynamic sectors, and $i\ge 0$ for the mixed sectors.

A non-obvious implementation is that of $i = 1$ for the fundamental nonhydrodynamic sectors, where the parameter $\boldsymbol{\alpha}$ must be tuned so as to allow a solution for $b_{1}^{(\boldsymbol{e_{k}})}$ which vanishes at $s=0$. For $\boldsymbol{\alpha} = \frac{\boldsymbol{A}}{6}$ this condition is satisfied for any finite value of $b_{1}^{(\boldsymbol{e_{k}})}(s=1)$ at the horizon. For $i \ge 1$, the value of $b_{i}^{(\boldsymbol{e_{k}})}$ at the horizon $(s=1)$ must be chosen so as to fix $b_{i+1}^{(\boldsymbol{e_{k}})}(s=0) = 0$. This was employed in our code via a shooting method.

\subsection{\label{subSec:enden}Energy density of the dual theory}

It was shown in \cite{Janik:2005zt} that imposing the symmetries of conformal  Bjorken flow and the conservation of energy and momentum leads to the following form of the energy-momentum tensor:
\begin{eqnarray}\label{eq:Tmunu}
T_{\mu\nu} & = & \text{diag}\left(\mathcal{E} , \tau^2 p_{L}, p_{T}, p_{T}\right)_{\mu \nu} \, , \\
& = &  \frac{3 N_c ^2}{8 \pi^2} \, \text{diag}\left(f(\tau), - \tau^{3} f'(\tau) - \tau^2 f(\tau) , f(\tau) + \frac{1}{2} \tau f'(\tau) , f(\tau) + \frac{1}{2} \tau f'(\tau)\right)_{\mu\nu} \, . \nonumber
\end{eqnarray}
The particular energy-momentum tensor expectation value in the state of the boundary theory which is dual to our bulk geometry can be evaluated via holographic renormalization \cite{deHaro:2000vlm}, which determines the function $f(\tau)$ appearing in \rf{eq:Tmunu} in terms of the expansion
\begin{equation}
H(r,\tau) = r^2\left( 1 - \frac{f(\tau)}{r^4} + \, ... \, \right) \, .
\end{equation}
Re-expressing our solution in terms of $u$ and $s$ variables we can write,
\begin{equation} \label{eq:endenSeries}
f(\tau)=\sum_{\boldsymbol{n}\in\mathbb{N}_{0}^{\infty}}\boldsymbol{\sigma}^{\boldsymbol{n}}\, \Omega_{\boldsymbol{n}}(u) \, \left( u^{-2} \sum_{i=0} \, f_{i}^{(\boldsymbol{n})} u^{-i} \right)
\end{equation}
where the coefficients $f_{i}^{(\boldsymbol{n})}$ can be read from our gravity solution as
\begin{equation} \label{eq:endenCoeffs}
f_{i}^{(\boldsymbol{n})} =   - \frac{1}{4!}\frac{d^4}{d s^4} h_{i}^{(\boldsymbol{n})}(s=0) \, .
\end{equation}
The factor of $\boldsymbol{\sigma}^{\boldsymbol{n}}$ (see \rf{eq:sigmas-def}) is introduced into Eq.
(\ref{eq:endenSeries}) so that the dependence on the choice of initial condition is explicit in the energy density.
For generic $\boldsymbol{n}$, some number of the first coefficients $f_{i}^{(\boldsymbol{n})}$ in the sum
(\ref{eq:endenCoeffs}) will be zero. For convenience we define $\varepsilon_{0}^{(\boldsymbol{n})}$ to be (up to
normalisation) the first non-zero coefficient of $f_{i}^{(\boldsymbol{n})}$ (with
$\varepsilon_{i}^{(\boldsymbol{n})},\,i>0$ given by the subsequent coefficients), and absorb the shift of the index into
the definition of the characteristic exponent, which will be given by $\beta_{\boldsymbol{n}}$. One then arrives at the
final formula for the energy density in the form of a transseries, given in \rf{edens}, which we repeat here for
convenience, %
\bel{edens2}
\mathcal{E}\left(u,\boldsymbol{\sigma}\right) = \sum_{\boldsymbol{n}\in\mathbb{N}_{0}^{\infty}} \boldsymbol{\sigma}^{\boldsymbol{n}}\, \mathrm{e}^{-\boldsymbol{n}\cdot\boldsymbol{A}\,u}\, \Phi_{\boldsymbol{n}}\left(u\right)~,
\ee
with
\begin{equation}
\Phi_{\boldsymbol{n}}\left(u\right)=u^{-\beta_{\boldsymbol{n}}} \sum_{k=0}^{+\infty}\varepsilon_{k}^{(\boldsymbol{n})}\,u^{-k}\,.
\end{equation}
These coefficients $\varepsilon_{k}^{(\boldsymbol{n})}$ have been included with
this submission for the sectors $\Phi_{\boldsymbol{n}}$ with
$\boldsymbol{n}=\boldsymbol{0},\,\boldsymbol{e}_{1},\,\boldsymbol{e}_{2},\,2
\boldsymbol{e}_{1},\,(\boldsymbol{e}_{1} + \overline{\boldsymbol{e}}_{1}$),
along with their corresponding exponential weights $A_i$ and characteristic
exponents $\beta_{\boldsymbol{n}}$. For the hydrodynamic sector
$\Phi_{\boldsymbol{0}}$ 
coefficients for the hydrodynamic expansion were taken from
\cite{Casalderrey-Solana:2017zyh}.

The normalisations for the hydrodynamic series $\Phi_{\boldsymbol{0}}$ and each
of the fundamental sectors
$\Phi_{\boldsymbol{e}_{k}},\,\Phi_{\overline{\boldsymbol{e}}_{k}}$ ( associated
to each $k^{\mathrm{th}}$ QNM frequency) are not fixed, and have been chosen
such that
\begin{equation}
\varepsilon_{0}^{(\boldsymbol{0})}=\pi^{-4},\,  \hspace{8pt} \varepsilon_{0}^{(\boldsymbol{e}_{k})}=1,  \hspace{8pt} \varepsilon_{0}^{(\overline{\boldsymbol{e}}_{k})}=1,
\hspace{8pt}
k\in\mathbb{N}~.
\end{equation}
\noindent
With this choice of normalisation, the mixed sectors have no freedom in their respective coefficients. The list of sectors we have included are as follows:\label{page:charac-exp}
\begin{itemize}

    \item  The first 250 coefficients of the \textit{fundamental} sector $\Phi_{\boldsymbol{e}_{1}}$, and $\beta_{\boldsymbol{e}_{1}} = -\frac{A_{1}}{6}+3$
    with $A_{1}=\mathrm{i}\,\frac{3}{2}\,\omega_{1}$,
where  $\omega_{1}\approx 3.1195-\mathrm{i}\,2.7467$ is the lowest nonhydro QNM frequency.

    \item The first 200 coefficients of the \textit{fundamental} sector $\Phi_{\boldsymbol{e}_{2}}$, and $\beta_{\boldsymbol{e}_{2}} = -\frac{A_{2}}{6}+3$
    with $A_{2}=\mathrm{i}\,\frac{3}{2}\,\omega_{2}$ where $\omega_{2} \approx 5.1695 - \mathrm{i}\,4.7636$ is the second lowest nonhydro QNM frequency.

    \item The first 100 coefficients of the \textit{mixed} sector $\Phi_{2 \boldsymbol{e}_{1}}$, with $\beta_{2\boldsymbol{e}_{1}}= -2 \frac{A_{1}}{6} + 4 = 2\beta_{\boldsymbol{e}_{1}}-2$.

    \item The first 100 coefficients of the \textit{mixed} sector $ \Phi_{\boldsymbol{e}_{1} + \overline{\boldsymbol{e}}_{1}}$,
    with $\beta_{\boldsymbol{e}_{1} + \overline{\boldsymbol{e}}_{1}}= -\frac{A_{1}}{6} - \frac{\overline{A_{1}}}{6} + 4 = \beta_{\boldsymbol{e}_{1}} + \beta_{\overline{\boldsymbol{e}}_{1}} -2$.

\end{itemize}
Note that the fundamental sectors $\Phi_{\overline{\boldsymbol{e}}_1}$,
$\Phi_{\overline{\boldsymbol{e}}_2}$ are just complex conjugate of the sectors
$\Phi_{\boldsymbol{e}_1}$ and $\Phi_{\boldsymbol{e}_2}$,
respectively. Furthermore, direct calculation shows that
$\Phi_{2\overline{\boldsymbol{e}}_1}$ is complex conjugate of
$\Phi_{2\boldsymbol{e}_1}$. Their respective characteristic exponents
$\beta_{\boldsymbol{n}}$ are also related by complex conjugation (as are their
exponential weights).\footnote{Curiously, the characteristic exponent appears
  to follow the pattern $\beta_{\boldsymbol{n}}= -\frac{\boldsymbol{n}\cdot\boldsymbol{A}}{6}+\boldsymbol{n}\cdot\boldsymbol{1}+\beta_{\boldsymbol{0}}$, with $\boldsymbol{1}=(1,1,1,\cdots)$.}

\section{The resurgent transseries of the energy density}
\label{sec:transeries}

The coefficients appearing in the different sectors of the transseries for the
energy density \eqref{eq:trans-en-dens} calculated in the previous section can
be seen to grow factorially at large-order. Thus, each sector is formally
defined as an asymptotic series.  It is well known that one can remove the
factorial growth by performing a Borel transform in each of the asymptotic
sectors $\Phi_{\boldsymbol{n}}$, which is the first step to the summation of
these series. To actually perform such summation one needs to know the
singularity structure of the analytic continuation of the Borel transform. In
the case at hand this large-order behaviour has an oscillatory component which
prevents the application of standard techniques used to extract  information
from it.  Some methods of approaching this problem were presented in
Ref.~\cite{Aniceto:2015mto}, but we have developed a new, powerful approach
explained in detail in \rfApp{app:app-res}, which we will apply below.

\subsection{The singularities in the Borel plane}

In general terms, the Borel transform corresponds to mapping a divergent series in the variable $u\gg 1$ to a series in the Borel variable $\xi$ via the map $u^{-\alpha}\mapsto \frac{\xi^{\alpha-1}}{\Gamma (\alpha)}$. The series $\Phi_{\boldsymbol{n}}$ (see \rf{phin}) is mapped to
\bel{borel}
\mathcal{B}\left[\Phi_{\boldsymbol{n}}\right](\xi) =  \xi^{\beta_{\boldsymbol{n}}-1}\sum_{k=0}^{+\infty} \f{\varepsilon_{k}^{(\boldsymbol{n})}}{\Gamma(k+\beta_{\boldsymbol{n}})}\,\xi^{k}\,.
\ee
 This series has a finite radius of convergence and we can analyse the singularity structure of its analytic continuation\footnote{This analytic continuation is performed by means of diagonal Padé approximants.}, which is necessary if we eventually wish to carry out the Borel summation procedure.

\rff{fig:BorelPlanes} shows the poles of the analytic continuation of the Borel transform for the leading asymptotic sectors of our transseries: the hydrodynamic sector $\Phi_{\boldsymbol{0}}$ (top plot), the fundamental sectors $\Phi_{\boldsymbol{e}_1},\,\Phi_{\boldsymbol{e}_2}$ (middle plots) and the mixed sectors $\Phi_{2\boldsymbol{e}_1}$ and $\Phi_{\boldsymbol{e}_1+\overline{\boldsymbol{e}}_1}$ (bottom plots). The condensation of poles is taken to be indicative of a branch point singularity. We can directly check there that the positions of the inferred branch points are  determined by the different exponential weights appearing in our transseries solution \eqref{eq:trans-en-dens}. We can see the appearance of the different fundamental sectors (shown in the figure as filled circles) as well as mixed sectors (shown as filled purple diamonds).


\begin{figure}[tp]
\centering
  \includegraphics[width=0.475\linewidth]{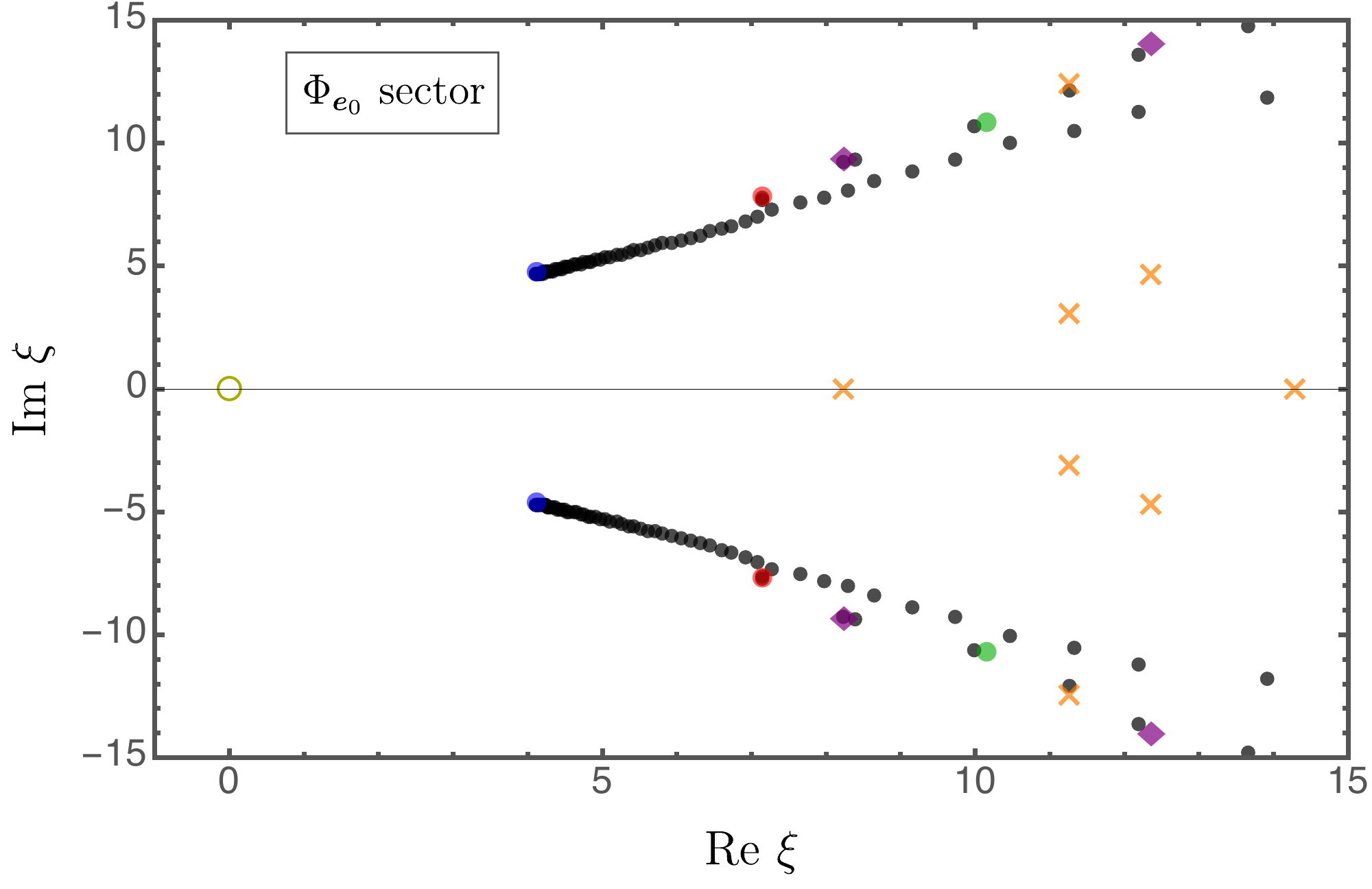} \\
  \includegraphics[width=0.475\linewidth]{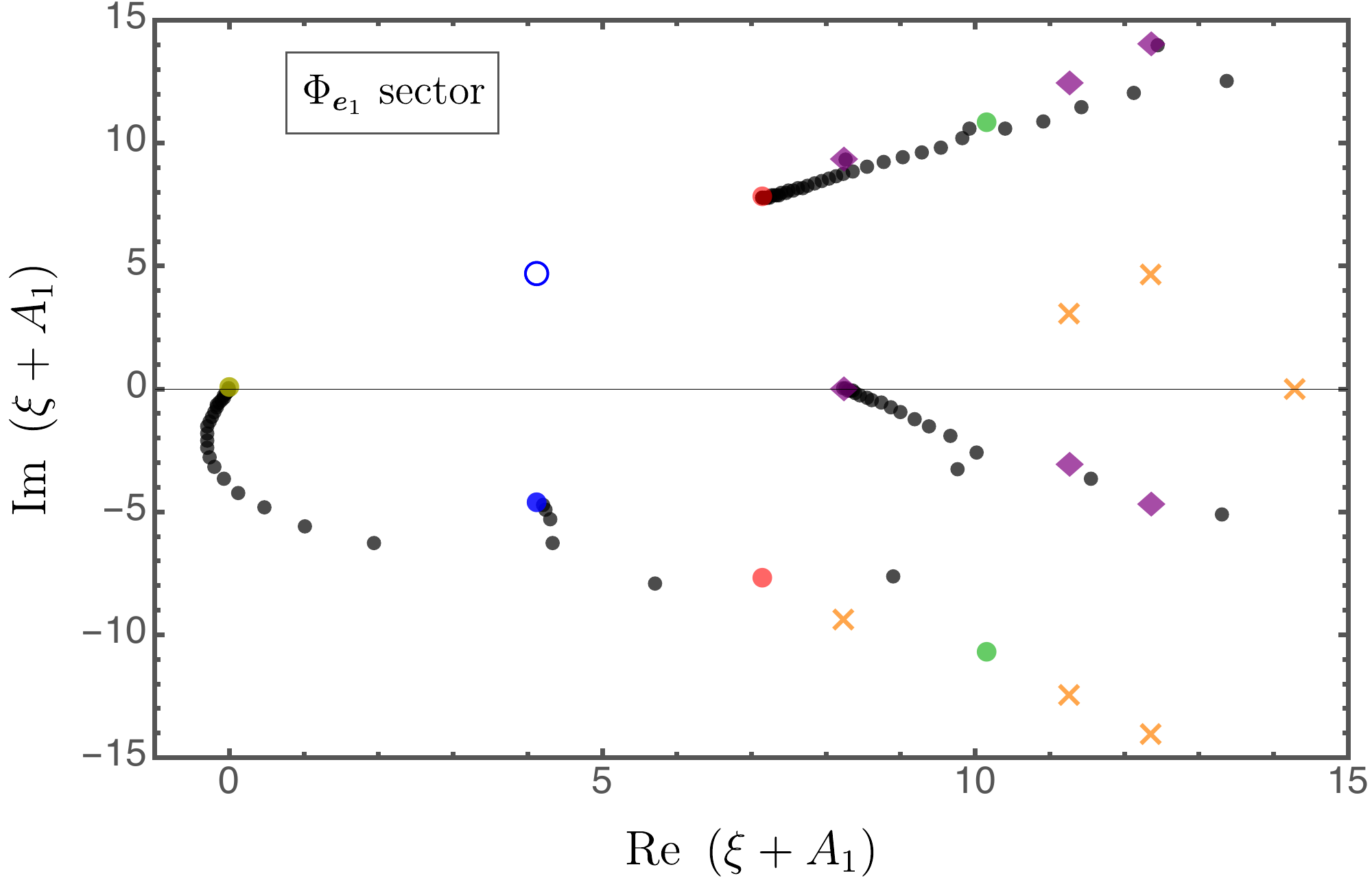} \hspace{15pt}
  \includegraphics[width=0.475\linewidth]{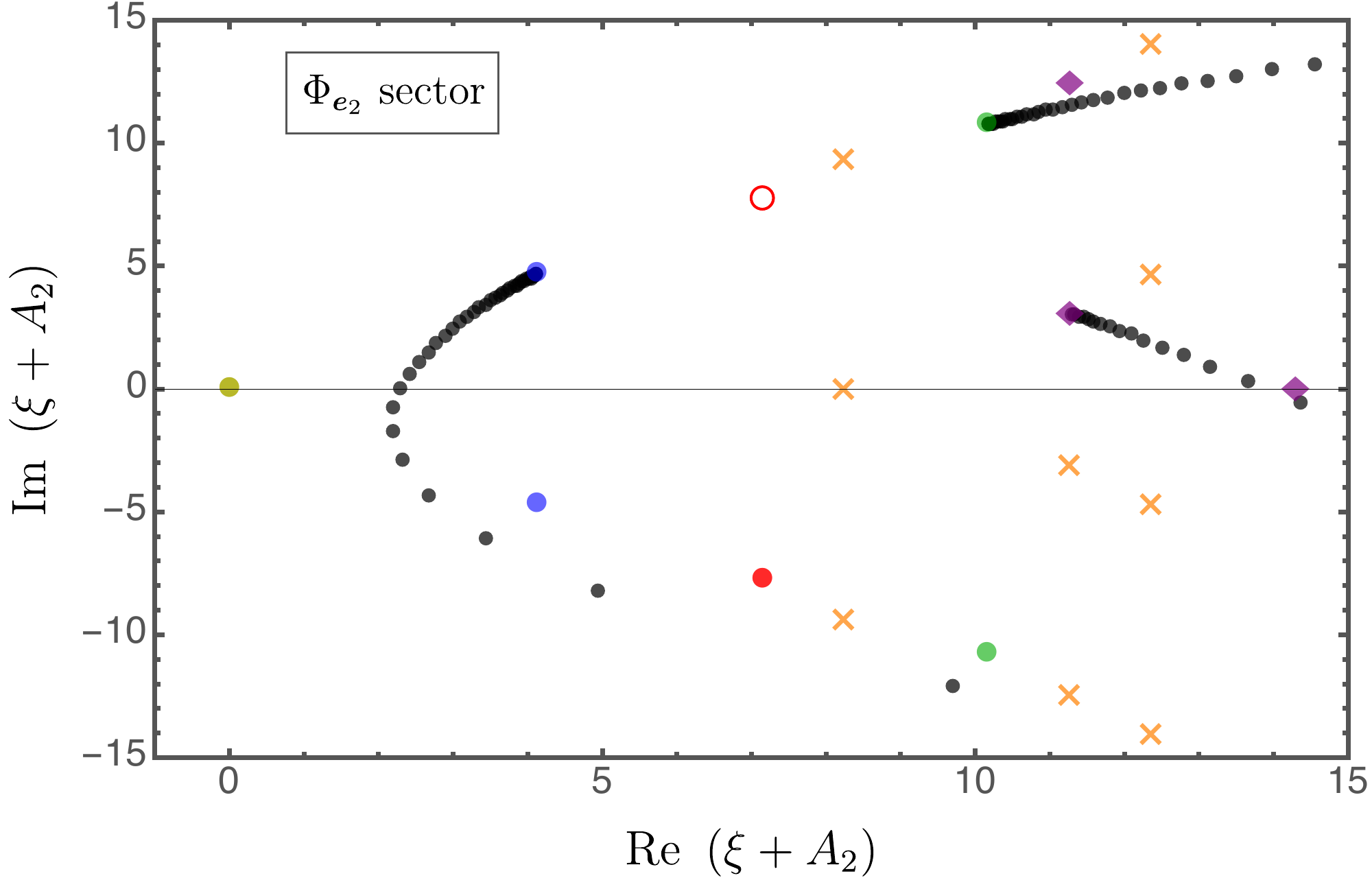} \\
  \includegraphics[width=0.475\linewidth]{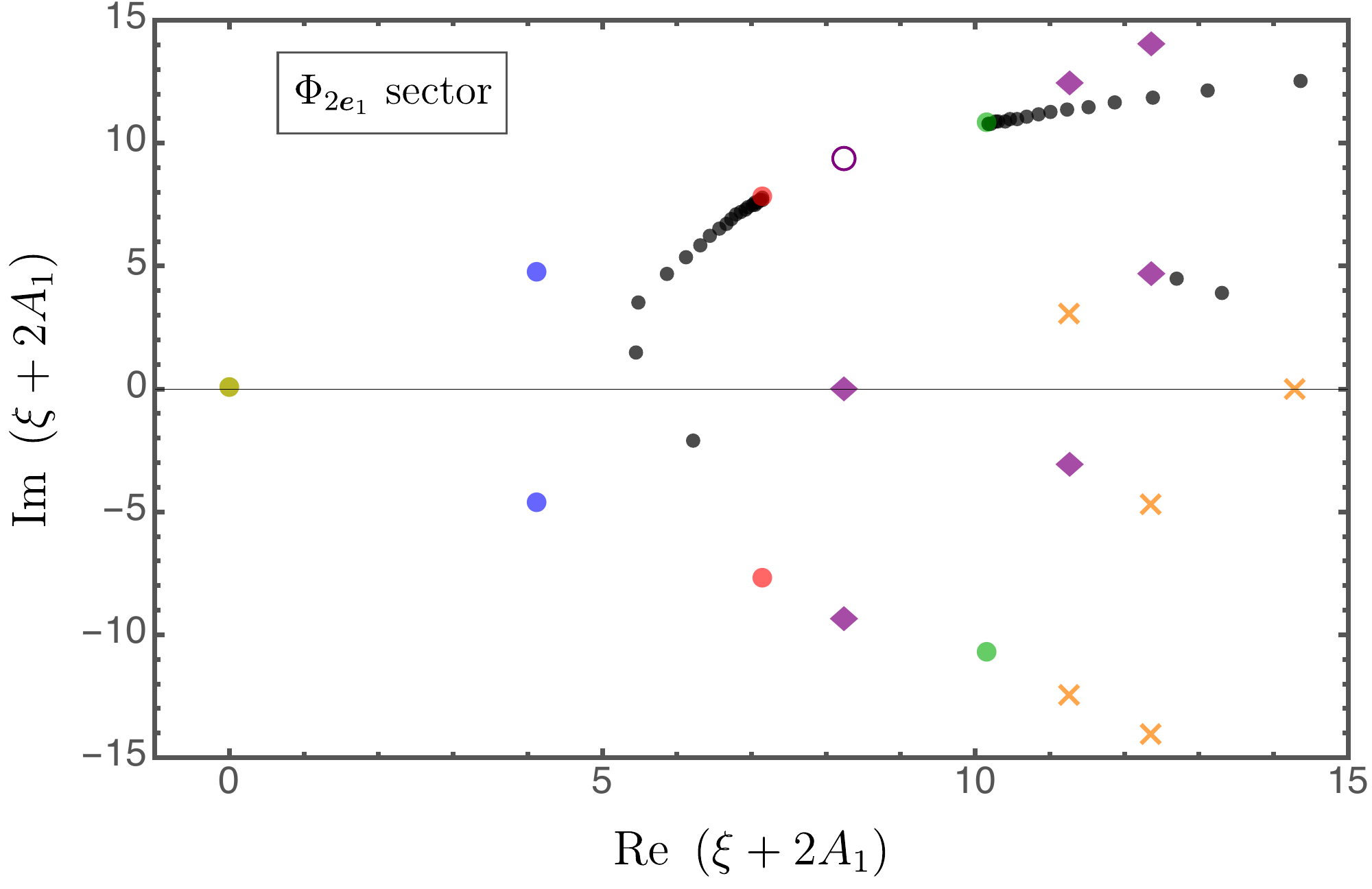}  \hspace{15pt}
\includegraphics[width=0.475\linewidth]{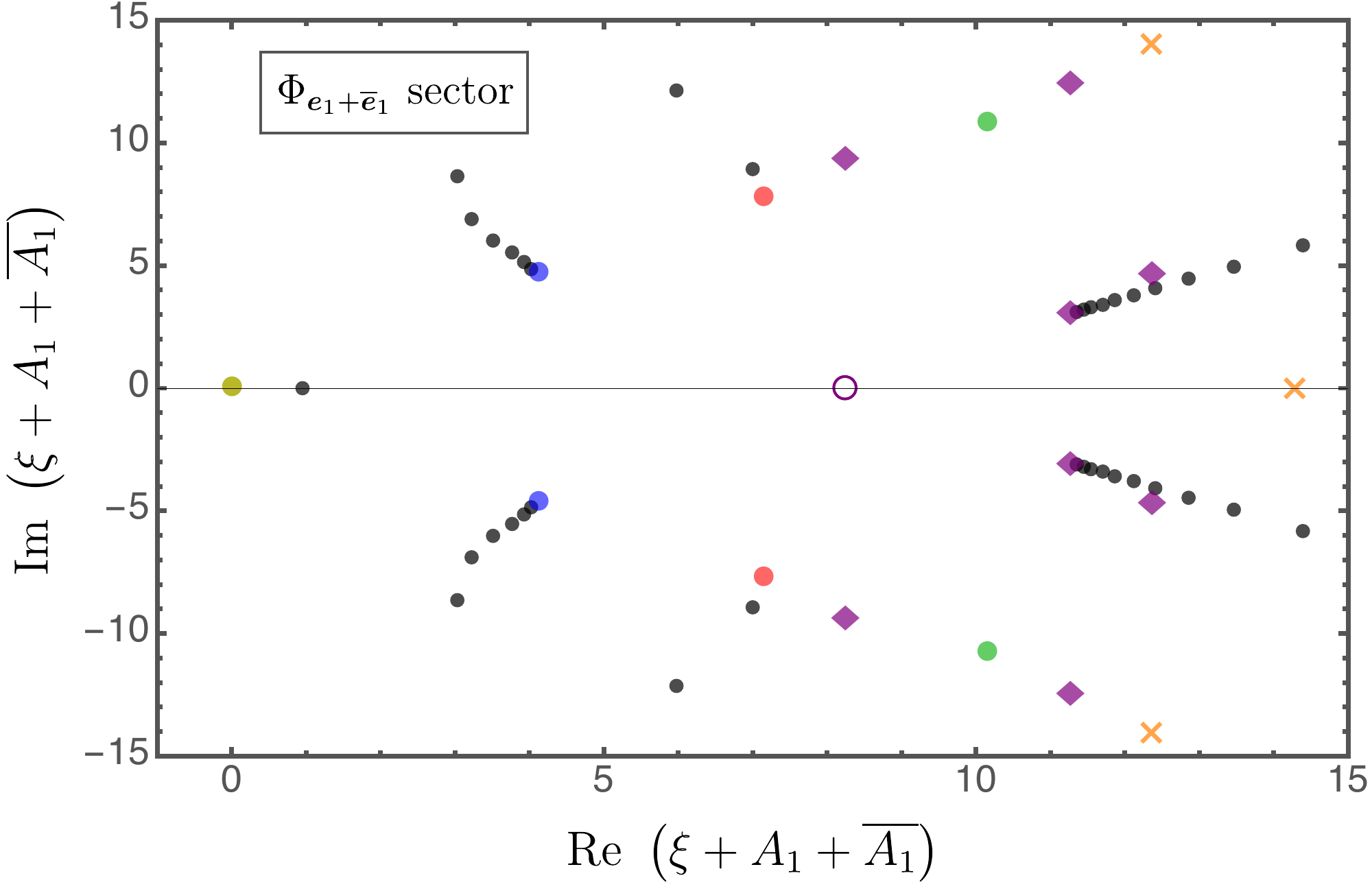}
  \caption{ \label{fig:BorelPlanes} The singularity structure (grey dots) in
    the complex $\xi$-plane (Borel plane) associated to the Borel transforms
    of sectors $\Phi_{\boldsymbol{n}}$, with
    $\boldsymbol{n}=\boldsymbol{0},\,\boldsymbol{e}_{1},\,\boldsymbol{e}_{2},\,2\boldsymbol{e}_{1},\boldsymbol{e}_{1}+\overline{\boldsymbol{e}}_{1}$. This
    singularity structure is obtained via the poles of the respective
    Borel-Pad\'{e} approximant (of order $N$) $\mathrm{BP}_N
    \left[\Phi_{\boldsymbol{n}}\right]$. The Borel-Padé approximants used
    were: $\mathrm{BP}_{189}\left[\Phi_{\boldsymbol{0}}\right]$,
    $\mathrm{BP}_{135}\left[\Phi_{\boldsymbol{e}_1}\right]$,
    $\mathrm{BP}_{129}\left[\Phi_{\boldsymbol{e}_2}\right]$,
    $\mathrm{BP}_{47}\left[\Phi_{2\boldsymbol{e}_1}\right]$ and
    $\mathrm{BP}_{47}\left[\Phi_{\boldsymbol{e}_1+\overline{\boldsymbol{e}}_1}\right]$. The
    Borel planes have been shifted to be represented on a common axis, with
    the expansion points shown by a hollow circle and the hydrodynamic
    expansion  located at the origin (in dark yellow). The predicted branch
    points for each sector (listed in \rfApp{app:app-sing}) have been
    superimposed on each Borel plane
  and are divided into fundamental (solid circles) and mixed sectors (purple diamonds). The fundamental sectors include: $A_{1}$ and $\overline{A_{1}}$ (blue), $A_{2}$ and $\overline{A_{2}}$ (red), $A_{3}$ and $\overline{A_{3}}$ (green). Any branch points associated to sectors of the transseries which do not give origin to branch points are shown as orange crosses.}
\end{figure}

However, not all the mixed sectors appearing in our transseries \eqref{eq:trans-en-dens} will contribute to the singularity structure of a given asymptotic sector. Applying resurgence techniques to our transseries, one can predict all the branch point singularities for the Borel transform of each sector. The procedure is thoroughly explained in section 5 of \cite{Aniceto:2018bis} (for multi-parameter transseries). The full list of the branch point singularities for each sector can be found in \rfApp{app:app-sing}, and these were accordingly marked in Fig. \ref{fig:BorelPlanes} in the form of circles (fundamental) or diamonds (mixed). Analysing the figure further, we note that there are several predicted branch points without any singular behaviour (grey dots). This is due to the fact that we are only taking a finite and limited number of terms from the original series and approximating the analytic continuation of the Borel transform by means of Padé approximants. Depending on the number of terms available the effectiveness with which we can detect branch point singularities in the Borel plane diminishes with distance as one moves away from the origin (or the empty circle in the shifted plots of Fig. \ref{fig:BorelPlanes}).

The Borel plane singularities shown in  \rff{fig:BorelPlanes} have an
unexpected feature, particularly evident in the middle left plot (the Borel
plane singularities of sector $\Phi_{\boldsymbol{e}_1}$): the hydrodynamic
series has a cut associated to it, at the origin of the plot. This is very
surprising: typically, in a resurgent transseries, the perturbative sector has
no transseries parameter associated with it (such as the $\sigma_{i}$ appearing
in \eqref{eq:trans-en-dens}). The appearance of the cut associated to the
hydrodynamic series suggests otherwise. The resolution of this puzzle is very
simple: the energy density as a function of proper time contains in fact a
dimensionful parameter, which we have set to a convenient value (following
\cite{Heller:2013fn}). This parameter is a vestige of the initial conditions and
as such it leads to a branch-point singularity and a cut in the Borel plane. We
have checked that there are no such cuts if instead of the energy density we
consider the pressure anisotropy (defined by $\mathcal{A} =
\frac{p_{T}-p_{L}}{\epsilon/3}$, see also Eq. \eqref{eq:Tmunu}) as a function of
$\tau {\cal E}^{1/4}$ whose hydrodynamic gradient expansion is known to be
universal at late times, and so does not involve any integration
constants.\footnote{From the independence of the pressure anisotropy on initial
conditions we can apply resurgence techniques to determine the \textit{exact}
behaviour of the branch cut appearing at the origin of the middle left plot of
\rff{fig:BorelPlanes}: the asymptotic series associated to it is  related but
not exactly equal to the hydrodynamic series. To do a full resurgent analysis of
the transseries describing the energy density, including the cut at the origin,
one would have to recover the dependence on this dimensionful parameter. For the
purposes of this work we will keep this parameter fixed, thus obtaining the
solution in Eq. \eqref{eq:trans-en-dens}.}

Each of the branch cuts appearing in the analytically continued Borel transforms
contains all the information pertaining to the sector associated with it.
Naturally, one can expect this information to be encoded in the asymptotic
behaviour of the the Borel transform of the hydrodynamic sector,
$\mathcal{B}\left[\Phi_{\boldsymbol{0}}\right]$, as shown on the top plot of
Fig. \ref{fig:BorelPlanes}. In other words, the large-order (factorially
divergent) behaviour of the coefficients of this series will encode information
about all of the fundamental sectors and mixed sectors. This can be most
efficiently and systematically studied through the tools of resurgence, which
when applied to our transseries solution \eqref{eq:trans-en-dens} give rise to
the so-called large-order relations, which express precise relations between the
coefficients of the different transseries sectors. These large-order relations
allow us to very accurately check the consistency of the the transseries ansatz \eqref{eq:trans-en-dens}, including the presence of mixed sectors
reflecting the non-linear QNM coupling. A comprehensive explanation of how to
determine these relations can be found in \cite{Aniceto:2018bis} (sections 2 and
5).

Taking as an example the hydrodynamic series, the large-order behaviour of its coefficients,
implied by the assumption that our transseries is resurgent, is schematically given by:
\begin{eqnarray}
\label{eq:large-order-pert-line1}
\varepsilon_k^{(\boldsymbol{0})} & \underset{k\gg 1}{\simeq} & -\frac{\mathsf{S}_{\boldsymbol{0}\rightarrow \boldsymbol{e}_1}}{2\pi\mathrm {i}}\,\frac{\Gamma(k+\beta_{\boldsymbol{0}}-\beta_{\boldsymbol{e}_1})}{A_1^{k+\beta_{\boldsymbol{0}}-\beta_{\boldsymbol{e}_1}}} \,\chi_{\boldsymbol{0}\rightarrow \boldsymbol{e}_1}
-\frac{\mathsf{S}_{\boldsymbol{0}\rightarrow \overline{\boldsymbol{e}}_1}}{2\pi\mathrm {i}}\,\frac{\Gamma(k+\beta_{\boldsymbol{0}}-\beta_{\overline{\boldsymbol{e}}_1})}{\left(\overline{A_1}\right)^{k+\beta_{\boldsymbol{0}}-\beta_{\overline{\boldsymbol{e}}_1}}} \,\chi_{\boldsymbol{0}\rightarrow \overline{\boldsymbol{e}}_1} + \\
  &  &  -\frac{\mathsf{S}_{\boldsymbol{0}\rightarrow \boldsymbol{e}_2}}{2\pi\mathrm {i}}\,\frac{\Gamma(k+\beta_{\boldsymbol{0}}-\beta_{\boldsymbol{e}_2})}{A_2^{k+\beta_{\boldsymbol{0}}-\beta_{\boldsymbol{e}_2}}} \,\chi_{\boldsymbol{0}\rightarrow \boldsymbol{e}_2}
-\frac{\mathsf{S}_{\boldsymbol{0}\rightarrow \overline{\boldsymbol{e}}_2}}{2\pi\mathrm {i}}\,\frac{\Gamma(k+\beta_{\boldsymbol{0}}-\beta_{\overline{\boldsymbol{e}}_2})}{\left(\overline{A_2}\right)^{k+\beta_{\boldsymbol{0}}-\beta_{\overline{\boldsymbol{e}}_2}}} \,\chi_{\boldsymbol{0}\rightarrow \overline{\boldsymbol{e}}_2} +
\label{eq:large-order-pert-line2} \\
  &  &  -\frac{\mathsf{S}_{\boldsymbol{0}\rightarrow 2\boldsymbol{e}_1}}{2\pi\mathrm {i}}\,\frac{\Gamma(k+\beta_{\boldsymbol{0}}-\beta_{2\boldsymbol{e}_1})}{(2A_1)^{n+\beta_{\boldsymbol{0}}-\beta_{2\boldsymbol{e}_1}}} \,\chi_{\boldsymbol{0}\rightarrow 2\boldsymbol{e}_1}
-\frac{\mathsf{S}_{\boldsymbol{0}\rightarrow 2\overline{\boldsymbol{e}}_1}}{2\pi\mathrm {i}}\,\frac{\Gamma(k+\beta_{\boldsymbol{0}}-\beta_{2\overline{\boldsymbol{e}}_1})}{\left(2\overline{A_1}\right)^{k+\beta_{\boldsymbol{0}}-\beta_{2\overline{\boldsymbol{e}}_1}}} \,\chi_{\boldsymbol{0}\rightarrow 2\overline{\boldsymbol{e}}_1} +
\label{eq:large-order-pert-line3}   \\
  &  &  +\cdots. \nonumber
\end{eqnarray}
Each line above corresponds to different exponentially suppressed contributions
(top line: leading; middle line: first exponentially suppressed; third line:
second exponentially suppressed). The proportionality constants
$\mathsf{S}_{\boldsymbol{0}\rightarrow \boldsymbol{n}}$ are the so-called Borel
residues, which are well known functions of the Stokes constants associated with
the problem (see \cite{Aniceto:2018bis} for the exact relations).\footnote{For
our purposes it is accurate enough to say that  every Stokes constant is equal
to (minus) the Borel residue of the hydro sector to a fundamental sector, given
by $S_{\boldsymbol{e}_i} = -\mathsf{S}_{\boldsymbol{0}\rightarrow
\boldsymbol{e}_{i}}$ and $S_{\overline{\boldsymbol{e}}_i} =
-\mathsf{S}_{\boldsymbol{0}\rightarrow \overline{\boldsymbol{e}}_{i}}$. For a
more general definition given by alien calculus see \cite{Aniceto:2018bis}.}
Naturally, if a particular sector $\Phi_{\boldsymbol{n}}$ does not appear as a
singular branch point in the Borel plane of
$\mathcal{B}\left[\Phi_{\boldsymbol{0}}\right]$, then it won't contribute to the
large-order behaviour of $\varepsilon_k^{(\boldsymbol{0})}$, because the
corresponding  Borel residue vanishes, $\mathsf{S}_{\boldsymbol{0}\rightarrow
\boldsymbol{n}}=0$. The expansions $\chi_{\boldsymbol{0}\rightarrow
\boldsymbol{n}}(k)$, which depend solely on the coefficients of the sector
$\Phi_{\boldsymbol{n}}$ and the leading power of the hydrodynamic expansion
$\beta_{\boldsymbol{0}}$, are asymptotic expansions in $k\gg 1$ and are
generally defined as
\begin{equation}
\chi_{\boldsymbol{m}\rightarrow \boldsymbol{n}}(k)  \simeq \sum_{h=0}^{+\infty}\,\frac{\Gamma(k+\beta_{\boldsymbol{m}}-\beta_{\boldsymbol{n}}-h)}{\Gamma(k+\beta_{\boldsymbol{m}}-\beta_{\boldsymbol{n}})}\,((\boldsymbol{n}-\boldsymbol{m})\cdot\boldsymbol{A})^h\,\varepsilon_h^{(\boldsymbol{n})} \, .
\label{eq:large-ord-asymptotics}
\end{equation}
\noindent
Expanding the above expression at large-order we thus obtain an asymptotic series, \textit{e.g.}
\begin{equation}
\chi_{\boldsymbol{0}\rightarrow \boldsymbol{e}_1}(k)  \simeq
\varepsilon^{(\boldsymbol{e}_1)}_0+\frac{A_1\,\varepsilon^{(\boldsymbol{e}_1)}_1}{k}+\frac{(A_1)^2\,\varepsilon^{(\boldsymbol{e}_1)}_2-A_1\,\left(\beta_{\boldsymbol{0}}-\beta_{\boldsymbol{e}_1}-1\right)\varepsilon^{(\boldsymbol{e}_1)}_1}{k^2}+\cdots .
\end{equation}

Recalling that the coefficients of the hydrodynamic series are all real, from these large-order relations we can predict certain conjugation properties of the Borel residues.
Consider, for instance, the leading contributions to the large-order behaviour \eqref{eq:large-order-pert-line1}: these come from the complex conjugate sectors $\Phi_{\boldsymbol{e}_1}$ and $\Phi_{\overline{\boldsymbol{e}}_1}$, whose contributions are of the same order. To sum these contributions into a real result, we need to have $\mathsf{S}_{\boldsymbol{0}\rightarrow \boldsymbol{e}_1}=-\overline{\mathsf{S}_{\boldsymbol{0}\rightarrow \overline{\boldsymbol{e}}_1}}$.

A question now arises: is it possible to retrieve information concerning the
sub-leading, exponentially suppressed contributions, such as
\eqref{eq:large-order-pert-line2} and \eqref{eq:large-order-pert-line3}? To
reach these exponentially smaller terms, we need to subtract the first line
(\textit{i.e.} \rf{eq:large-order-pert-line2}) from the asymptotic behaviour of the
original hydro series coefficients $\varepsilon_k^{(\boldsymbol{0})}$. However,
as mentioned before,  $\chi_{\boldsymbol{0}\rightarrow \boldsymbol{n}}(k)$ is
itself an asymptotic series and we need to perform a resummation of it for each
value of $k$ needed to carry out the subtraction. We do so via the so-called
Écalle-Borel-Padé procedure, wherein we determine its Borel transform,
analytically continue it via a Padé approximant $\mathrm{BP}_N
\left[\chi_{\boldsymbol{0}\rightarrow\boldsymbol{n}} \right]$, and then perform
the inverse transform\footnote{The Borel transform is equivalent to applying an
inverse Laplace transform to each term of a series. Thus the inverse transform
will be a Laplace transform applied to the Borel-Padé approximant, with
integration contour over the real axis. Given that this Padé approximant turns
out to have poles along the real axis, to avoid them we need to shift the
integration contour slightly above this axis: this is called a \textit{lateral
resummation}.} to obtain the summed result
$\mathcal{S}_{0^{+}}\chi_{\boldsymbol{0}\rightarrow\boldsymbol{n}}(k)$:
\begin{equation}
\label{eq:large-order-resum}
\mathcal{S}_{0^{+}}\chi_{\boldsymbol{0}\rightarrow\boldsymbol{n}}(k)\equiv \int_0^{\mathrm{e}^{\mathrm{i}\epsilon}\,\infty}\,
\mathrm{d}\xi \,\mathrm{e}^{-k\,\xi}\,\mathrm{BP}_N \left[\chi_{\boldsymbol{0}\rightarrow\boldsymbol{n}}\right](\xi),
\end{equation}

\noindent where we assumed $\epsilon>0$ small, and $k\ge1$ (we cannot use this
procedure to determine the summation at $k=0$). We can then define the
subtracted coefficients as\footnote{The numerical lateral summation
$\mathcal{S}_{0^+}\chi_{\boldsymbol{0}\rightarrow\boldsymbol{e}_1}$ used a
diagonal Padé approximant of order $N=96$.}
\begin{equation}
\label{eq:subtracted-coeffs}
\delta_1\varepsilon_k^{(\boldsymbol{0})}\equiv \varepsilon_k^{(\boldsymbol{0})}+
\frac{\mathsf{S}_{\boldsymbol{0}\rightarrow \boldsymbol{e}_1}}{2\pi\mathrm {i}}\,\frac{\Gamma(k+\beta_{\boldsymbol{0}}-\beta_{\boldsymbol{e}_1})}{A_1^{k+\beta_{\boldsymbol{0}}-\beta_{\boldsymbol{e}_1}}} \,\mathcal{S}_{0^+}\chi_{\boldsymbol{0}\rightarrow \boldsymbol{e}_1}
+\frac{\mathsf{S}_{\boldsymbol{0}\rightarrow \overline{\boldsymbol{e}}_1}}{2\pi\mathrm {i}}\,\frac{\Gamma(k+\beta_{\boldsymbol{0}}-\beta_{\overline{\boldsymbol{e}}_1})}{\left(\overline{A_1}\right)^{k+\beta_{\boldsymbol{0}}-\beta_{\overline{\boldsymbol{e}}_1}}} \,\mathcal{S}_{0^+}\chi_{\boldsymbol{0}\rightarrow \overline{\boldsymbol{e}}_1} \, .
\end{equation}

\begin{figure}[ht!]
  \begin{center}
    \includegraphics[width=0.48\linewidth
    ]{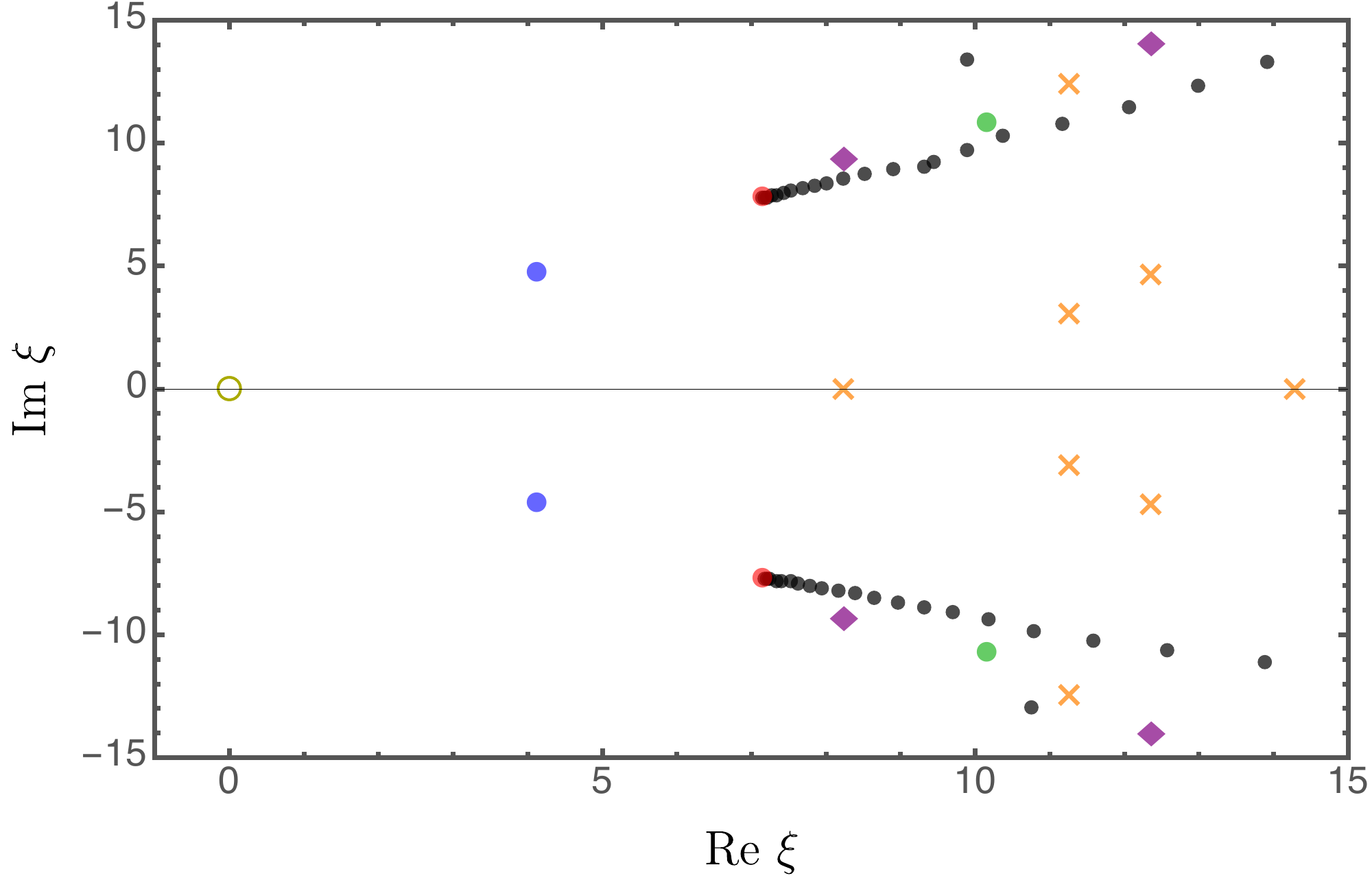}
    \caption{Poles of the Borel-Padé approximant
      $\mathrm{BP}_{90}\left[\delta_{1}\Phi_{\boldsymbol{0}}\right]$. Comparing
      the figure above to the top plot of Fig. \ref{fig:BorelPlanes} we can
      see that the singularity structure of the leading sector can be
      consistently removed from the coefficients of $\Phi_{\boldsymbol{0}}$,
      and the information of the higher sectors can be recovered. The slight
      asymmetry in the figure above is due to the particular choice of
      \emph{lateral resummation} where we have integrated along the real line
      from above.}
	\label{fig:Pade-resummed-hydro-1}
	\end{center}
\end{figure}

The subtracted series defined by these coefficients
$\delta_1\Phi_{\boldsymbol{0}} (u)\equiv
u^{-\beta_{\boldsymbol{0}}}\sum_{k=1}^{+\infty}\delta_1\varepsilon_k^{(\boldsymbol{0})}\,u^{-k}$
is again asymptotic, and the singularity structure in the respective Borel plane
can be seen in Fig. \ref{fig:Pade-resummed-hydro-1}. As before, in blue are
shown the singular points associated to the actions $A_{1},\overline{A_{1}}$,
while in red we can see the singular branch points associated to the actions
$A_{2},\overline{A_{2}}$. Comparing to the top plot of Fig.
\ref{fig:BorelPlanes}, it is evident that indeed the singularities associated to
$s=A_{1},\overline{A_{1}}$ have been subtracted, and now the leading
singularities are at $s=A_{2},\overline{A_{2}}$.


One can now numerically check that the large-order behaviour of the subtracted coefficients $\delta_1\varepsilon_k^{(\boldsymbol{0})}$ is just given by the second (leading) and third lines (sub-leading) \eqref{eq:large-order-pert-line2} and \eqref{eq:large-order-pert-line3}. From explicitly examining the numerical coefficients, one can notice that $\mathrm{Re}\,\delta_1\varepsilon_k^{(\boldsymbol{0})}\sim \frac{\Gamma(k)}{\left|A_2\right|^k}$ while $\mathrm{Im}\,\delta_1\varepsilon_k^{(\boldsymbol{0})}\sim \frac{\Gamma(k)}{\left|2 A_1 \right|^k}$, which means we only need the real part of this summation procedure to analyse the contributions from the fundamental sectors $\Phi_{\boldsymbol{e}_2}$, $\Phi_{\overline{\boldsymbol{e}}_2}$, appearing in \eqref{eq:large-order-pert-line2}. Given that the  sectors $\Phi_{\boldsymbol{e}_2}$ and $\Phi_{\overline{\boldsymbol{e}}_2}$ are complex conjugates, with $\beta_{\boldsymbol{e}_2} = \overline{\beta_{\overline{\boldsymbol{e}}_2}}$, then we again expect $\mathsf{S}_{\boldsymbol{0}\rightarrow \boldsymbol{e}_2}=-\overline{\mathsf{S}_{\boldsymbol{0}\rightarrow \overline{\boldsymbol{e}}_2}}$. This conjugacy relation between Borel residues should extend to every fundamental sector:
\begin{equation}
\mathsf{S}_{\boldsymbol{0}\rightarrow \boldsymbol{e}_k}=-\overline{\mathsf{S}_{\boldsymbol{0}\rightarrow \overline{\boldsymbol{e}}_k}}\,.
\label{eq:borel-res-fund-relations}
\end{equation}

The resurgent properties of our transseries also impose several constraints on  Stokes constants and consequently on the Borel residues. Of particular interest to our analysis is the expected constraint $\mathsf{S}_{\boldsymbol{0}\rightarrow {n\,\boldsymbol{e}_k}}=\,(-1)^{n+1}\left(\mathsf{S}_{\boldsymbol{0}\rightarrow {\boldsymbol{e}_k}}\right)^n$ (see \cite{Aniceto:2018bis} for more details). Given the conjugacy relations just mentioned between Borel residues of conjugate fundamental sectors, we obtain
\begin{equation}
\mathsf{S}_{\boldsymbol{0}\rightarrow 2\boldsymbol{e}_1}=-\left(\mathsf{S}_{\boldsymbol{0}\rightarrow \boldsymbol{e}_1}\right)^2=-\overline{\left(\mathsf{S}_{\boldsymbol{0}\rightarrow \overline{\boldsymbol{e}}_1}\right)^2}=\overline{\mathsf{S}_{\boldsymbol{0}\rightarrow 2\overline{\boldsymbol{e}}_1}}.
\end{equation}

\noindent From direct gravity calculations we know that
$\Phi_{2\boldsymbol{e}_1}$ and $\Phi_{2\overline{\boldsymbol{e}}_1}$ are complex
conjugates of each other. We could then worry that the large-order contribution
coming from the third line \eqref{eq:large-order-pert-line3} is not real, as one
could have expected. However, the resummations from the previous lines
\eqref{eq:large-order-pert-line1} and \eqref{eq:large-order-pert-line2} will
give a non trivial imaginary contribution of the same order.\footnote{This can
be traced to the fact that we had to choose a particular lateral resummation to
subtract the leading results. Because this summation prescription is the
\textit{same} for complex conjugate sectors, we obtain non-trivial imaginary
contributions.}

The above discussion was framed in the context of the hydrodynamic sector, but
an analogous large-order analysis can be done for any sector of the transseries.
The upshot of this analysis is that despite the appearance of an infinite number
of fundamental QNM sectors one can really analyse each contribution individually
due to the hierarchy of exponential damping. We have thus verified that our
original transseries ansatz is consistent at this level. In the following
section we will develop the techniques necessary for a quantitative and
systematic testing of the resurgence relations between coefficients appearing in
different sectors.

\subsection{A Borel analysis of large-order relations}
In many cases one can directly use the large-order relations to determine the coefficients of exponentially suppressed sectors from the values of the hydrodynamic gradient series using numerical acceleration methods such as Richardson transforms~\cite{Aniceto:2015mto,Basar:2015ava}. However, in the case of interest here the oscillatory nature of our problem, due to the presence of complex conjugate sectors with complex characteristic exponents $\beta_{\boldsymbol{n}}$, complicates the task exceedingly. This type of issue has already been encountered in the analysis of large-order behaviour of the hydrodynamic model of~\cite{Heller:2014wfa} which was studied in Ref.~\cite{Aniceto:2015mto}. The equations considered there were designed specifically to model the leading QNMs of \symm\ and they served as a testing ground for the analysis undertaken here. Due to the lack of suitable techniques Ref.~\cite{Aniceto:2015mto} tested the large-order relations using values of the expansion coefficients appearing in the various sectors. The methods used there started from the large-order relations such as Eq. \eqref{eq:large-order-pert-line1}, and performed the resummation of the expansions $\chi_{\boldsymbol{0}\rightarrow \boldsymbol{e}_1},\,\chi_{\boldsymbol{0}\rightarrow \overline{\boldsymbol{e}}_1}$ (using Eq. \eqref{eq:large-order-resum}). This way one obtains
\be
\varepsilon_k^{(\boldsymbol{0})}\simeq -
\frac{\mathsf{S}_{\boldsymbol{0}\rightarrow \boldsymbol{e}_1}}{2\pi\mathrm {i}}\,\frac{\Gamma(k+\beta_{\boldsymbol{0}}-\beta_{\boldsymbol{e}_1})}{A_1^{k+\beta_{\boldsymbol{0}}-\beta_{\boldsymbol{e}_1}}} \,\mathcal{S}_{0^+}\chi_{\boldsymbol{0}\rightarrow \boldsymbol{e}_1}
-\frac{\mathsf{S}_{\boldsymbol{0}\rightarrow \overline{\boldsymbol{e}}_1}}{2\pi\mathrm {i}}\,\frac{\Gamma(k+\beta_{\boldsymbol{0}}-\beta_{\overline{\boldsymbol{e}}_1})}{\left(\overline{A_1}\right)^{k+\beta_{\boldsymbol{0}}-\beta_{\overline{\boldsymbol{e}}_1}}} \,\mathcal{S}_{0^+}\chi_{\boldsymbol{0}\rightarrow \overline{\boldsymbol{e}}_1}+\cdots\, .
\ee
as the leading contributions to the large-order behaviour. In the above equation
everything is known except the Borel residues
$\mathsf{S}_{\boldsymbol{0}\rightarrow
\boldsymbol{e}_1},\,\mathsf{S}_{\boldsymbol{0}\rightarrow
\overline{\boldsymbol{e}}_1}$, which we know to be related by Eq.
\eqref{eq:borel-res-fund-relations} due to the reality condition which our
solution satisfies. We can therefore solve for the modulus and argument of the
Borel residue $\mathsf{S}_{\boldsymbol{0}\rightarrow \boldsymbol{e}_1}$. While
the result formally depends on $k$, the large-order relations imply that it
should saturate at large values of $k$, which can easily be verified, leading to
an accurate estimate of the Borel residue. The result of such a procedure --
implementing precisely the techniques of Ref.~\cite{Aniceto:2015mto} but now
using the series calculated for \symm\ leads to the results shown in
\rff{fig:modstokes}. This provides further support for our approach.


\begin{figure}[ht!]
	\begin{center}
	\includegraphics[width=0.48\linewidth]{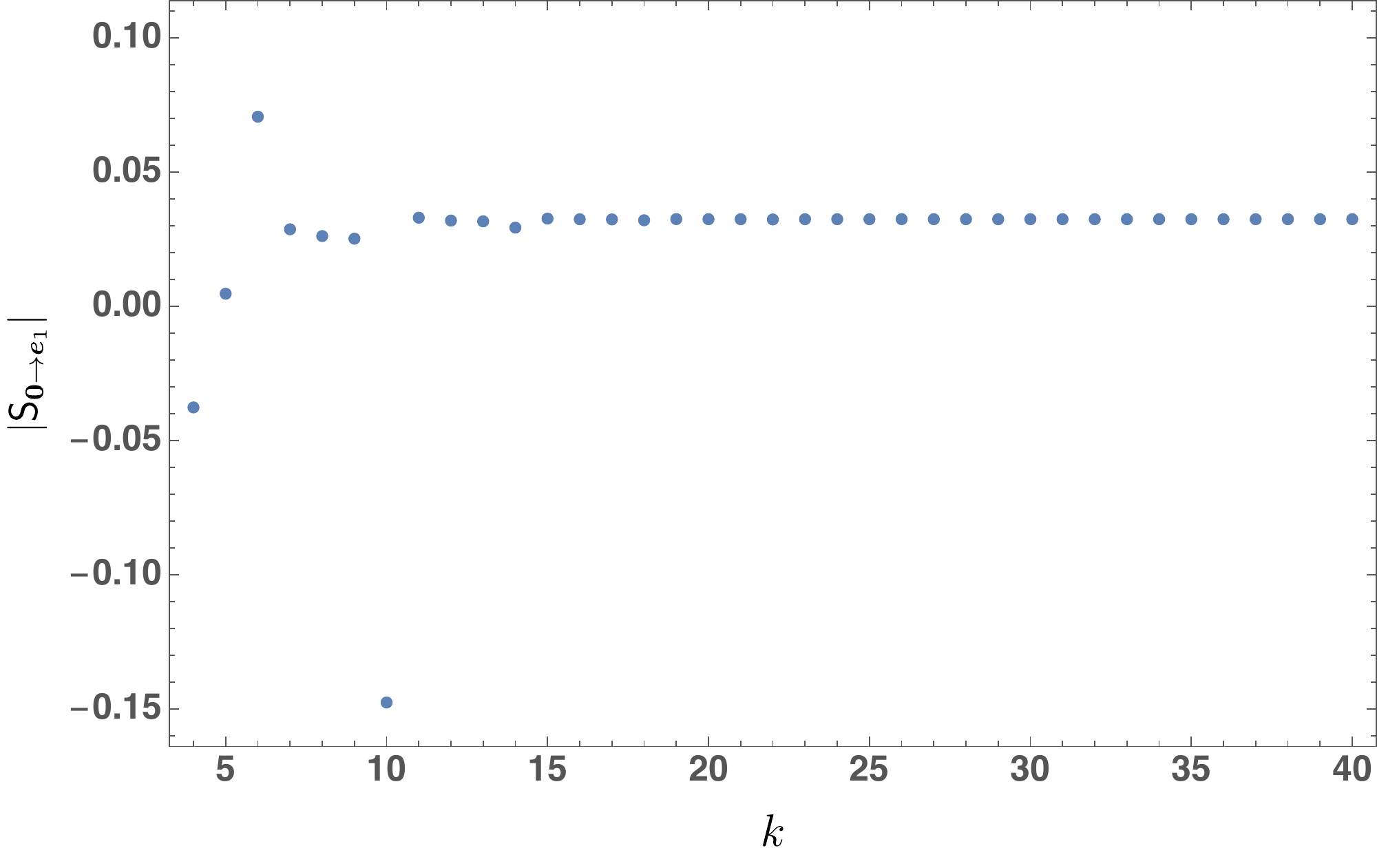}\quad 
	\includegraphics[width=0.48\linewidth]{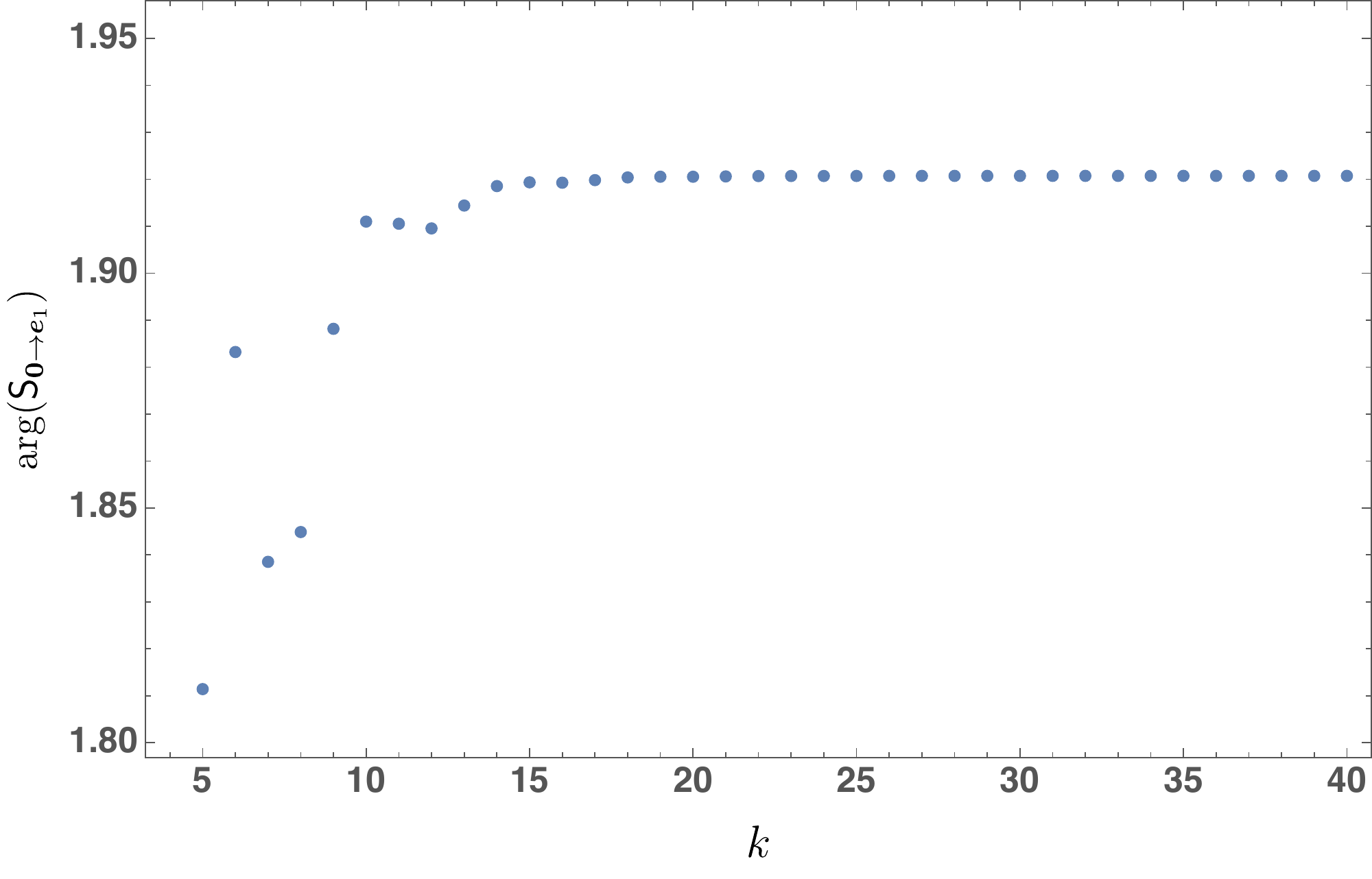}
	\caption{The plots show the convergence of the modulus and argument of
          Stokes constant (or rather, the appropriate Borel residue). The
          quantities plotted are $k$-dependent, but the large-order relations
          imply that for large enough values of $k$ they should saturate,
          approaching the values of the modulus and argument of the Stokes
          constant   (for details see \rfc{Aniceto:2015mto}). We see that this
          happens already at moderate values of $k$. At $k=250$ the variation
          is of order $10^{-54}$.}
	\label{fig:modstokes}
	\end{center}
\end{figure}


While the arguments presented so far provide strong evidence that we are dealing with a resurgent transseries, we would still like to extract values of the coefficients appearing in the exponentially suppressed transseries sectors from the hydrodynamic gradient expansion alone. The difficulties discussed in the previous paragraph preclude the use of standard techniques, but nevertheless, an equivalent analysis can be carried out at the level of the respective Borel transforms. This new method is  presented in some detail in \rfApp{app:app-res}, 
but we briefly summarise it below for use in the following Sections.

We consider the hydrodynamic series,  $\Phi_{\boldsymbol{0}}(u)$, whose coefficients $\varepsilon_{k}^{(\boldsymbol{0})}$ have the asymptotic large-order behaviour shown in Eqs. \eqref{eq:large-order-pert-line1} -- \eqref{eq:large-order-pert-line3}. As we saw previously, this large-order behaviour is intimately linked to the existence of branch cut singularities on the Borel plane, and resurgence analysis explicitly reveals the analytic properties of this link. It is  well known\footnote{We are within the realm of simple resurgent functions of Gevrey-1 type, see \textit{e.g.} \cite{Sauzin:2016fxp,Aniceto:2018bis}.} that if we multiply our asymptotic series $\Phi_{\boldsymbol{0}}(u)$ by an overall factor $u^{-\beta}$ for a specific value of $\beta$ such that the Borel transform removes the \textit{exact} leading factorial growth of its coefficients, then this Borel transform will present a logarithmic cut at its leading singularity. For the particular case of the hydrodynamic sector and the leading singularity at $\xi=A_1$ (corresponding to the first large-order contribution of \eqref{eq:large-order-pert-line1}) we can write
 \be\label{eq:boreltrans-behaviour-log}
 \left.\mathcal{B}\left[u^{\beta_{\boldsymbol{e}_1}}\,\Phi_{\boldsymbol{0}}\right](\xi)\right|_{\xi=A_1}=\mathsf{S}_{\boldsymbol{0}\rightarrow\boldsymbol{e}_1}\,\mathcal{B}\left[u^{\beta_{\boldsymbol{e}_1}}\,\Phi_{\boldsymbol{e}_1}\right](\xi-A_1)\,\frac{\log(\xi-A_1)}{2\pi\,\mathrm{i}}+\textrm{regular terms}\,.
 \ee
This means that if we analyse each branch cut of the Borel plane separately, we
can recover the coefficients of the sector associated to that branch cut (in
this case $\Phi_{\boldsymbol{e}_1}$). It is vital that we know the factorial
growth corresponding to each of the branch points (as given in the large-order
relations) with as high an accuracy as possible, as this determines the exact
type of branch cut one finds in the Borel plane. This is necessary to be able to
transform it into the logarithmic cut shown in \rf{eq:boreltrans-behaviour-log}
above.

From the knowledge of Eq.~\eqref{eq:boreltrans-behaviour-log}, we can transform
the logarithmic behaviour into a leading square root branch cut by further
multiplying the original sector by $u^{-1/2}$
 \begin{eqnarray}
 \label{eq:boreltrans-behaviour-sqrt}
 \left.\mathcal{B}\left[u^{\beta_{\boldsymbol{e}_1}-1/2}\,\Phi_{\boldsymbol{0}}\right](\xi)\right|_{\xi=A_1} & = & \frac{\mathsf{S}_{\boldsymbol{0}\rightarrow\boldsymbol{e}_1}}{2}\,\mathcal{B}\left[u^{\beta_{\boldsymbol{e}_1}-1/2}\,\Phi_{\boldsymbol{e}_1}\right](\xi-A_1)+\textrm{regular terms} \\
 &\hspace{-40pt} = & \hspace{-20pt}\frac{\mathsf{S}_{\boldsymbol{0}\rightarrow\boldsymbol{e}_1}}{2\sqrt{\xi-A_1}}\,\left( \frac{\varepsilon_{0}^{(\boldsymbol{e}_1)}}{\Gamma(1/2)}+\varepsilon_{1}^{(\boldsymbol{e}_1)}\, \frac{(\xi-A_1)}{\Gamma(3/2)}+\varepsilon_{2}^{(\boldsymbol{e}_1)}\, \frac{(\xi-A_1)^{2}}{\Gamma(5/2)}+\cdots\right).\nonumber
 \end{eqnarray}
This result can be demonstrated  through a simple extension of an analogous
proof contained in Section 4 of \cite{Aniceto:2018bis} (for the case of the
quartic integral), see also \rfApp{app:app-res}.

Various changes of variables have been used in the literature to more effectively analyse the Borel plane singularity structure, in particular through the use of conformal maps, see \textit{e.g.} \cite{Jentschura:2000fm,ZinnJustin:2004cg,Caliceti:2007ra}, as well as recent applications \cite{Costin:2017ziv}.\footnote{We would like to thank O. Costin for useful discussions on this subject, as well as on on-going work on applications of conformal maps as an efficient method of retrieving asymptotic information on the Borel plane \cite{Costin:upcoming}.} To further simplify our results, we can now perform a change of variables effectively transforming the branch cut into a simple pole. To this end we define $\xi=A_1-(\zeta-A_1)^2$. In terms of this new variable we can write the above equation as
 \be
 \label{eq:boreltrans-behaviour-pole}
 \left.\mathcal{B}\left[u^{\beta_{\boldsymbol{e}_1}-1/2}\,\Phi_{\boldsymbol{0}}\right](\zeta)\right|_{\zeta=A_1}  =
 \frac{\mathsf{S}_{\boldsymbol{0}\rightarrow\boldsymbol{e}_1}}{2\mathrm{i}\,(\zeta-A_1)}\,\left( \frac{\varepsilon_{0}^{(\boldsymbol{e}_1)}}{\Gamma(1/2)}-\varepsilon_{1}^{(\boldsymbol{e}_1)}\, \frac{(\zeta-A_1)^2}{\Gamma(3/2)}+\cdots\right).
 \ee
From this expression we can see that by calculating the Borel transform
$\mathcal{B}\left[u^{\beta_{\boldsymbol{e}_1}-1/2}\,\Phi_{\boldsymbol{0}}(u)\right](\zeta)$
and analysing its residue at $\zeta=A_1$ we have \textbf{direct access} to the
Borel residue
$\mathsf{S}_{\boldsymbol{0}\rightarrow\boldsymbol{e}_1}$.\footnote{In fact we
have access to the combination
$\mathsf{S}_{\boldsymbol{0}\rightarrow\boldsymbol{e}_1}\,\varepsilon_{0}^{(\boldsymbol{e}_1)}$,
but we normalise the fundamental sectors such that their first non-zero
coefficient is $1$.} Moreover, by subtracting the simple pole from the Borel
transform:
\be
\label{eq:boreltrans-behaviour-subleading}
 \left.\mathcal{B}\left[u^{\beta_{\boldsymbol{e}_1}-1/2}\,\Phi_{\boldsymbol{0}}\right](\zeta)\right|_{\zeta=A_1} - \frac{\mathsf{S}_{\boldsymbol{0}\rightarrow\boldsymbol{e}_1}}{2\mathrm{i}\,(\zeta-A_1)}\, \frac{\varepsilon_{0}^{(\boldsymbol{e}_1)}}{\Gamma(1/2)} =
 -\varepsilon_{1}^{(\boldsymbol{e}_1)}\,\frac{\mathsf{S}_{\boldsymbol{0}\rightarrow\boldsymbol{e}_1}}{2\mathrm{i}\,\Gamma(3/2)}\, (\zeta-A_1)+\cdots,
 \ee
and multiplying everything by $(\zeta-A_1)^{-2}$, we can once again use the
residue theorem to predict the next coefficient
$\varepsilon_{1}^{(\boldsymbol{e}_1)}$ of the nonhydro sector
$\Phi_{\boldsymbol{e}_1}$. This procedure can be systematised to predict an
arbitrary number of coefficients of the nonhydro sectors, as is shown in
\rfApp{app:app-res}.

In the following we will consider in some detail the singular behaviour of
sectors $\Phi_{\boldsymbol{0}}$, $\Phi_{\boldsymbol{e}_1}$, and how this
behaviour is governed by the other fundamental sectors, as well as mixed
sectors. We will observe the remarkable accuracy of the predictions obtained via
resurgence from the analysis of the Borel planes. In \rfs{sec:res-fund-sectors}
we look at leading contributions coming from fundamental sectors (associated
directly with the AdS black-brane QNMs): we calculate the coefficients of
fundamental sectors $\Phi_{\boldsymbol{e_1}}$ and  $\Phi_{\boldsymbol{e_2}}$
from the Borel analysis of the hydrodynamic sector $\Phi_{\boldsymbol{0}}$ and
compare them with  the values obtained directly from the
exponentially-suppressed contributions to the bulk gravity solution of
\rfs{sec:bulksolution}. In \rfs{sec:res-mixed-sector} we will show that the
mixed sectors, coming from non-linear couplings between QNMs, contribute
non-trivially to the singular behaviour of the transseries sectors and cannot be
ignored in a full description of the energy density. Our analysis will focus on
the Borel analysis of sector $\Phi_{\boldsymbol{e_1}}$ and the corresponding
prediction of coefficients from sectors $\Phi_{2\boldsymbol{e_1}}$ and
$\Phi_{\boldsymbol{e_1}+\overline{\boldsymbol{e}}_1}$.

The results presented in Sections \ref{sec:res-fund-sectors} and
\ref{sec:res-mixed-sector} provide significant further evidence that the energy
density of \symm\ can at late times be described by the resurgent transseries
solution given in  \rf{eq:trans-en-dens} and discussed in this section.

\section{Resurgence of QNMs in the hydrodynamic expansion}
\label{sec:res-fund-sectors}

We will now use the techniques explained in the previous section to analyse the resurgence relations between different sectors in the transseries shown in Eq. \eqref{eq:trans-en-dens}. We will focus on the hydrodynamic gradient expansion and the appearance of the QNMs in the large-order behaviour of its coefficients. The appearance of the these modes was already evident from the visual Borel analysis shown in the top plot of Fig. \ref{fig:BorelPlanes}.

\begin{figure}[t]
  \includegraphics[width=0.48\linewidth]{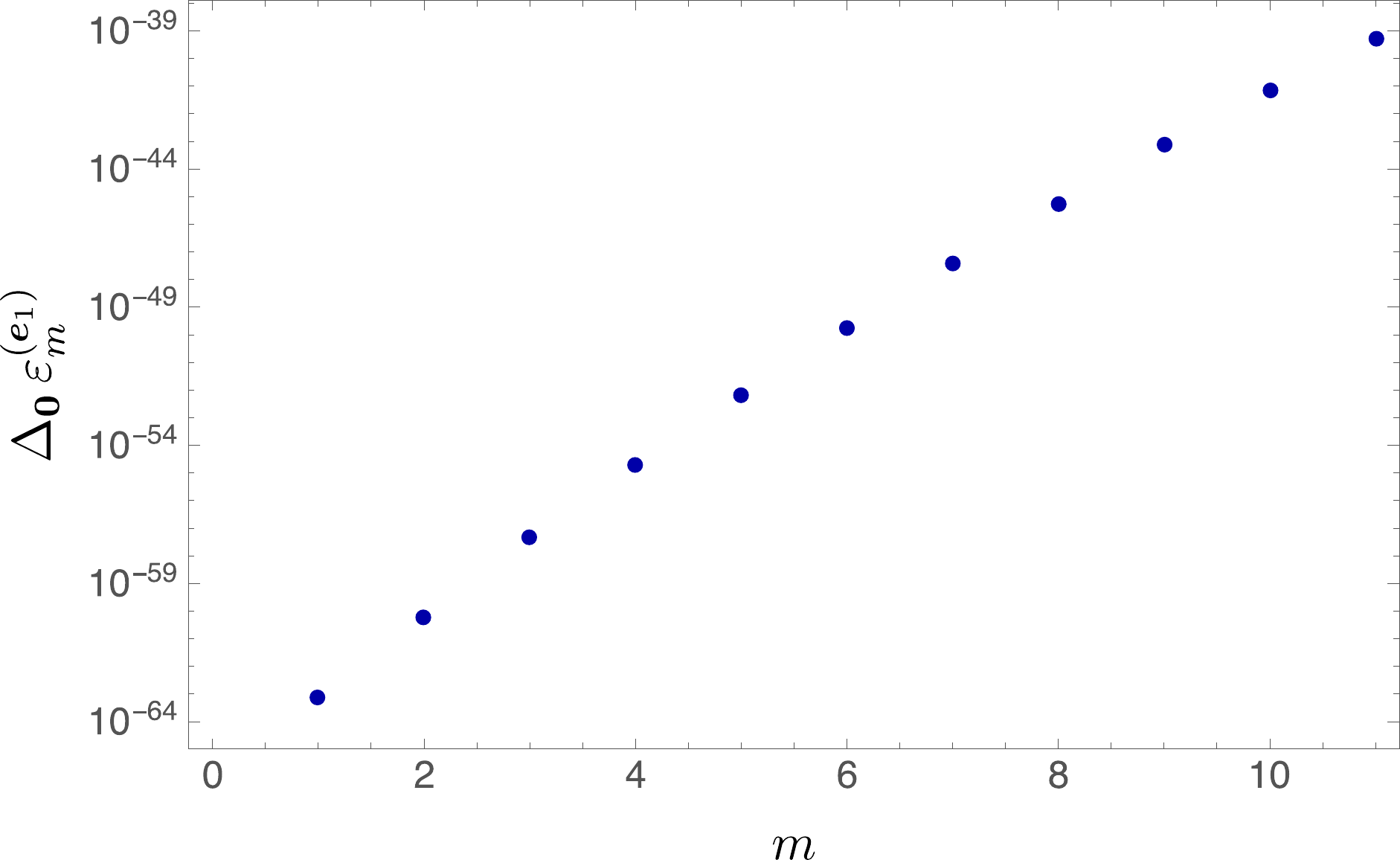}  \quad
  \includegraphics[width=0.48\linewidth]{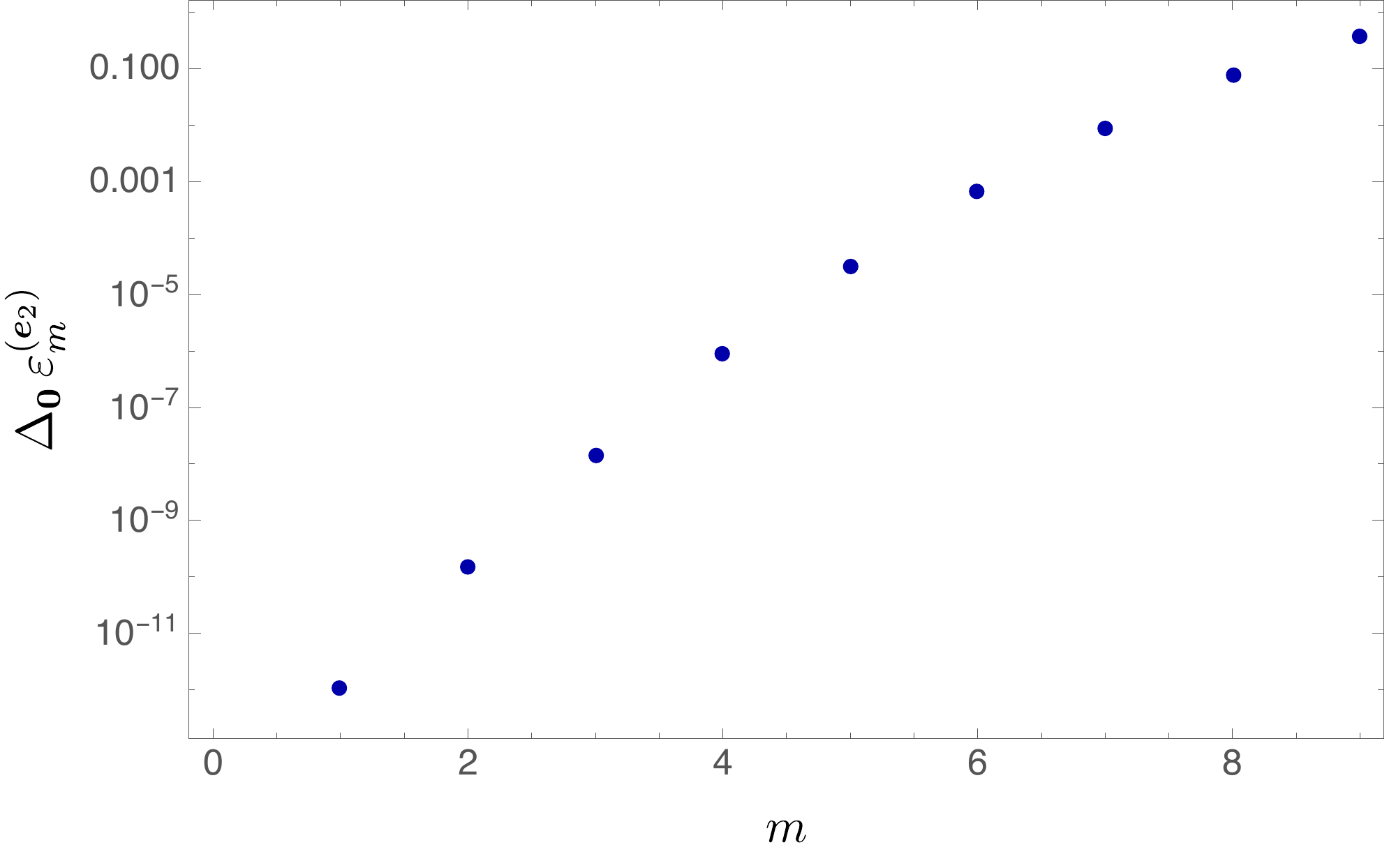}
  \caption{ \label{fig:predictionError}  Comparison of the predicted results
    from resurgence techniques and the numerical ones obtained from solving
    Einstein equations. In the above plots we show this comparison for the
    coefficients of the fundamental sectors $\Phi_{\boldsymbol{e}_1}$ and
    $\Phi_{\boldsymbol{e}_2}$, as predicted by the large-order behaviour of
    the coefficients of the perturbative sector $\Phi_{\boldsymbol{0}}$ (see
    Eq. \eqref{eq:coeffs-comparison}).}
\end{figure}

By applying the methodology introduced in the last section (and thoroughly
explained  in \rfApp{app:app-res}) to the perturbative sector
$\Phi_{\boldsymbol{0}}$, we can predict the values of the coefficients of the
sectors associated the leading (least damped) QNMs, associated to the
transseries fundamental sectors $\Phi_{\boldsymbol{e}_1}$ and
$\Phi_{\boldsymbol{e}_2}$. In Fig. \ref{fig:predictionError} we can find the
normalised error of the resurgence prediction and numerical gravity calculation
of the first coefficients associated to these two sectors. This error is given
by
\be
\Delta_{\boldsymbol{n}}\varepsilon_k^{(\boldsymbol{m})}\equiv
\frac{\varepsilon_k^{(\boldsymbol{m})}\left.\right|_{\boldsymbol{n}-\mathrm{predicted}}-\varepsilon_k^{(\boldsymbol{m})}\left.\right|_{\mathrm{numerical}}}{\varepsilon_k^{(\boldsymbol{m})}\left.\right|_{\mathrm{numerical}}}\, ,\,\, k\ge 1,
\label{eq:coeffs-comparison}
\ee
where we defined $\varepsilon_
k^{(\boldsymbol{m})}\left.\right|_{\mathrm{numerical}}$ as the coefficients
associated to the sector $\Phi_{\boldsymbol{m}}$ numerically determined from the
gravity calculation of \rfs{sec:bulksolution}, and
$\varepsilon_k^{(\boldsymbol{m})}\left.\right|_{\boldsymbol{n}-\mathrm{predicted}}$
being the same coefficients as predicted by the resurgence analysis of sector
$\Phi_{\boldsymbol{n}}$, which is outlined by Eqs.
\eqref{eq:boreltrans-behaviour-pole} and
\eqref{eq:boreltrans-behaviour-subleading}.  The comparison of predicted and
numerical coefficients for $k\ge 1$  follows the calculation of the respective
Borel residues $\mathsf{S}_{\boldsymbol{0}\rightarrow\boldsymbol{e}_1}$ and
$\mathsf{S}_{\boldsymbol{0}\rightarrow\boldsymbol{e}_2}$, as detailed below.


\subsection{Behaviour of the hydrodynamic series at the leading Borel singularity}

As mentioned earlier, the singularity structure of the Borel transform of the hydrodynamic (perturbative) series $\Phi_{\boldsymbol{0}}(u)$ can be found in the top plot of Fig. \ref{fig:BorelPlanes}, where we have plotted the zeros of the denominator of the (diagonal) Borel-Padé approximant $\mathrm{BP}_{N}\left[\Phi_{\boldsymbol{0}}\right]$, of
order $N=189$. We clearly observe pole condensation indicative of branch point singularities at $\xi=A_{i},\,\overline{A_{i}}$ with $i=1,2,3$ shown in blue, red and green (filled) circles, respectively. These are the branch points associated to the first fundamental sectors $\Phi_{\boldsymbol{e}_i},\,\Phi_{\overline{\boldsymbol{e}}_i}$ with $i=1,2,3$. The purple diamonds correspond to mixed sectors, which will be the subject of \rfs{sec:res-mixed-sector}.\footnote{Branch points associated to more exponentially suppressed sectors are not shown due to the loss of accuracy of the Padé approximant.}

The two leading nonhydrodynamic, fundamental sectors appearing in this plot are
$\Phi_{\boldsymbol{e}_{1}},\Phi_{\overline{\boldsymbol{e}}_{1}}$, responsible
for the singular points $\xi=A_{1},\overline{A_{1}}$. We can then perform the
Borel plane analysis introduced in the previous section and thoroughly described
in \rfApp{app:app-res} for any of these singular points. Consider for instance
the sector associated to $\xi=A_{1}$, given in Eq. \eqref{phin} with the
characteristic exponent $\beta_{\boldsymbol{e}_1}$ given on page
\pageref{page:charac-exp}. The leading growth of the coefficients of the
hydrodynamic series $\varepsilon_k^{(\boldsymbol{0})}$ associated to this sector
is given by $\beta=\beta_{\boldsymbol{0}}-\beta_{\boldsymbol{e}_{1}}$, as shown
in the first line of the large-order relations Eq.
\eqref{eq:large-order-pert-line1}. Then from a direct application of Eq.
\eqref{eq:boreltrans-behaviour-pole} we can determine the Borel residue
associated with this singular point:
\begin{eqnarray}
\mathsf{S}_{\boldsymbol{0}\rightarrow\boldsymbol{e}_{1}} & = & -0.011131682120023118507124753501675427870232109210448+\nonumber\\
 &  & +0.030501348613000820066073498678990910567099376605738\,\mathrm{i}\,.
\end{eqnarray}
The convergence plots of Fig. \ref{fig:modstokes} show exactly this result. But
we can now go further and analyse the sub-leading behaviour of the perturbative
series at the singular point $\xi=A_{1}$, as predicted by the procedure shown in
Eq.~\eqref{eq:boreltrans-behaviour-subleading}.\footnote{More explicitly, the
sub-leading coefficients can be predicted from Eq.
\eqref{eq:res-subleading-pert}.} The comparison between values predicted using
resurgence and what was numerically calculated directly from the bulk gravity
solution in sector $\Phi_{\boldsymbol{e}_1}$ can be inspected on the left plot
of Fig. \ref{fig:predictionError} for the first 11 coefficients
$\varepsilon_{m}^{(\boldsymbol{e}_{1})}$. We can see that the error remains
extremely small, thus confirming the resurgent properties of our transseries
solution \eqref{eq:trans-en-dens}.

As we have seen in the previous section, the Borel residue associated to
$\mathsf{S}_{\boldsymbol{0}\rightarrow\overline{\boldsymbol{e}}_{1}}$
corresponding with the singularity $\xi=\overline{A_{1}}$ is related to
$\mathsf{S}_{\boldsymbol{0}\rightarrow\boldsymbol{e}_{1}}$ by Eq.
\eqref{eq:borel-res-fund-relations}:
\begin{equation}
\mathsf{S}_{\boldsymbol{0}\rightarrow\overline{\boldsymbol{e}}_{1}}=-\overline{\mathsf{S}_{\boldsymbol{0}\rightarrow\boldsymbol{e}_{1}}}\,.
\end{equation}
This was directly checked by repeating the above calculation for the singularity $\xi=\overline{A_{1}}$.

\subsection{Sub-leading exponentially suppressed behaviour of the hydrodynamic series}

We have now verified the resurgent behaviour of the perturbative series in the
nonhydrodynamic sector corresponding to the least damped QNM. We can ask if we
can go further and analyse the next   pair of complex-conjugate singularities of
the hydrodynamic series at $\xi=A_{2},\,\overline{A_{2}}$, which correspond to
the second QNM. The sectors associated with these singularities are
$\Phi_{\boldsymbol{e}_{2}}$ and $\Phi_{\overline{\boldsymbol{e}}_{2}}$,
respectively, given in Eq. \eqref{phin} with the characteristic exponents given
on page \pageref{page:charac-exp}. As explained before, in order to analyse the
resurgent behaviour at these singularities we need to first subtract the leading
contributions coming from the first QNM. This can done by subtracting the
resummed leading large-order behaviour (through a Borel-Padé-Écalle lateral
summation), shown in  \eqref{eq:large-order-pert-line1}, from the coefficients
of the hydrodynamic series. We obtain the subtracted coefficients given in Eq.
\eqref{eq:subtracted-coeffs}, which now  obey the large-order relation given in
Eqs. \eqref{eq:large-order-pert-line2}, \eqref{eq:large-ord-asymptotics}:
\be
\delta_{1}\varepsilon_{k}^{(\boldsymbol{0})}
\simeq-\frac{S_{\boldsymbol{0}\rightarrow\boldsymbol{e}_2}}{2\pi\mathrm{i}}\frac{\Gamma\left(k+\beta_{\boldsymbol{0}}-\beta_{\boldsymbol{e}_{2}}\right)}{A_{2}^{k+\beta_{\boldsymbol{0}}-\beta_{\boldsymbol{e}_{2}}}}\,\chi_{\boldsymbol{0}\rightarrow\boldsymbol{e}_2}(k)+\mbox{c.c.}\,.
\ee
These large-order relations allow us to predict the coefficients
$\varepsilon_{k}^{(\boldsymbol{e}_{2})}$, through the same Borel analysis
applied before to the leading singularities, but now applied to the series
\begin{equation}
\delta_{1}\Phi_{\boldsymbol{0}}(u)\simeq u^{-\beta_{\boldsymbol{0}}}\sum_{k=1}^{+\infty}\delta_{1}\varepsilon_{k}^{(\boldsymbol{0})}\,u^{-k}=u^{-\beta_{\boldsymbol{0}}-1}\sum_{k=0}^{+\infty}\delta_{1}\varepsilon_{k+1}^{(\boldsymbol{0})}\,u^{-k}.
\end{equation}
The singularity structure on the Borel plane for this asymptotic series was
already shown in Fig. \ref{fig:Pade-resummed-hydro-1}, where we can clearly
notice the absence of the leading singularities at $\xi=A_1,\,\overline{A_1}$.

Note that we have an extra shift of the characteristic exponent
$\beta_{\boldsymbol{0}}$: this was because we cannot define a resummation
procedure for $k=0$ to determine $\delta_{1}\varepsilon_{0}^{(\boldsymbol{0})}$.
Nevertheless, the asymptotic properties are encoded in the large-orders, and
taking this shift into account, we can easily determine the Borel residue
associated to the second exponential weight (\textit{i.e.} the second QNM):
\begin{eqnarray}
\mathsf{S}_{\boldsymbol{0}\rightarrow\boldsymbol{e}_{2}} & = & 0.17002438360768242627609799156749507590247289563179+\nonumber \\
 &  & +0.09746084799999641974938699072923874879800635398686\,\mathrm{i}\,.
\end{eqnarray}
Also, from the direct analysis of the branch cut at $\xi=\overline{A_2}$ we have
checked that the relation between Borel residues
$\mathsf{S}_{\boldsymbol{0}\rightarrow\overline{\boldsymbol{e}}_{2}}=-\overline{\mathsf{S}_{\boldsymbol{0}\rightarrow\boldsymbol{e}_{2}}}$
is satisfied as expected.

We can now apply the method proposed in the previous section to predict the
subleading resurgent behaviour of the hydrodynamic series' coefficients at the
singular point $\xi=A_{2}$. This corresponds to the resurgent prediction of the
coefficients $\varepsilon_{m}^{(\boldsymbol{e}_{2})}$. \footnote{These
coefficients are explicitly predicted by Eq. (\ref{eq:res-subleading-pert}).}

The right plot of Fig. \ref{fig:predictionError} shows the comparison between
prediction based on resurgence and the direct numerical calculation starting
from the bulk gravity solution for the first 8 of these coefficients. We can see
that the error remains small (although we have lost accuracy due to the
resummation process), confirming once more the resurgent properties of the
transseries solution~\eqref{eq:trans-en-dens}.

\section{Nonhydrodynamic sectors and effects of QNM coupling}
\label{sec:res-mixed-sector}

In the previous section we have analysed the resurgence relations revealed by
the asymptotic hydrodynamic expansion of the energy density and the
exponentially suppressed fundamental sectors, directly associated with QNMs of
the AdS black brane. This analysis already showed very strong evidence that the
energy density has a representation as the resurgent transseries
\eqref{eq:trans-en-dens}. However, a much more striking part of this transseries
is the existence of mixed sectors, corresponding to non-trivial non-linear
coupling between (different) QNMs, such as $\Phi_{2\boldsymbol{e}_1}$ or
$\Phi_{\boldsymbol{e}_1+\overline{\boldsymbol{e}}_1}$. The occurrence of these
mixed sectors has already been seen in the gravity calculation of
\rfs{sec:bulksolution}; here we show that they are indeed essential to complete
the resurgent picture of our transseries for the energy density.

The easiest way to show the non-linear coupling between quasinormal modes is by
studying the large-order behaviour of the coefficients of the fundamental sector
associated to the first QNM $\Phi_{\boldsymbol{e}_1}$. The singularity structure
of its Borel transform can be found on the middle left plot of Fig.
\ref{fig:BorelPlanes}. Here we already have direct evidence of the contribution
of fundamental sectors such as
$\Phi_{\boldsymbol{e}_2},\,\Phi_{\overline{\boldsymbol{e}}_1}$ (via the branch
cut starting at $\xi+A_1=A_2$ and $\xi+A_1=\overline{A_1}$, respectively), as
well as the mixed sectors $\Phi_{2\boldsymbol{e}_1}$ and
$\Phi_{\boldsymbol{e}_1+\overline{\boldsymbol{e}}_1}$ (via the branch cuts
starting at $\xi+A_1=2A_1$ and $\xi+A_1=A_1+\overline{A_1}$). The resurgent
properties of these contributions can be explicitly written in the form of the
large-order behaviour of coefficients $\varepsilon_k^{(\boldsymbol{e}_1)}$
(analogous to Eq. \eqref{eq:large-ord-asymptotics}, see \cite{Aniceto:2018bis}):
\begin{eqnarray}
\label{eq:large-order-e1-line1}
\varepsilon_k^{(\boldsymbol{e}_1)} & \underset{k\gg 1}{\simeq} & -\frac{\mathsf{S}_{\boldsymbol{e}_1\rightarrow \boldsymbol{e}_2}}{2\pi\mathrm {i}}\,\frac{\Gamma(k+\beta_{\boldsymbol{e}_1}-\beta_{\boldsymbol{e}_2})}{\left(A_2-A_1\right)^{k+\beta_{\boldsymbol{e}_1}-\beta_{\boldsymbol{e}_2}}} \,\chi_{\boldsymbol{e}_1\rightarrow \boldsymbol{e}_2}+\\
  &  &  \hspace*{-30pt}-\frac{\mathsf{S}_{\boldsymbol{e}_1\rightarrow 2\boldsymbol{e}_1}}{2\pi\mathrm {i}}\,\frac{\Gamma(k+\beta_{\boldsymbol{e}_1}-\beta_{2\boldsymbol{e}_1})}{A_1^{k+\beta_{\boldsymbol{e}_1}-\beta_{2\boldsymbol{e}_1}}} \,\chi_{\boldsymbol{e}_1\rightarrow 2\boldsymbol{e}_1}
-\frac{\mathsf{S}_{\boldsymbol{e}_1\rightarrow \boldsymbol{e}_1+\overline{\boldsymbol{e}}_1}}{2\pi\mathrm {i}}\,\frac{\Gamma(k+\beta_{\boldsymbol{e}_1}-\beta_{\boldsymbol{e}_1+\overline{\boldsymbol{e}}_1})}{\left(\overline{A_1}\right)^{k+\beta_{\boldsymbol{e}_1}-\beta_{\boldsymbol{e}_1+\overline{\boldsymbol{e}}_2}}} \,\chi_{\boldsymbol{e}_1\rightarrow \boldsymbol{e}_1+\overline{\boldsymbol{e}}_2} +\nonumber  \\
  &  &  +\cdots. \nonumber
\end{eqnarray}
The values of the characteristic exponents appearing above can be found on page \pageref{page:charac-exp}. The first line of Eq. \eqref{eq:large-order-e1-line1} gives the leading growth of these coefficients, governed solely by the sector $\Phi_{\boldsymbol{e}_2}$ (closest singularity on the Borel plane), while the second line shows the sub-leading exponentially suppressed growth governed by the mixed sectors $\Phi_{2\boldsymbol{e}_1}$ and $\Phi_{\boldsymbol{e}_1+\overline{\boldsymbol{e}}_1}$.\footnote{At the same exponentially suppressed level, the branch cut related to the hydrodynamic series will also contribute, but our focus will be on the contribution of mixed sectors.}

\begin{figure}[t]
	\begin{center}
  \includegraphics[width=0.48\linewidth]{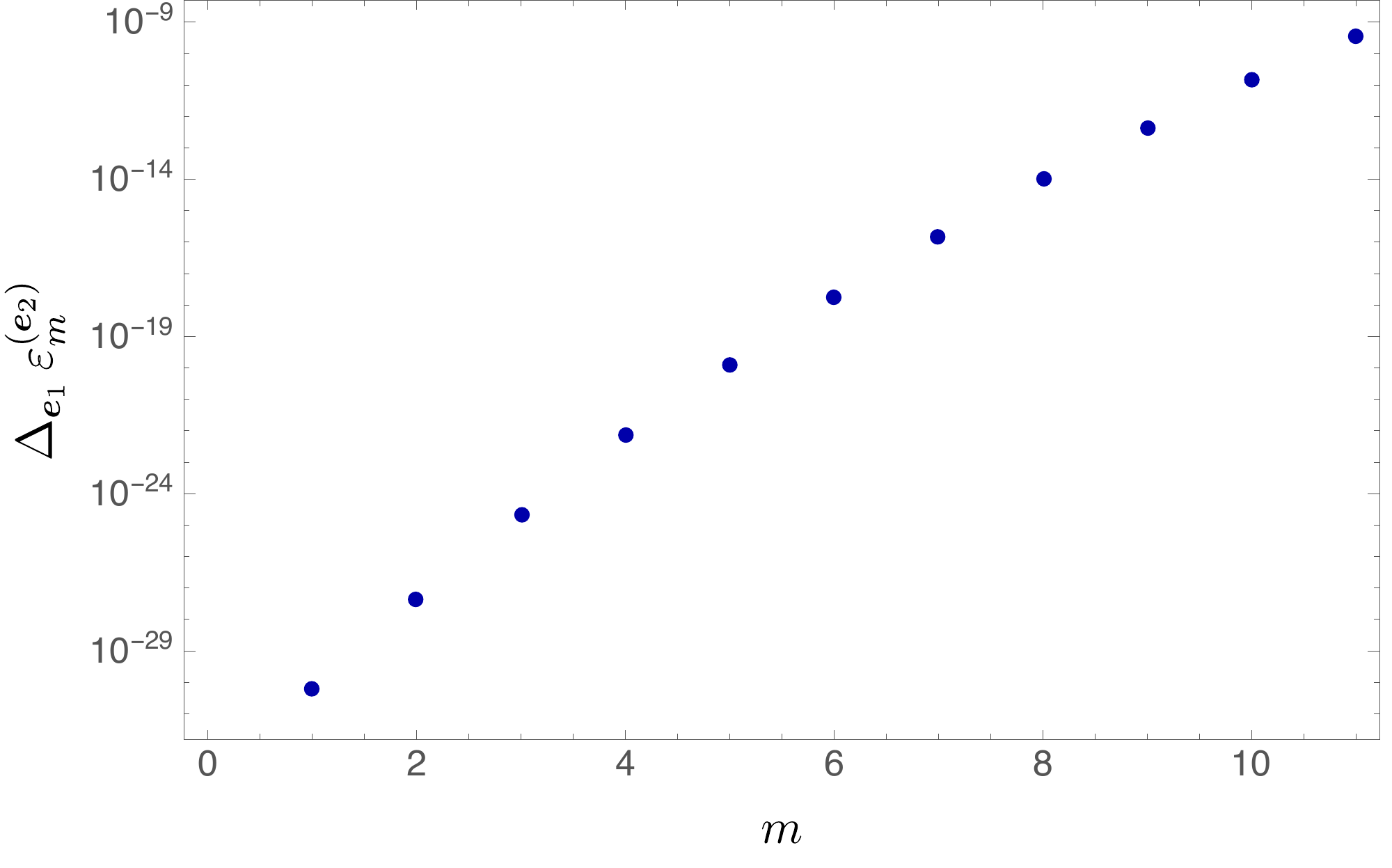} \\
  \includegraphics[width=0.48\linewidth]{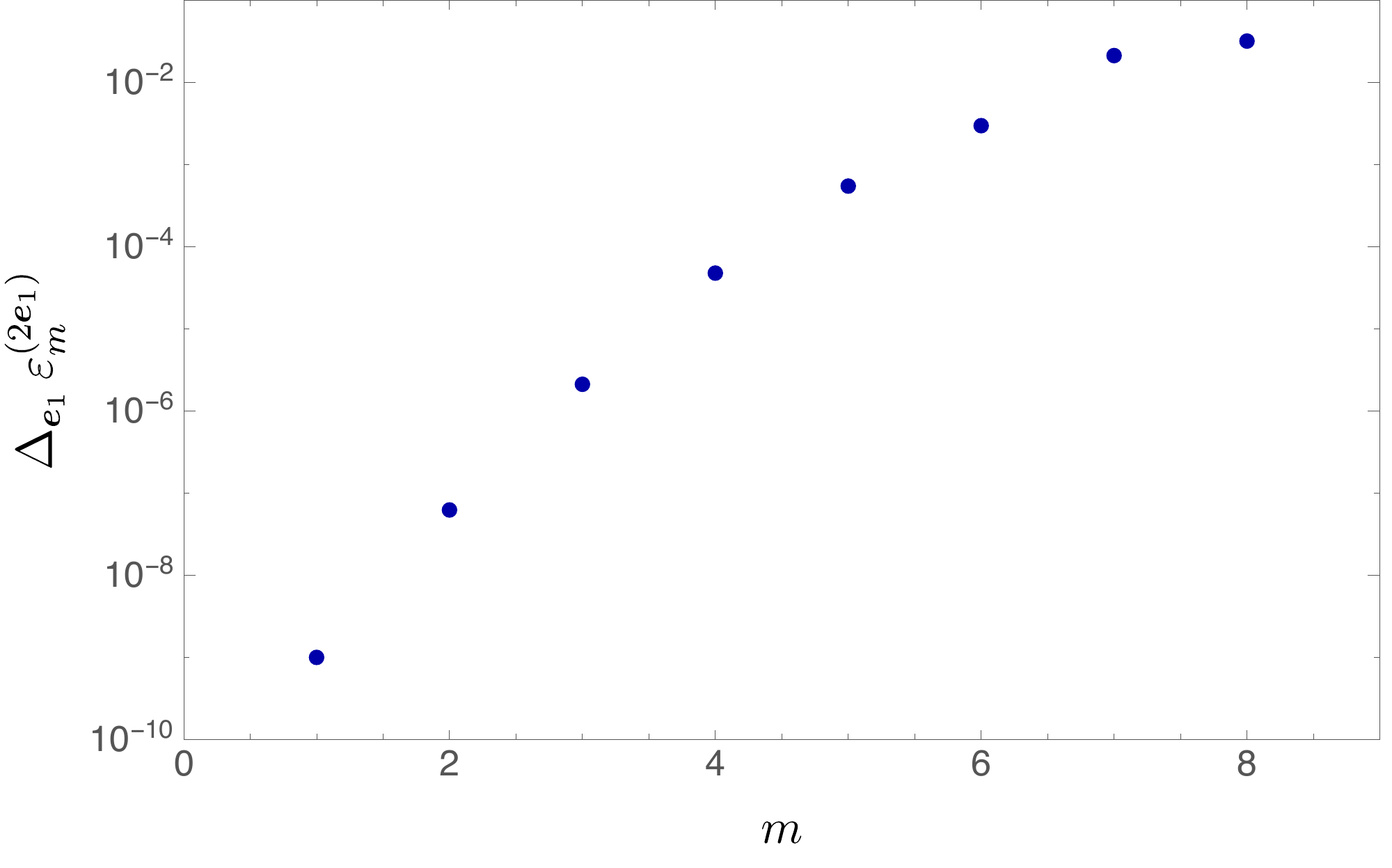} \quad
  \includegraphics[width=0.48\linewidth]{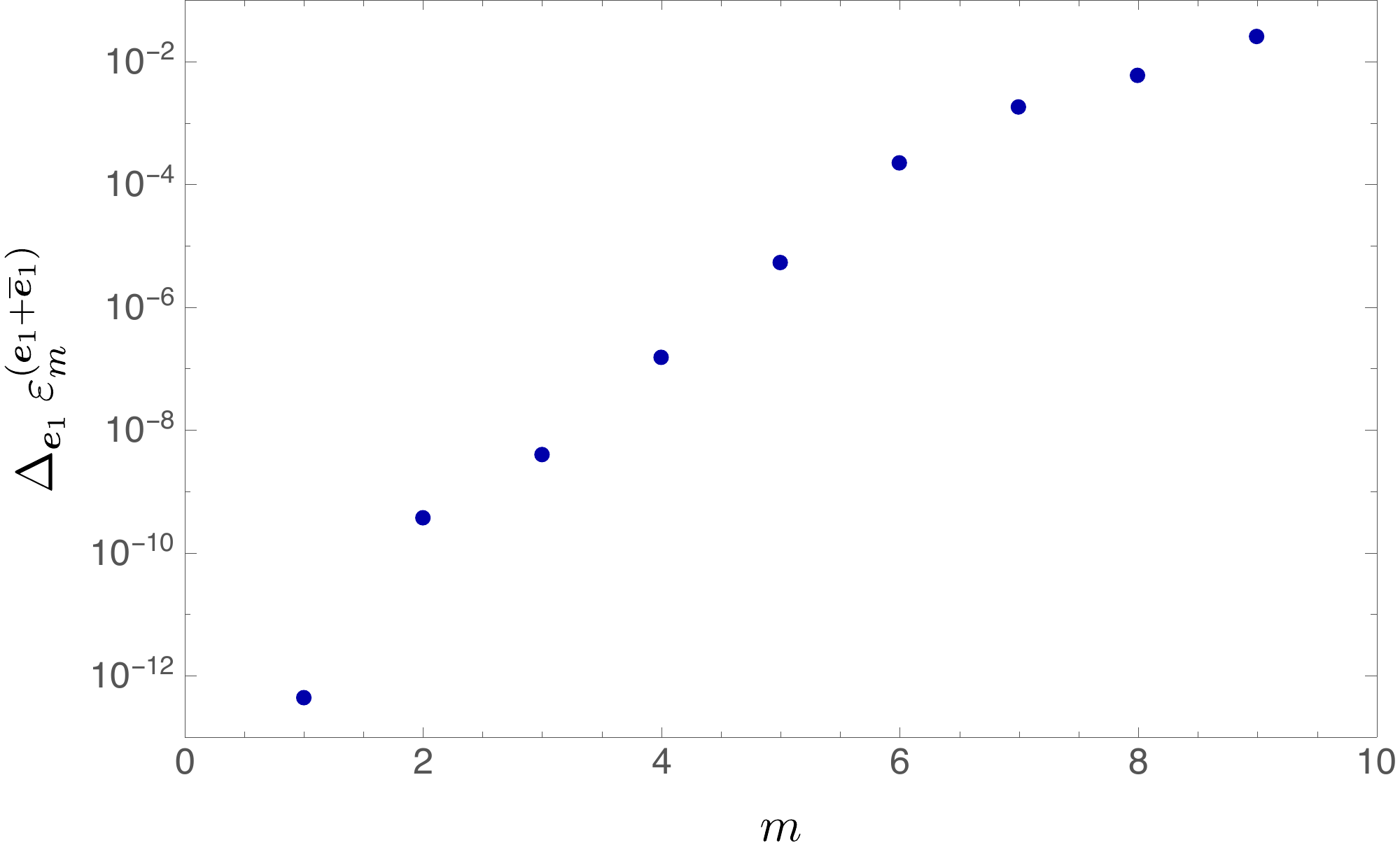}
  \caption{ \label{fig:predictionError-nl} Comparison of the predicted results from resurgence techniques and the numerical ones obtained from solving Einstein equations. In the above plots we show this comparison for the coefficients of the fundamental sector $\Phi_{\boldsymbol{e}_2}$ and mixed sectors $\Phi_{2\boldsymbol{e}_1}$ and $\Phi_{\boldsymbol{e}_1+\overline{\boldsymbol{e}}_1}$, as predicted by their relation to sector $\Phi_{\boldsymbol{e}_1}$ (following Eq. \eqref{eq:coeffs-comparison}).}
 \end{center}
\end{figure}

Applying the methodology introduced in \rfs{sec:transeries} (and explained in detail in \rfApp{app:app-res}) to the coefficients of sector $\Phi_{\boldsymbol{e}_1}$ we have predicted the values of the coefficients of the the fundamental sector $\Phi_{\boldsymbol{e}_2}$,  as well as of the mixed sectors $\Phi_{2\boldsymbol{e}_1},\, \Phi_{\boldsymbol{e}_1+\overline{\boldsymbol{e}}_1}$. The comparison between the resurgence predictions and the numerical gravity calculations can be found on the plots of Fig. \ref{fig:predictionError-nl}, where we have used the notation for the comparison of coefficients introduced in Eq. \eqref{eq:coeffs-comparison}. The following subsections discuss these developments in more detail.


\subsection{Resurgence of different fundamental sectors}
The first evidence of non-trivial relations between different QNMs comes from the fact that the leading large-order behaviour of the fundamental sector $\Phi_{\boldsymbol{e}_1}$ associated to the least damped QNM is governed
by the coefficients of the fundamental sector $\Phi_{\boldsymbol{e}_2}$, associated to the second QNM. The precise
relation can be analysed by focusing on the leading singularity $\xi=A_{2}-A_{1}$ of the Borel transform of
$\Phi_{\boldsymbol{e}_1}$. Applying an approach almost identical to Eq. \eqref{eq:boreltrans-behaviour-pole} we can
determine the Borel residue associated with this singular point:
\begin{eqnarray}
\mathsf{S}_{\boldsymbol{e}_{1}\rightarrow\boldsymbol{e}_2} & = & 2.6127578014515638725739302086318633648979300592843-\nonumber\\
 &  & -10.6770578911184045100721836940875685761278306458495\,\mathrm{i}.
\end{eqnarray}
This Borel residue appears as the proportionality constant for the leading large-order growth shown in Eq. \eqref{eq:large-order-e1-line1}.

Using the procedure shown in Eq. \eqref{eq:boreltrans-behaviour-subleading} we can further analyse the sub-leading large-order behaviour of the  series $\Phi_{\boldsymbol{e}_1}$ at the singular Borel plane
point $\xi=A_{2}-A_1$ and obtain a prediction of the coefficients of sector
$\Phi_{\boldsymbol{e}_2}$.
The comparison between the values of the first 11 coefficients $\varepsilon_{m}^{(\boldsymbol{e}_{2})}$ predicted using resurgence and their values calculated from the bulk solution is found in the top plot of Fig.
\ref{fig:predictionError-nl}. The error observed there is very small, confirming
the presence of nontrivial resurgence relations between different fundamental sectors, and
consequently different QNMs. This provides further evidence for the correctness
of the picture based on resurgence theory.

 \subsection{Mixed sectors}
Until this moment we have dealt solely with the resurgent relations between
fundamental sectors appearing in our transseries for the energy density, Eq.
\eqref{eq:trans-en-dens}. However, the full transseries \eqref{eq:trans-en-dens}
also includes mixed sectors which do not correspond to single QNMs: they are in
fact expansions in sectors whose exponential weight is given by linear
combinations of the QNM frequencies as a result of the coupling between
different fundamental sectors. The presence of such mixed sectors was already
shown by the direct gravity calculation of \rfs{sec:bulksolution}, and it can be
further justified via their resurgent relations to the fundamental sectors.

To determine the sub-leading, exponentially suppressed, behaviour of the
$\Phi_{\boldsymbol{e}_{1}}$ sector  shown on the second line of Eq.
\eqref{eq:large-order-e1-line1}, we want to analyse the  singularities on the
Borel plane at  $\xi+A_1=2A_{1}$ and $\xi+A_1=A_1+\overline{A_1}$. As explained
in \rfs{sec:transeries} and applied in \rfs{sec:res-fund-sectors}, we can do so
by performing a resummation of the contributions from leading singularity
$\xi+A_1=A_{2}$, shown on the first line of Eq. \eqref{eq:large-order-e1-line1},
and determine a subtracted series:
\be
\delta_{1}\Phi_{\boldsymbol{e}_{1}} \simeq u^{-\beta_{\boldsymbol{e}_1}}\sum_{k=1}^{+\infty}\delta_{1}\varepsilon_{k}^{(\boldsymbol{e}_1)}\,u^{-k}\,,
\ee
where the subtracted coefficients are given by the laterally resummed series
\be
\delta_1\varepsilon_k^{(\boldsymbol{e}_1)}\equiv \varepsilon_k^{(\boldsymbol{e}_1)}+
\frac{\mathsf{S}_{\boldsymbol{e}_1\rightarrow \boldsymbol{e}_2}}{2\pi\mathrm {i}}\,\frac{\Gamma(k+\beta_{\boldsymbol{e}_1}-\beta_{\boldsymbol{e}_2})}{A_1^{k+\beta_{\boldsymbol{e}_1}-\beta_{\boldsymbol{e}_2}}} \,\mathcal{S}_{0^+}\chi_{\boldsymbol{e}_1\rightarrow \boldsymbol{e}_2} \, ,
\ee
whose large-order behaviour is now given by the second line of Eq.
\eqref{eq:large-order-e1-line1}. Through the analysis of the leading
singularities of $\mathcal{B}\left[\delta_1 \Phi_{\boldsymbol{e}_1}\right](\xi)$
at $\xi+A_1=2A_1,\,A_1+\overline{A_1}$, we can now determine the Borel residues
associated with contributions of the mixed sectors
$\Phi_{2\boldsymbol{e}_1},\,\Phi_{\boldsymbol{e}_1+\overline{\boldsymbol{e}}_1}$
to the large-order behaviour of $\Phi_{\boldsymbol{e}_1}$:
\be
\mathsf{S}_{\boldsymbol{e}_{1}\rightarrow2\boldsymbol{e}_{1}}  = 2\mathsf{S}_{\boldsymbol{0}\rightarrow\boldsymbol{e}_{1}}\quad;\qquad
\mathsf{S}_{\boldsymbol{e}_{1}\rightarrow\boldsymbol{e}_{1}+\overline{\boldsymbol{e}}_{1}}  =  \mathsf{S}_{\boldsymbol{0}\rightarrow\overline{\boldsymbol{e}}_{1}} = -\overline{\mathsf{S}_{\boldsymbol{0}\rightarrow\boldsymbol{e}_{1}}}~.
\ee
These relations between Borel residues were expected as they are some of the
many such relations which can be found from the direct resurgent analysis of the
transseries in Eq. \eqref{eq:trans-en-dens} (see Secs. 2 and 5 of the review
\cite{Aniceto:2018bis} for more details).

We can continue our procedure further and predict the coefficients
$\varepsilon_{k}^{(2\boldsymbol{e}_{1})}$ and
$\varepsilon_{k}^{(\boldsymbol{e}_{1}+\overline{\boldsymbol{e}}_1)}$ of mixed
sectors $\Phi_{2
\boldsymbol{e}_1},\,\Phi_{\boldsymbol{e}_1+\overline{\boldsymbol{e}}_1}$,
respectively. The two bottom plots of Fig. \ref{fig:predictionError-nl} show the
comparison between resurgence prediction and numerical calculation for the first
8/9 of these coefficients.
While the errors are somewhat larger here (with some loss of accuracy due to the resummation process), they are still small enough to be confident that the presented picture, including the presence of mixed sectors in the full solution for the energy density \eqref{eq:trans-en-dens}, is fully consistent with expectations based on the theory of resurgence.

\section{Outlook}
\label{sec:outlook}

Through the AdS/CFT correspondence the problem of plasma equilibration can be
viewed equivalently as the process by which a dynamical black object evolves
toward its static state. The dissipative character of the black-brane horizon
translates into the dissipative properties of the late-time asymptotic solution
whose leading term was found in \rfc{Janik:2005zt}. Some of the subleading terms
were determined analytically in
Refs.~\cite{Janik:2006ft,Heller:2007qt,Booth:2009ct} and many more numerically
in Refs.~\cite{Heller:2013fn,Casalderrey-Solana:2017zyh} through a power series in the variable $u=\tau^{2/3}$. In this paper we presented a novel transseries solution for the bulk geometry which supplements the asymptotic power-series by including contributions which are non-perturbative (exponentially damped) in the late time expansion. These additional transseries sectors can be interpreted either as linearised perturbations of the late-time bulk solution or as non-linear couplings of these perturbations. They become most relevant at early times and naturally encode information about the initial data, which is ``lost'' as the system evolves toward equilibrium. These perturbations can be identified with quasinormal modes of the static AdS black-brane solution, modified by gradient corrections. We were able to explicitly compute the bulk solution expanded around several transseries sectors to high orders in the late-time expansion. Since the AdS/CFT correspondence maps QNMs of the black brane to nonhydrodynamic modes of the dual gauge theory plasma,  from the bulk solution we derived a transseries for the energy density of \symm ~plasma undergoing Bjorken flow.

On the field theory side, our bulk solution provides an asymptotic form of the
energy density as a function of proper time together with corrections which
depend on the initial state. The physical meaning of this object is the same as
at the level of hydrodynamics~\cite{Heller:2015dha,Aniceto:2015mto}. Even though
the energy density depends on a scale parameter (which we have fixed as in
Ref.~\cite{Heller:2013fn}), it is independent of other features of the initial
state, which enter only through the exponentially suppressed contributions.
Thus, the transseries for the energy density has a similar interpretation as the
transseries for quantities which are universal in the sense of
Ref.~\cite{Spalinski:2018mqg} -- it describes the dissipation of initial state
information in the process of
hydrodynamization~\cite{Heller:2015dha,Aniceto:2015mto,Heller:2018qvh,Spalinski:2018mqg}.

From a mathematical perspective our results provide a novel example where
resurgence is at work: the late-time expansion of a strongly coupled QFT
computed in a microscopic theory. The intricate large-order relations which
express  the resurgent properties of our transseries solution provide strong
links between the original hydrodynamic series and \textit{all} the
non-hydrodynamic sectors. The predictive power of these relations is such that
one can in principle recover all nonhydro sectors directly from the large-order
behaviour of the coefficients of the hydro sector (the limiting factor being the
actual number of coefficients determined). One can ask how is this possible,
given that the Borel plane of the hydro sector (top plot of Fig.
\ref{fig:BorelPlanes}) does not directly show all the mixed sectors? The answer
is that this is an iterative process: the hydro series returns full information
about all fundamental sectors, which in turn provide all the information about
the mixed sectors which was previously not visible.

The asymptotic series which arise in this way show similarities with those
arising in the context of the hydrodynamic model of Ref.~\cite{Heller:2014wfa},
whose asymptotic behaviour was investigated in Ref.~\cite{Aniceto:2015mto}. In
particular, the large-order behaviour of these series involve multiple,
competing factorial growths dependent of different complex parameters
(exponential weights and characteristic exponents), which make it impossible to
apply standard tools used in many recent studies of asymptotic series. We have
however succeeded in developing an equivalent method for extracting information
from asymptotic series directly at the level of the Borel plane singularities,
which has made it possible to effectively calculate Borel residues (Stokes
constants) and verify large-order relations. In Secs. \ref{sec:res-fund-sectors}
and \ref{sec:res-mixed-sector} presented extremely accurate numerical evidence
that both nonhydro fundamental and mixed sectors need to be included in the full
solution in order to have a complete picture of the fluid's energy density.
Although natural from the point of view of resurgent transseries, the existence of these mixed sectors, interpreted as
the result of coupling between different QNMs, is completely novel.

Finally, we comment on the applicability of our approach to early time dynamics.
Our transseries solution, once supplemented by initial data, provides a full
non-perturbative description of the energy density of the fluid. Thus it
naturally extends the regime of validity of the original hydrodynamic expansion.
Although the transseries was formally defined for large-proper times, one can
use a summation prescription on every (asymptotic) sector to obtain the energy
density for smaller values of proper time. Naturally the smaller the proper
time, the more nonhydro sectors one will need to consider to obtain an accurate
result. Of possible summation prescriptions the most widely used in the
so-called Borel-Pad\'{e}-\'{E}calle summation procedure (see
\cite{Aniceto:2018bis}) as discussed, for example, in
\cite{Cherman:2014xia,Grassi:2014cla,Heller:2014wfa,Couso-Santamaria:2015wga,Aniceto:2015rua,Couso-Santamaria:2016vwq,Casalderrey-Solana:2017zyh}.
Such results may help to shed light on the question which initial conditions
correspond to the early-time attractor in \symm
\cite{Romatschke:2017vte,Spalinski:2017mel}. Alternative summation procedures
can also be adopted to analyse the early-time regime, such as trans-asymptotics
and analytic transseries summation
\cite{Costin:1995ea,Costin:1998bs,Aniceto:toapp}.

Finally, let us note that the  explicit identification of nonhydrodynamic modes
with the quasinormal mode spectrum suggests that a generalisation of the
gradient expansion through the fluid-gravity duality~\cite{Bhattacharyya:2008jc}
should exist. Such a solution would systematically map exponentially suppressed
corrections in the bulk to nonhydrodynamic mode contributions in the dual
Yang-Mills theory even without the imposition of the symmetries of Bjorken flow.

\section*{Acknowledgements}
We would like to thank Jorge Casalderrey-Solana, Ovidiu Costin, Michał Heller,
Romuld Janik, Marcel Vonk and Przemysław Witaszczyk for useful discussions. IA has been
supported in part by the Polish National Science Centre (NCN) grants
2012/06/A/ST2/00396 and 2015/19/B/ST2/02824, and the  UK Engineering and
Physical Sciences Research Council (EPSRC) Early Career Fellowship EP/S004076/1.
JJ  was  supported by the NCN grant 2016/23/D/ST2/03125. BM is a Commonwealth
Scholar and is also supported by the Oppenheimer Fund Scholarship. MS is
supported by the NCN grant 2015/19/B/ST2/02824.

\newpage
\begin{appendices}

\section{Leading singularities in the Borel plane}
\label{app:app-sing}
The theory of resurgence can help us predict which singularities will appear in
the complex Borel plane associated to each asymptotic sector. This can be done
by following the procedure explained in Section 5 of \cite{Aniceto:2018bis},
applicable to a multi-parameter resurgent transseries with many fundamental
sectors (many Stokes directions) as well as the corresponding mixed sectors. The
techniques of resurgence theory thus allow us to predict which singular branch
points will be visible in the Borel planes of different asymptotic sectors of
the transseries for the energy density given by \rf{eq:trans-en-dens}.

In this appendix we list the expected branch points appearing in the Borel plane
$\xi$  of the asymptotic sectors discussed in the main text:

\begin{itemize}
    \item hydrodynamic sector $\Phi_{\boldsymbol{0}}$: $\xi=A_k,\,\overline{A_k},\,k=1,\cdots$ (singularities associated to fundamental sectors); $\xi=n\,A_k,\,\overline{A_k},\,k=1,\cdots,\,n=2,\cdots$ (singularities associated to mixed sectors);
    \item fundamental sector $\Phi_{\boldsymbol{e}_1}$: $\xi+A_1=\overline{A_1},\,A_k,\,\overline{A_k},\,k=2,\cdots$ (singularities associated to fundamental sectors); $\xi=n\,A_1,\,A_1+k\,\overline{A_1},\,A_1+k\,A_n,\,A_1+k\,\overline{A_n},\,k=1,\cdots,\,n=2,\cdots$ (singularities associated to mixed sectors);
    \item fundamental sector $\Phi_{\boldsymbol{e}_2}$: $\xi+A_2=A_1,\,\overline{A_1},\,\overline{A_2},\,A_k,\,\overline{A_k},\,k=3,\cdots$ (singularities associated to fundamental sectors); $\xi=(n-1)\,A_2,\,A_2+k\,A_1,\,A_2+k\,\overline{A_1},\,A_2+k\,\overline{A_2},\,A_2+k\,A_n,\,A_2+k\,\overline{A_n},\,k=1,\cdots,\,n=3,\cdots$ (singularities associated to mixed sectors);
    \item mixed sector $\Phi_{2\boldsymbol{e}_1}$: $\xi+2A_1=A_k,\,\overline{A_k},\,k=1,\cdots$ (singularities associated to fundamental sectors); $\xi=(n+1)A_1,\,2\overline{A_1},\,2A_n,\,2\overline{A_n},\,A_1+\overline{A_1},\,A_1+A_n,\,A_1+\overline{A_n},\,2A_1+k\,\overline{A_1},\,2A_1+k\,A_n,\,2A_1+k\,\overline{A_n},\,k=1,\cdots,\,n=2,\cdots$ (singularities associated to mixed sectors).
\end{itemize}

\section{Large-order predictions from Borel plane residues}
\label{app:app-res}

In this appendix we will present some of the details of the large-order analysis
of transseries sectors directly from the Borel plane point of view. This
approach leads to a systematic procedure where by taking residues of
singularities associated with a particular sector one can predict the
coefficients of the other non-perturbative sectors. This method is especially
useful in situations where we have complex-conjugate singularities in the Borel
plane $A,\overline{A}$ and multiple contributions characterised by different
factorial growths $\Gamma\left(k+\beta_i\right)$ for $k\gg1$ with
$\beta_i\in\mathbb{C}$. This leads to intricate oscillatory large-order
behaviour of the coefficients of our asymptotic series, from which the usual
acceleration methods (such as Richardson transforms) cannot be effectively
applied. This  is a major complication if one wishes to calculate the Stokes
constants or expansion coefficients appearing in nontrivial transseries sectors.

\subsection{Behaviour of an asymptotic series at the first Borel singularity}
\label{sec:app-A-first-sing}

We assume an asymptotic series expansion (for $u\gg1$) of the form
\begin{equation}
\Phi^{(0)}(u)\simeq u^{-\beta_{0}}\sum_{n=0}^{+\infty}F_{n}^{(0)}\,\f{1}{u^{n}},
\label{eq:res-pert-series}
\end{equation}
whose coefficients $F_{n}^{(0)}$ show large-order factorial growth. This growth
will contain possibly multiple contributions, each of which growing as
$\Gamma(n+\beta)\,A^{-n-\beta}$ for $n\gg1$ for some possibly complex parameters
$\beta_{0},\beta,A$. The parameters $\beta$ are the so-called characteristic
exponents, defining the type of branch point singularities that we will find in
the Borel plane, while $A$ determines location of these singularities. In the
context of the present paper we could just write $\varepsilon_n$ instead of
$F_n$ here, but the actual method is clearly more general than the present
context.

Given the factorial growth of the coefficients in the hydrodynamic (in general: perturbative) series in
\rf{eq:res-pert-series}, it is convenient to define a one-parameter family of
series\footnote{This definition is appropriate for analysing the hydrodynamic
sector, but it can be trivially modified so it can be used for the
nonhydrodynamic sectors, as done in \rfs{sec:res-mixed-sector}.}
\begin{equation}
\Phi_{\alpha}^{(k)}(u)=u^{-\alpha-\beta+\beta_{0}}\Phi^{(k)}(u)\, ,
\label{eq:res-general-beta}
\end{equation}
whose Borel transform (defined via the rule $u^{-\delta}\mapsto
\xi^{\delta-1}/\Gamma(\delta)$) will have a singularity structure dependent on
the choice of $\alpha$ (see Section 4 of \cite{Aniceto:2018bis} for more on this
subject). In the vicinity of the singular point $\xi=A$ the Borel transform for
the particular case of $\alpha=0$ has the standard behaviour of simple resurgent
functions:\footnote{Depending on the value of $\beta$ one might need the exclude
any constant term in  the series, as the Borel transform is only defined for
positive powers of $1/u$. Such a constant term should then be added after the
summation procedure is performed.}
\begin{equation}
\mathcal{B}\left[\Phi_{0}^{(0)}\right](\xi) = \sum_{n=0}^{+\infty}\frac{F_{n}^{(0)}}{\Gamma\left(n+\beta\right)}\xi^{n+\beta-1}\sim
\mathsf{S}_{0\rightarrow 1}\,\mathcal{B}\left[\Phi^{(1)}_{0}\right](\xi-A)\,\frac{\log(\xi-A)}{2\pi\mathrm{i}}+\mathrm{regular}.
\label{eq:res-Borel-simple-resurgent}
\end{equation}
In this equation $\mathsf{S}_{0\rightarrow 1}$ is a a Borel residue, given by the first Stokes constant of the problem, and $\Phi^{(1)}_0$ is defined via \rf{eq:res-general-beta} with
\begin{equation}
\Phi^{(1)}(u)\simeq u^{-\beta_{1}}\sum_{n=0}^{+\infty}F_{n}^{(1)}\f{1}{u^{n}}.
\label{eq:np-series-gen}
\end{equation}
This asymptotic series will be associated with the first nonhydrodynamic (in general: nonperturbative) sector appearing in our transseries (with $\left|A_2\right|>\left|A\right|$)
\begin{equation}
F\left(u,\sigma_i \right)=\Phi^{(0)}(u)+\sigma_1\, \mathrm{e}^{-A\, u}\,\Phi^{(1)}(u)+\sigma_2\, \mathrm{e}^{-A_2\, u}\,\Phi^{(2)}(x)+\cdots.
\label{eq:trans-example}
\end{equation}

The key point of this procedure is that given the behaviour in \rf{eq:res-Borel-simple-resurgent}, we can transform the logarithmic behaviour into a square root behaviour. To see this,
it will be very convenient to define
\bel{psidef}
\Psi^{(0)}\left(\xi\right) \equiv \xi^{\frac{1}{2}-\beta}\mathcal{B}\left[\Phi_{\frac{1}{2}}^{(0)}\right](\xi),
\ee
which, as we now proceed to show, has a square root branch point singularity at $\xi = A$.

We begin by observing that in the vicinity of that point one has
\begin{equation}
\mathcal{B}\left[\Phi_{\frac{1}{2}}^{(0)}\right](\xi)|_{\xi=A} =
\frac{\mathsf{S}_{0\rightarrow 1}}{2}\mathcal{B}\left[\Phi^{(1)}_{\frac{1}{2}}\right](\xi-A)+\mathrm{regular}~.\label{eq:res-Borel-pert-1/2}
\end{equation}
The validity of this relation can be verified using a simple generalisation of
the methods described in Section 4 of the review \cite{Aniceto:2018bis} for the
case of the quartic integral.\footnote{In general, one can obtain Eq.
(\ref{eq:res-Borel-simple-resurgent}) from Eq. (\ref{eq:res-Borel-pert-1/2}) by
evaluating the semi derivative associated with
$\mathcal{B}\left[\Phi_{\frac{1}{2}}^{(0)}\right](\xi)$ term by term, expanding
the result about $(\xi - A)$ and isolating the contributions that are linear in
$\log{(\xi-A)}$. The reverse will also be true, so Eq.
(\ref{eq:res-Borel-pert-1/2}) will always hold.} On the other hand, by
definition,
\begin{equation}
\mathcal{B}\left[\Phi^{(1)}_{\frac{1}{2}}\right](\xi)=
\xi^{\beta_{1}-\beta_{0}+\beta-\frac{1}{2}}\sum_{n=0}^{+\infty}\frac{F_{n}^{(1)}}{\Gamma\left(n+\beta_{1}-\beta_{0}+\beta+\frac{1}{2}\right)}\,\xi^{n}~,
\end{equation}
for general values of $\beta_{i}$ this will have a complicated leading behaviour
at $\xi=0$. However, when $\Phi^{(0)}$ and $\Phi^{(1)}$ are two different
sectors of a transseries solution of the form of Eq. \eqref{eq:trans-example},
the factorial growth of the coefficients of the original series
\eqref{eq:res-pert-series} will obey $\beta=\beta_{0}-\beta_{1}$. In this case
the previous expression becomes
\begin{equation}
\label{borsimp}
\mathcal{B}\left[\Phi^{(1)}_{\frac{1}{2}}\right](\xi)=
\xi^{-\frac{1}{2}}\sum_{n=0}^{+\infty}\frac{F_{n}^{(1)}}{\Gamma\left(n+\frac{1}{2}\right)}\,\xi^{n}~.
\end{equation}
Using \rf{borsimp} and \rf{eq:res-Borel-pert-1/2} we find that in the vicinity of $\xi=A$
\begin{eqnarray}
\label{eq:res-pert-Borel-1/2-expanded-in-s}
\Psi^{(0)}\left(\xi\right) |_{\xi=A} =  \frac{\mathsf{S}_{0\rightarrow 1}\,A^{\frac{1}{2}-\beta}}{2\sqrt{\xi-A}}\sum_{n=0}^{+\infty}\,\left(\xi-A\right)^{n}
 \times\sum_{k=0}^{n}\frac{(-1)^{k}}{k!\,(2A)^{k}}\,\frac{F_{n-k}^{(1)}}{\Gamma\left(n-k+\frac{1}{2}\right)}\,\prod_{j=0}^{k-1}\left(2\beta+2j-1\right)
\end{eqnarray}
where we also used that
\begin{equation}
\left.\xi^{\frac{1}{2}-\beta}\right|_{\xi=A}= A^{\frac{1}{2}-\beta}\sum_{k=0}^{+\infty}\frac{(-1)^{k}}{k!\,(2A)^{k}}\left(\xi-A\right)^{k}\,\prod_{j=0}^{k-1}\left(2\beta+2j-1\right).
\end{equation}
As anticipated, \rf{eq:res-pert-Borel-1/2-expanded-in-s} has a square root branch point singularity.

The next step involves changing the Borel plane variable in order to more easily analyse the behaviour around each singular branch point. Such changes of variables have been used in the literature, in particular through the use of conformal maps, see \textit{e.g.} \cite{Jentschura:2000fm,ZinnJustin:2004cg,Caliceti:2007ra}, as well as recent applications \cite{Costin:2017ziv}.
In our case, it will be useful to change variables to transform the square root behaviour into a pole. This in turn will make it possible to extract information from \rf{eq:res-pert-series} by calculating residues. The required change of variable is
\bel{zetadef}
\xi = A+\mathrm{e}^{\mathrm{i}\pi}\left(\zeta-A\right)^{2}.
\ee
Note that the the point $\xi=A$ corresponds to $\zeta=A$. With this change of variables it follows from (\ref{eq:res-pert-Borel-1/2-expanded-in-s}) that
\be
\Psi^{(0)}\left(\zeta\right) |_{\zeta=A} & = & \frac{ \mathsf{S}_{0\rightarrow 1}}{2\mathrm{i}}\,\frac{A^{\frac{1}{2}-\beta}}{(\zeta-A)}\,\sum_{n=0}^{+\infty}\,\left(\zeta-A\right)^{2n}\,C_{n}^{(0\rightarrow 1)}\,,\label{eq:res-pert-Borel-1/2-expanded-in-t}
\ee
where
\begin{equation}
C_{n}^{(0\rightarrow 1)}\equiv \sum_{k=0}^{n}\frac{(-1)^{n-k}}{k!\,(2A)^{k}}\,\frac{F_{n-k}^{(1)}}{\Gamma\left(n-k+\frac{1}{2}\right)}\,\prod_{j=0}^{k-1}\left(2\beta+2j-1\right).\label{eq:res-pert-Borel-1/2-expanded-in-t-coeffs}
\end{equation}
The $C_{n}^{(0\rightarrow 1)}$ depend solely on the coefficients
of the non-perturbative series \eqref{eq:np-series-gen}, with each $C_{n}^{(0\rightarrow 1)}$ depending on all the coefficients $F_{k}^{(1)}$ for $k\le n$, for example:
\begin{eqnarray}
C_{0}^{(0\rightarrow 1)}& = & \frac{F_{0}^{(1)}}{\Gamma(1/2)}\,;\nonumber\\
C_{1}^{(0\rightarrow 1)}& = &
-\frac{F_{1}^{(1)}}{\Gamma\left(3/2\right)}+\frac{F_{0}^{(1)}}{\Gamma\left(1/2\right)}\,\frac{\left(2\beta-1\right)}{2A}\,.
\label{eq:res-sums-of-coeffs}
\end{eqnarray}

It is now straightforward to see that evaluating residues of \rf{psidef} will
give us direct access to these coefficients, as well as the Borel residue (or,
equivalently, the Stokes constant). To effectively perform this analysis, we
need to use the hydrodynamic expansion coefficients to calculate the series
$\Psi^{(0)}\left(\xi\right)$ and perform the change of variable in \rf{zetadef}.
Expanding the result around $\zeta_{0}=A-\sqrt{A}$ (which is the regular point
of the Borel transform corresponding to $\xi=0$) one finds
\begin{eqnarray}
\Psi^{(0)}\left(\zeta\right)
&=&\sum_{n=0}^{+\infty}\frac{F_{n}^{(0)}\,\left(2A^{1/2}\right)^{n}}{\Gamma\left(n+\beta+\frac{1}{2}\right)}\sum_{k=0}^{+\infty}\frac{\left(-1\right)^{k}}{\left(2A^{1/2}\right)^{k}}\frac{n!}{k!(n-k)!}\left(\zeta-\zeta_{0}\right)^{k+n}\nonumber \\  & = & \sum_{n=0}^{+\infty}\left(\zeta-\zeta_{0}\right)^{n}\,\sum_{k=0}^{n}\frac{F_{n-k}^{(0)}\,\left(2A^{1/2}\right)^{n-2k}\left(-1\right)^{k}}{\Gamma\left(n-k+\beta+\frac{1}{2}\right)}\frac{(n-k)!}{k!(n-2k)!}.
\end{eqnarray}
Assuming that we have $N$ terms for this series, we define an analytic
continuation of this truncation using the Padé approximant
$\mathrm{BP}_{\frac{N}{2}}\left[\Psi^{(0)}\right]\left(\zeta\right)$. By virtue of the
arguments presented earlier, this quantity should have an isolated pole at
$\zeta = A$. This can be verified directly, and we can then calculate the
residue at that point. The result can be compared with the value obtained from
the asymptotic calculation (\ref{eq:res-pert-Borel-1/2-expanded-in-t}), which
leads to the conclusion that
\begin{equation}
\mathrm{Res}_{\zeta=A}\mathrm{BP}_{N}\left[\Psi^{(0)}\right](\zeta)=
\frac{A^{\frac{1}{2}-\beta}}{2\,\mathrm{i}\,\Gamma(1/2)}\,\mathsf{S}_{0\rightarrow 1}\,F_{0}^{(1)}\, .
\label{eq:res-first-Stokes-const}
\end{equation}
This gives explicit access to the Borel residue $\mathsf{S}_{0\rightarrow 1}$,
since all the remaining quantities appearing in \rf{eq:res-first-Stokes-const}
are known (recall that we normalize the leading coefficients in the fundamental
nonhydrodynamic sectors to $1$).

Once the Borel residue is known, we can gain access to higher coefficients $F_{m}^{(1)}$ of the non-perturbative series $\Phi^{(1)}$ (which first appear in $C_{m}^{(0\rightarrow 1)}$ of (\ref{eq:res-pert-Borel-1/2-expanded-in-t})), by calculating the residue of
\begin{equation}
B_{m}\Phi^{(0)}\equiv\left\{ \frac{2\,\mathrm{i}\,A^{\beta-\frac{1}{2}}}{\mathsf{S}_{0\rightarrow 1}}\mathrm{BP}_{N}\left[\Psi^{(0)}\right]-\sum_{n=0}^{m-1}\,\left(\zeta-A\right)^{2n-1}\,C_{n}^{(0\rightarrow 1)}\right\} \left(\zeta-A\right)^{-2m}.
\end{equation}
Indeed this residue will give
\begin{equation}
\mathrm{Res}_{\zeta=A}B_{m}\Phi^{(0)}=C_{m}^{(0\rightarrow 1)}.\label{eq:res-subleading-pert}
\end{equation}
The case $m=0$, corresponds to the calculation of the Borel residue found in Eq. \eqref{eq:res-first-Stokes-const}. 
With the value of this Borel residue one can predict the value of the coefficients $F_m^{(1)}$, with $m>0$. For example, for $m=1$ we can use Eq. \eqref{eq:res-sums-of-coeffs} to find
\be
F_{1}^{(1)}=F_{0}^{(1)}\,\frac{\Gamma\left(3/2\right)}{\Gamma\left(1/2\right)}\,\frac{\left(2\beta-1\right)}{2A}-\Gamma\left(3/2\right)\,\mathrm{Res}_{\zeta=A}B_{1}\Phi^{(0)}\,.
\ee
%

\subsection{Behaviour of an asymptotic series at its second Borel singularity}
\label{sec:app-A-second-sing}

We have analysed the properties of the leading branch cut (closest to the
origin) appearing in the Borel plane of the perturbative series. The position of
the branch point was $\xi=A$. In order to differentiate between different branch
point in the Borel plane, let us redefine $A=A_{1}$. We would now like to
analyse the behaviour around the the singularities in the Borel plane further
away from the origin, such as the one at $\xi=A_{2}$ (where
$\left|A_2\right|>\left|A_1\right|$). As it was explained in
\rfs{sec:transeries}, in order to analyse the branch cut at this singularity, we
need to first subtract the contributions from the closer one at $\xi=A_{1}$. We
saw that this can done by removing the leading large-order behaviour from the
coefficients of the perturbative series. In much the same way as in Eqs.
\eqref{eq:large-order-pert-line1}--\eqref{eq:large-order-pert-line3}, this
leading behaviour is given by (see \cite{Aniceto:2018bis} for further details):
\begin{equation}
F_{n}^{(0)}\simeq-\frac{\mathsf{S}_{0\rightarrow 1}}{2\pi\mathrm{i}}\frac{\Gamma\left(n+\beta_{0}-\beta\right)}{A_{1}^{n+\beta_{0}-\beta}}\chi_{0\rightarrow 1}(n)+\mathcal{O}\left(\left|A_{2}\right|^{-n}\right)\, ,
\end{equation}
 where
\begin{equation}
\chi_{0\rightarrow1}(n)\simeq\,\sum_{h=0}^{+\infty}\,\frac{\Gamma\left(n+\beta_{0}-\beta-h\right)}{\Gamma\left(n+\beta_{0}-\beta\right)}\,A_{1}^{h}\,F_{h}^{(1)}.
\end{equation}
The sum appearing in $\chi_{0\rightarrow1}(n)$ is to be taken as a series in
$n^{-1}$ for large-order $n$, and is asymptotic. We can then perform a
Borel-Padé-Écalle resummation for each value of $n$.\footnote{We determine its
Borel transform, approximate it via the method of Padé approximants, and do a
Borel summation of the result along the real axis, as $n\in\mathbb{N}$. See the
reviews \cite{Aniceto:2018bis,Aniceto:2017me}.} However, given that the
singularities of the corresponding Borel transform also fall on the path of
integration (the real axis), we will need to perform a lateral summation
$\mathcal{S}_{0^{+}}\chi_{0\rightarrow1}(n)$, as defined in Eq.
\eqref{eq:large-order-resum} (for a Padé approximant of order $N$). The
subtracted coefficients
\be
\delta_{1}F_{n}^{(0)}  \equiv F_{n}^{(0)}+\frac{\mathsf{S}_{0\rightarrow 1}}{2\pi\mathrm{i}}\frac{\Gamma\left(n+\beta_{0}-\beta\right)}{A_{1}^{n+\beta_{0}-\beta}}\,\mathcal{S}_{0^{+}}\chi_{0\rightarrow1}(n)
\ee
obey new large-order relations governed by the coefficients $F_{k}^{(2)}$ of the next asymptotic sector  $\Phi^{(2)}\left(u\right)$ in the transseries solution to our problem \eqref{eq:trans-example}.

To check the large-order behaviour of these subtracted coefficients we perform the analysis described above, but now applied to Borel transform of the series
\begin{equation}
\delta_{1}\Phi^{(0)}(u)\simeq \,u^{-\beta_{0}-1}\sum_{n=0}^{+\infty}\delta_{1}F_{n+1}^{(0)}\,u^{-n}.
\end{equation}
and compare the residues $\mathrm{Res}_{\zeta=A_2}\, B_m \delta_1\Phi^{(0)}$ to
the corresponding coefficients $C_m^{(0\rightarrow 2)}$. Note that the the
resummation process can only be done for positive $n$, and we cannot determine
the coefficient  $\delta_{1}F_{0}^{(0)}$. Thus, the leading term of the
subtracted series $\delta_{1}\Phi^{(0)}(u)$ is now $u^{-\beta_{0}-1}$.

\end{appendices}

\newpage
\bibliography{resurge}

\providecommand{\href}[2]{#2} \providecommand{\beforedoihref}{}
  \providecommand{\afterdoihref}{}\begingroup\raggedright\begin{thebibliography}{10}

\bibitem{Florkowski:2017olj}
W.~Florkowski, M.~P. Heller and M.~Spali{\'n}ski, {\it {New theories of
  relativistic hydrodynamics in the LHC era}},
  \beforedoihref\href{http://dx.doi.org/10.1088/1361-6633/aaa091}{Rept. Prog.
  Phys.}\afterdoihref\  {\bf 81} (2018), no.~4 046001
  [\href{http://arXiv.org/abs/1707.02282}{{arXiv:1707.02282}}].

\bibitem{Romatschke:2017ejr}
P.~Romatschke and U.~Romatschke, {\it {Relativistic Fluid Dynamics Out of
  Equilibrium}},  \href{http://arXiv.org/abs/1712.05815}{{arXiv:1712.05815}}.

\bibitem{Chesler:2009cy}
P.~M. Chesler and L.~G. Yaffe, {\it {Boost invariant flow, black hole
  formation, and far-from-equilibrium dynamics in N = 4 supersymmetric
  Yang-Mills theory}},
  \beforedoihref\href{http://dx.doi.org/10.1103/PhysRevD.82.026006}{Phys.Rev.}\afterdoihref\
  {\bf D82} (2010) 026006
  [\href{http://arXiv.org/abs/0906.4426}{{arXiv:0906.4426}}].

\bibitem{Heller:2011ju}
M.~P. Heller, R.~A. Janik and P.~Witaszczyk, {\it {The characteristics of
  thermalization of boost-invariant plasma from holography}},
  \beforedoihref\href{http://dx.doi.org/10.1103/PhysRevLett.108.201602}{Phys.Rev.Lett.}\afterdoihref\
  {\bf 108} (2012) 201602
  [\href{http://arXiv.org/abs/1103.3452}{{arXiv:1103.3452}}].

\bibitem{Jankowski:2014lna}
J.~Jankowski, G.~Plewa and M.~Spali{\'n}ski, {\it {Statistics of thermalization
  in Bjorken Flow}},
  \beforedoihref\href{http://dx.doi.org/10.1007/JHEP12(2014)105}{JHEP}\afterdoihref\
  {\bf 12} (2014) 105
  [\href{http://arXiv.org/abs/1411.1969}{{arXiv:1411.1969}}].

\bibitem{Spalinski:2016fnj}
M.~Spali{\'n}ski, {\it {Small systems and regulator dependence in relativistic
  hydrodynamics}},
  \beforedoihref\href{http://dx.doi.org/10.1103/PhysRevD.94.085002}{Phys.
  Rev.}\afterdoihref\  {\bf D94} (2016), no.~8 085002
  [\href{http://arXiv.org/abs/1607.06381}{{arXiv:1607.06381}}].

\bibitem{Romatschke:2016hle}
P.~Romatschke, {\it {Do nuclear collisions create a locally equilibrated
  quark–gluon plasma?}},
  \beforedoihref\href{http://dx.doi.org/10.1140/epjc/s10052-016-4567-x}{Eur.
  Phys. J.}\afterdoihref\  {\bf C77} (2017), no.~1 21
  [\href{http://arXiv.org/abs/1609.02820}{{arXiv:1609.02820}}].

\bibitem{Heller:2013fn}
M.~P. Heller, R.~A. Janik and P.~Witaszczyk, {\it {Hydrodynamic Gradient
  Expansion in Gauge Theory Plasmas}},
  \beforedoihref\href{http://dx.doi.org/10.1103/PhysRevLett.110.211602}{Phys.Rev.Lett.}\afterdoihref\
  {\bf 110} (2013), no.~21 211602
  [\href{http://arXiv.org/abs/1302.0697}{{arXiv:1302.0697}}].

\bibitem{Heller:2015dha}
M.~P. Heller and M.~Spali{\'n}ski, {\it {Hydrodynamics Beyond the Gradient
  Expansion: Resurgence and Resummation}},
  \beforedoihref\href{http://dx.doi.org/10.1103/PhysRevLett.115.072501}{Phys.
  Rev. Lett.}\afterdoihref\  {\bf 115} (2015), no.~7 072501
  [\href{http://arXiv.org/abs/1503.07514}{{arXiv:1503.07514}}].

\bibitem{Romatschke:2017vte}
P.~Romatschke, {\it {Relativistic Fluid Dynamics Far From Local Equilibrium}},
  \beforedoihref\href{http://dx.doi.org/10.1103/PhysRevLett.120.012301}{Phys.
  Rev. Lett.}\afterdoihref\  {\bf 120} (2018), no.~1 012301
  [\href{http://arXiv.org/abs/1704.08699}{{arXiv:1704.08699}}].

\bibitem{Spalinski:2017mel}
M.~Spali{\'n}ski, {\it {On the hydrodynamic attractor of Yang–Mills plasma}},
   \beforedoihref\href{http://dx.doi.org/10.1016/j.physletb.2017.11.059}{Phys.
  Lett.}\afterdoihref\  {\bf B776} (2018) 468--472
  [\href{http://arXiv.org/abs/1708.01921}{{arXiv:1708.01921}}].

\bibitem{Maldacena:1997re}
J.~M. Maldacena, {\it {The Large N limit of superconformal field theories and
  supergravity}},  Adv.Theor.Math.Phys. {\bf 2} (1998) 231--252
  [\href{http://arXiv.org/abs/hep-th/9711200}{{arXiv:hep-th/9711200}}].

\bibitem{CasalderreySolana:2011us}
J.~Casalderrey-Solana, H.~Liu, D.~Mateos, K.~Rajagopal and U.~A. Wiedemann,
  {\it {Gauge/String Duality, Hot QCD and Heavy Ion Collisions}},
  \href{http://arXiv.org/abs/1101.0618}{{arXiv:1101.0618}}.

\bibitem{Bjorken:1982qr}
J.~Bjorken, {\it {Highly Relativistic Nucleus-Nucleus Collisions: The Central
  Rapidity Region}},
  \beforedoihref\href{http://dx.doi.org/10.1103/PhysRevD.27.140}{Phys.Rev.}\afterdoihref\
  {\bf D27} (1983) 140--151.

\bibitem{Janik:2005zt}
R.~A. Janik and R.~B. Peschanski, {\it {Asymptotic perfect fluid dynamics as a
  consequence of Ads/CFT}},
  \beforedoihref\href{http://dx.doi.org/10.1103/PhysRevD.73.045013}{Phys.
  Rev.}\afterdoihref\  {\bf D73} (2006) 045013
  [\href{http://arXiv.org/abs/hep-th/0512162}{{arXiv:hep-th/0512162}}].

\bibitem{Casalderrey-Solana:2017zyh}
J.~Casalderrey-Solana, N.~I. Gushterov and B.~Meiring, {\it {Resurgence and
  Hydrodynamic Attractors in Gauss-Bonnet Holography}},
  \href{http://arXiv.org/abs/1712.02772}{{arXiv:1712.02772}}.

\bibitem{Baier:2007ix}
R.~Baier, P.~Romatschke, D.~T. Son, A.~O. Starinets and M.~A. Stephanov, {\it
  {Relativistic viscous hydrodynamics, conformal invariance, and holography}},
  \beforedoihref\href{http://dx.doi.org/10.1088/1126-6708/2008/04/100}{JHEP}\afterdoihref\
  {\bf 04} (2008) 100
  [\href{http://arXiv.org/abs/0712.2451}{{arXiv:0712.2451}}].

\bibitem{Heller:2014wfa}
M.~P. Heller, R.~A. Janik, M.~Spali{\'n}ski and P.~Witaszczyk, {\it {Coupling
  hydrodynamics to nonequilibrium degrees of freedom in strongly interacting
  quark-gluon plasma}},
  \beforedoihref\href{http://dx.doi.org/10.1103/PhysRevLett.113.261601}{Phys.Rev.Lett.}\afterdoihref\
  {\bf 113} (2014), no.~26 261601
  [\href{http://arXiv.org/abs/1409.5087}{{arXiv:1409.5087}}].

\bibitem{Aniceto:2015mto}
I.~Aniceto and M.~Spali{\'n}ski, {\it {Resurgence in Extended Hydrodynamics}},
  \beforedoihref\href{http://dx.doi.org/10.1103/PhysRevD.93.085008}{Phys.
  Rev.}\afterdoihref\  {\bf D93} (2016), no.~8 085008
  [\href{http://arXiv.org/abs/1511.06358}{{arXiv:1511.06358}}].

\bibitem{Edgar:2008tr}
G.~A. Edgar, {\it Transseries for beginners},  Real Analysis Exchange {\bf 35}
  (2010), no.~2 253--310
  [\href{http://arXiv.org/abs/0801.4877}{{arXiv:0801.4877}}].

\bibitem{Aniceto:2018bis}
I.~Aniceto, G.~Ba{\c s}ar and R.~Schiappa, {\it {A Primer on Resurgent
  Transseries and Their Asymptotics}},
  \href{http://arXiv.org/abs/1802.10441}{{arXiv:1802.10441}}.

\bibitem{Motl:2003cd}
L.~Motl and A.~Neitzke, {\it {Asymptotic black hole quasinormal frequencies}},
  \beforedoihref\href{http://dx.doi.org/10.4310/ATMP.2003.v7.n2.a4}{Adv. Theor.
  Math. Phys.}\afterdoihref\  {\bf 7} (2003), no.~2 307--330
  [\href{http://arXiv.org/abs/hep-th/0301173}{{arXiv:hep-th/0301173}}].

\bibitem{Andersson:2003fh}
N.~Andersson and C.~J. Howls, {\it {The Asymptotic quasinormal mode spectrum of
  nonrotating black holes}},
  \beforedoihref\href{http://dx.doi.org/10.1088/0264-9381/21/6/021}{Class.
  Quant. Grav.}\afterdoihref\  {\bf 21} (2004) 1623--1642
  [\href{http://arXiv.org/abs/gr-qc/0307020}{{arXiv:gr-qc/0307020}}].

\bibitem{Natario:2004jd}
J.~Natario and R.~Schiappa, {\it {On the classification of asymptotic
  quasinormal frequencies for d-dimensional black holes and quantum gravity}},
  \beforedoihref\href{http://dx.doi.org/10.4310/ATMP.2004.v8.n6.a4}{Adv. Theor.
  Math. Phys.}\afterdoihref\  {\bf 8} (2004), no.~6 1001--1131
  [\href{http://arXiv.org/abs/hep-th/0411267}{{arXiv:hep-th/0411267}}].

\bibitem{Ecalle:8185}
J.~\'{E}calle, {\it {Les Fonctions R\'{e}surgentes}},  Pr\'{e}pub. Math.
  Universit\'{e} Paris-Sud {\bf 81-05 \textnormal{(1981)}, 81-06
  \textnormal{(1981)}, 85-05} (1985).

\bibitem{Garoufalidis:2010ya}
S.~Garoufalidis, A.~Its, A.~Kapaev and M.~Mari{\~n}o, {\it {Asymptotics of the
  Instantons of Painlev{\'e} I}},  Int. Math. Res. Notices {\bf 2012} (2012)
  561 [\href{http://arXiv.org/abs/1002.3634}{{arXiv:1002.3634}}].

\bibitem{Aniceto:2011nu}
I.~Aniceto, R.~Schiappa and M.~Vonk, {\it {The Resurgence of Instantons in
  String Theory}},
  \beforedoihref\href{http://dx.doi.org/10.4310/CNTP.2012.v6.n2.a3}{Commun.Num.Theor.Phys.}\afterdoihref\
  {\bf 6} (2012) 339--496
  [\href{http://arXiv.org/abs/1106.5922}{{arXiv:1106.5922}}].

\bibitem{Schiappa:2013opa}
R.~Schiappa and R.~Vaz, {\it {The Resurgence of Instantons: Multi-Cut Stokes
  Phases and the Painlev\'{e} II Equation}},
  \beforedoihref\href{http://dx.doi.org/10.1007/s00220-014-2028-7}{Commun.Math.Phys.}\afterdoihref\
  {\bf 330} (2014) 655--721
  [\href{http://arXiv.org/abs/1302.5138}{{arXiv:1302.5138}}].

\bibitem{Aniceto:2015rua}
I.~Aniceto, {\it {The Resurgence of the Cusp Anomalous Dimension}},
  \beforedoihref\href{http://dx.doi.org/10.1088/1751-8113/49/6/065403}{J.
  Phys.}\afterdoihref\  {\bf A49} (2016) 065403
  [\href{http://arXiv.org/abs/1506.03388}{{arXiv:1506.03388}}].

\bibitem{Dorigoni:2015dha}
D.~Dorigoni and Y.~Hatsuda, {\it {Resurgence of the Cusp Anomalous Dimension}},
   \beforedoihref\href{http://dx.doi.org/10.1007/JHEP09(2015)138}{JHEP}\afterdoihref\
  {\bf 09} (2015) 138
  [\href{http://arXiv.org/abs/1506.03763}{{arXiv:1506.03763}}].

\bibitem{Arutyunov:2016etw}
G.~Arutyunov, D.~Dorigoni and S.~Savin, {\it {Resurgence of the dressing phase
  for AdS$_{5} ×$ S$^{5}$}},
  \beforedoihref\href{http://dx.doi.org/10.1007/JHEP01(2017)055}{JHEP}\afterdoihref\
  {\bf 01} (2017) 055
  [\href{http://arXiv.org/abs/1608.03797}{{arXiv:1608.03797}}].

\bibitem{Couso-Santamaria:2016vcc}
R.~Couso-Santamaría, R.~Schiappa and R.~Vaz, {\it {On asymptotics and
  resurgent structures of enumerative Gromov–Witten invariants}},
  \beforedoihref\href{http://dx.doi.org/10.4310/CNTP.2017.v11.n4.a1}{Commun.
  Num. Theor. Phys.}\afterdoihref\  {\bf 11} (2017) 707--790
  [\href{http://arXiv.org/abs/1605.07473}{{arXiv:1605.07473}}].

\bibitem{Bender:1990pd}
C.~M. Bender and T.~T. Wu, {\it {Anharmonic oscillator. 2: A Study of
  perturbation theory in large order}},
  \beforedoihref\href{http://dx.doi.org/10.1103/PhysRevD.7.1620}{Phys.
  Rev.}\afterdoihref\  {\bf D7} (1973) 1620--1636.

\bibitem{Marino:2007te}
M.~Mari{\~n}o, R.~Schiappa and M.~Weiss, {\it {Nonperturbative Effects and the
  Large--Order Behavior of Matrix Models and Topological Strings}},  Commun.
  Num. Theor. Phys. {\bf 2} (2008) 349
  [\href{http://arXiv.org/abs/0711.1954}{{arXiv:0711.1954}}].

\bibitem{Basar:2013eka}
G.~Ba{\c s}ar, G.~V. Dunne and M.~{\"U}nsal, {\it {Resurgence theory,
  ghost-instantons, and analytic continuation of path integrals}},
  \beforedoihref\href{http://dx.doi.org/10.1007/JHEP10(2013)041}{JHEP}\afterdoihref\
  {\bf 10} (2013) 041
  [\href{http://arXiv.org/abs/1308.1108}{{arXiv:1308.1108}}].

\bibitem{Couso-Santamaria:2014iia}
R.~Couso-Santamaría, J.~D. Edelstein, R.~Schiappa and M.~Vonk, {\it {Resurgent
  Transseries and the Holomorphic Anomaly: Nonperturbative Closed Strings in
  Local ${\mathbb{C}\mathbb{P}^2}$}},
  \beforedoihref\href{http://dx.doi.org/10.1007/s00220-015-2358-0}{Commun.
  Math. Phys.}\afterdoihref\  {\bf 338} (2015), no.~1 285--346
  [\href{http://arXiv.org/abs/1407.4821}{{arXiv:1407.4821}}].

\bibitem{Basar:2015ava}
G.~Ba{\c s}ar and G.~V. Dunne, {\it {Hydrodynamics, resurgence, and
  transasymptotics}},
  \beforedoihref\href{http://dx.doi.org/10.1103/PhysRevD.92.125011}{Phys.
  Rev.}\afterdoihref\  {\bf D92} (2015), no.~12 125011
  [\href{http://arXiv.org/abs/1509.05046}{{arXiv:1509.05046}}].

\bibitem{Aniceto:2017me}
I.~Aniceto, {\it Asymptotics, ambiguities and resurgence},  in {\em Resurgence,
  Physics and Numbers} (F.~Fauvet, D.~Manchon, S.~Marmi and D.~Sauzin, eds.),
  (Pisa), pp.~1--66, Scuola Normale Superiore, 2017.

\bibitem{Codesido:2016dld}
S.~Codesido and M.~Mari{\~n}o, {\it {Holomorphic Anomaly and Quantum
  Mechanics}},
  \beforedoihref\href{http://dx.doi.org/10.1088/1751-8121/aa9e77}{J.
  Phys.}\afterdoihref\  {\bf A51} (2018), no.~5 055402
  [\href{http://arXiv.org/abs/1612.07687}{{arXiv:1612.07687}}].

\bibitem{Demulder:2016mja}
S.~Demulder, D.~Dorigoni and D.~C. Thompson, {\it {Resurgence in
  $\eta$-deformed Principal Chiral Models}},
  \beforedoihref\href{http://dx.doi.org/10.1007/JHEP07(2016)088}{JHEP}\afterdoihref\
  {\bf 07} (2016) 088
  [\href{http://arXiv.org/abs/1604.07851}{{arXiv:1604.07851}}].

\bibitem{Gukov:2016njj}
S.~Gukov, M.~Mari{\~n}o and P.~Putrov, {\it {Resurgence in complex Chern-Simons
  theory}},  \href{http://arXiv.org/abs/1605.07615}{{arXiv:1605.07615}}.

\bibitem{Dorigoni:2017smz}
D.~Dorigoni and P.~Glass, {\it {The grin of Cheshire cat resurgence from
  supersymmetric localization}},
  \beforedoihref\href{http://dx.doi.org/10.21468/SciPostPhys.4.2.012}{SciPost
  Phys.}\afterdoihref\  {\bf 4} (2018) 012
  [\href{http://arXiv.org/abs/1711.04802}{{arXiv:1711.04802}}].

\bibitem{Codesido:2017jwp}
S.~Codesido, M.~Marino and R.~Schiappa, {\it {Non-Perturbative Quantum
  Mechanics from Non-Perturbative Strings}},
  \href{http://arXiv.org/abs/1712.02603}{{arXiv:1712.02603}}.

\bibitem{Balian:1978pv}
R.~Balian, G.~Parisi and A.~Voros, {\it Quartic oscillator},  \textnormal{in}
  \textit{Marseille Workshop on Feynman Path Integrals} (1978).

\bibitem{ZinnJustin:1980uk}
J.~Zinn-Justin, {\it {Perturbation Series at Large Orders in Quantum Mechanics
  and Field Theories: Application to the Problem of Resummation}},
  \beforedoihref\href{http://dx.doi.org/10.1016/0370-1573(81)90016-8}{Phys.
  Rept.}\afterdoihref\  {\bf 70} (1981) 109.

\bibitem{Janik:2006ft}
R.~A. Janik, {\it {Viscous plasma evolution from gravity using AdS/CFT}},
  \beforedoihref\href{http://dx.doi.org/10.1103/PhysRevLett.98.022302}{Phys.
  Rev. Lett.}\afterdoihref\  {\bf 98} (2007) 022302
  [\href{http://arXiv.org/abs/hep-th/0610144}{{arXiv:hep-th/0610144}}].

\bibitem{Heller:2007qt}
M.~P. Heller and R.~A. Janik, {\it {Viscous hydrodynamics relaxation time from
  AdS/CFT}},
  \beforedoihref\href{http://dx.doi.org/10.1103/PhysRevD.76.025027}{Phys.
  Rev.}\afterdoihref\  {\bf D76} (2007) 025027
  [\href{http://arXiv.org/abs/hep-th/0703243}{{arXiv:hep-th/0703243}}].

\bibitem{Booth:2009ct}
I.~Booth, M.~P. Heller and M.~Spali{\'n}ski, {\it {Black brane entropy and
  hydrodynamics: The Boost-invariant case}},
  \beforedoihref\href{http://dx.doi.org/10.1103/PhysRevD.80.126013}{Phys.
  Rev.}\afterdoihref\  {\bf D80} (2009) 126013
  [\href{http://arXiv.org/abs/0910.0748}{{arXiv:0910.0748}}].

\bibitem{Heller:2008mb}
M.~P. Heller, P.~Surowka, R.~Loganayagam, M.~Spali{\'n}ski and S.~E. Vazquez,
  {\it {Consistent Holographic Description of Boost-Invariant Plasma}},
  \beforedoihref\href{http://dx.doi.org/10.1103/PhysRevLett.102.041601}{Phys.
  Rev. Lett.}\afterdoihref\  {\bf 102} (2009) 041601
  [\href{http://arXiv.org/abs/0805.3774}{{arXiv:0805.3774}}].

\bibitem{Kinoshita:2008dq}
S.~Kinoshita, S.~Mukohyama, S.~Nakamura and K.-y. Oda, {\it {A Holographic Dual
  of Bjorken Flow}},
  \beforedoihref\href{http://dx.doi.org/10.1143/PTP.121.121}{Prog. Theor.
  Phys.}\afterdoihref\  {\bf 121} (2009) 121--164
  [\href{http://arXiv.org/abs/0807.3797}{{arXiv:0807.3797}}].

\bibitem{Chesler:2013lia}
P.~M. Chesler and L.~G. Yaffe, {\it {Numerical solution of gravitational
  dynamics in asymptotically anti-de Sitter spacetimes}},
  \beforedoihref\href{http://dx.doi.org/10.1007/JHEP07(2014)086}{JHEP}\afterdoihref\
  {\bf 07} (2014) 086
  [\href{http://arXiv.org/abs/1309.1439}{{arXiv:1309.1439}}].

\bibitem{Janik:2006gp}
R.~A. Janik and R.~B. Peschanski, {\it {Gauge/gravity duality and
  thermalization of a boost-invariant perfect fluid}},
  \beforedoihref\href{http://dx.doi.org/10.1103/PhysRevD.74.046007}{Phys.
  Rev.}\afterdoihref\  {\bf D74} (2006) 046007
  [\href{http://arXiv.org/abs/hep-th/0606149}{{arXiv:hep-th/0606149}}].

\bibitem{Janik:2015waa}
R.~A. Janik, G.~Plewa, H.~Soltanpanahi and M.~Spali{\'n}ski, {\it {Linearized
  nonequilibrium dynamics in nonconformal plasma}},
  \beforedoihref\href{http://dx.doi.org/10.1103/PhysRevD.91.126013}{Phys.
  Rev.}\afterdoihref\  {\bf D91} (2015), no.~12 126013
  [\href{http://arXiv.org/abs/1503.07149}{{arXiv:1503.07149}}].

\bibitem{Kovtun:2005ev}
P.~K. Kovtun and A.~O. Starinets, {\it {Quasinormal modes and holography}},
  \beforedoihref\href{http://dx.doi.org/10.1103/PhysRevD.72.086009}{Phys.Rev.}\afterdoihref\
  {\bf D72} (2005) 086009
  [\href{http://arXiv.org/abs/hep-th/0506184}{{arXiv:hep-th/0506184}}].

\bibitem{Grandclement:2007sb}
P.~Grandclement and J.~Novak, {\it {Spectral methods for numerical
  relativity}},
  \beforedoihref\href{http://dx.doi.org/10.12942/lrr-2009-1}{Living Rev.
  Rel.}\afterdoihref\  {\bf 12} (2009) 1
  [\href{http://arXiv.org/abs/0706.2286}{{arXiv:0706.2286}}].

\bibitem{deHaro:2000vlm}
S.~de~Haro, S.~N. Solodukhin and K.~Skenderis, {\it {Holographic reconstruction
  of space-time and renormalization in the AdS / CFT correspondence}},
  \beforedoihref\href{http://dx.doi.org/10.1007/s002200100381}{Commun. Math.
  Phys.}\afterdoihref\  {\bf 217} (2001) 595--622
  [\href{http://arXiv.org/abs/hep-th/0002230}{{arXiv:hep-th/0002230}}].

\bibitem{Sauzin:2016fxp}
D.~Sauzin, {\it {Introduction to 1-Summability and Resurgence, in “Divergent
  Series, Summability and Resurgence I"}},
  \beforedoihref\href{http://dx.doi.org/10.1007/978-3-319-28736-2}{Lect. Notes
  Math.}\afterdoihref\  {\bf 2153} (2016)
  [\href{http://arXiv.org/abs/1405.0356}{{arXiv:1405.0356}}].

\bibitem{Jentschura:2000fm}
U.~D. Jentschura and G.~Soff, {\it {Improved conformal mapping of the Borel
  plane}},
  \beforedoihref\href{http://dx.doi.org/10.1088/0305-4470/34/7/316}{J.
  Phys.}\afterdoihref\  {\bf A34} (2001) 1451--1457
  [\href{http://arXiv.org/abs/hep-ph/0006089}{{arXiv:hep-ph/0006089}}].

\bibitem{ZinnJustin:2004cg}
J.~Zinn-Justin and U.~D. Jentschura, {\it {Multi-instantons and exact results
  II: Specific cases, higher-order effects, and numerical calculations}},
  \beforedoihref\href{http://dx.doi.org/10.1016/j.aop.2004.04.003}{Annals
  Phys.}\afterdoihref\  {\bf 313} (2004) 269--325
  [\href{http://arXiv.org/abs/quant-ph/0501137}{{arXiv:quant-ph/0501137}}].

\bibitem{Caliceti:2007ra}
E.~Caliceti, M.~Meyer-Hermann, P.~Ribeca, A.~Surzhykov and U.~D. Jentschura,
  {\it {From useful algorithms for slowly convergent series to physical
  predictions based on divergent perturbative expansions}},
  \beforedoihref\href{http://dx.doi.org/10.1016/j.physrep.2007.03.003}{Phys.
  Rept.}\afterdoihref\  {\bf 446} (2007) 1--96
  [\href{http://arXiv.org/abs/0707.1596}{{arXiv:0707.1596}}].

\bibitem{Costin:2017ziv}
O.~Costin and G.~V. Dunne, {\it {Convergence from Divergence}},
  \beforedoihref\href{http://dx.doi.org/10.1088/1751-8121/aa9e30}{J.
  Phys.}\afterdoihref\  {\bf A51} (2018), no.~4 04LT01
  [\href{http://arXiv.org/abs/1705.09687}{{arXiv:1705.09687}}].

\bibitem{Costin:upcoming}
O.~Costin and G.~Dunne, {\it \textnormal{in preparation}}, .

\bibitem{Spalinski:2018mqg}
M.~Spali{\'n}ski, {\it {Universal behaviour, transients and attractors in
  supersymmetric Yang–Mills plasma}},
  \beforedoihref\href{http://dx.doi.org/10.1016/j.physletb.2018.07.003}{Phys.
  Lett.}\afterdoihref\  {\bf B784} (2018) 21--25
  [\href{http://arXiv.org/abs/1805.11689}{{arXiv:1805.11689}}].

\bibitem{Heller:2018qvh}
M.~P. Heller and V.~Svensson, {\it {How does relativistic kinetic theory
  remember about initial conditions?}},
  \beforedoihref\href{http://dx.doi.org/10.1103/PhysRevD.98.054016}{Phys.
  Rev.}\afterdoihref\  {\bf D98} (2018), no.~5 054016
  [\href{http://arXiv.org/abs/1802.08225}{{arXiv:1802.08225}}].

\bibitem{Cherman:2014xia}
A.~Cherman, P.~Koroteev and M.~{\"U}nsal, {\it {Resurgence and Holomorphy: From
  Weak to Strong Coupling}},
  \beforedoihref\href{http://dx.doi.org/10.1063/1.4921155}{J. Math.
  Phys.}\afterdoihref\  {\bf 56} (2015), no.~5 053505
  [\href{http://arXiv.org/abs/1410.0388}{{arXiv:1410.0388}}].

\bibitem{Grassi:2014cla}
A.~Grassi, M.~Mari{\~n}o and S.~Zakany, {\it {Resumming the string perturbation
  series}},
  \beforedoihref\href{http://dx.doi.org/10.1007/JHEP05(2015)038}{JHEP}\afterdoihref\
  {\bf 05} (2015) 038
  [\href{http://arXiv.org/abs/1405.4214}{{arXiv:1405.4214}}].

\bibitem{Couso-Santamaria:2015wga}
R.~Couso-Santamaría, R.~Schiappa and R.~Vaz, {\it {Finite N from Resurgent
  Large N}},
  \beforedoihref\href{http://dx.doi.org/10.1016/j.aop.2015.02.019}{Annals
  Phys.}\afterdoihref\  {\bf 356} (2015) 1--28
  [\href{http://arXiv.org/abs/1501.01007}{{arXiv:1501.01007}}].

\bibitem{Couso-Santamaria:2016vwq}
R.~Couso-Santamaría, M.~Mari{\~n}o and R.~Schiappa, {\it {Resurgence Matches
  Quantization}},
  \beforedoihref\href{http://dx.doi.org/10.1088/1751-8121/aa5e01}{J.
  Phys.}\afterdoihref\  {\bf A50} (2017), no.~14 145402
  [\href{http://arXiv.org/abs/1610.06782}{{arXiv:1610.06782}}].

\bibitem{Costin:1995ea}
O.~Costin, {\it {Exponential Asymptotics, Transseries, and Generalized Borel
  Summation for Analytic Rank One Systems of ODE’s}},  Inter. Math. Res.
  Notices {\bf 8} (1995) 377
  [\href{http://arXiv.org/abs/math/0608414}{{arXiv:math/0608414}}].

\bibitem{Costin:1998bs}
O.~Costin, {\it {On Borel Summation and Stokes Phenomena for Rank-1 Nonlinear
  Systems of Ordinary Differential Equations}},  Duke Math. J. {\bf 93} (1998)
  289 [\href{http://arXiv.org/abs/math/0608408}{{arXiv:math/0608408}}].

\bibitem{Aniceto:toapp}
I.~Aniceto, R.~Schiappa and M.~Vonk, {\it Painlev\'e resurgent transseries},
  \textnormal{to appear} (2018)
  (\texttt{http://online.kitp.ucsb.edu/online/resurgent$\_$c17/vonk}).

\bibitem{Bhattacharyya:2008jc}
S.~Bhattacharyya, V.~E. Hubeny, S.~Minwalla and M.~Rangamani, {\it {Nonlinear
  Fluid Dynamics from Gravity}},
  \beforedoihref\href{http://dx.doi.org/10.1088/1126-6708/2008/02/045}{JHEP}\afterdoihref\
  {\bf 0802} (2008) 045
  [\href{http://arXiv.org/abs/0712.2456}{{arXiv:0712.2456}}].

\end{thebibliography}\endgroup
\bibliographystyle{my-JHEP-4}

\end{document}